\documentclass[a4paper,11pt]{article}
\usepackage{jheppub} % for details on the use of the package, please see the JINST-author-manual
\usepackage{lineno}
\usepackage{amsfonts}
\usepackage{physics,amsmath}
\usepackage{amssymb}
\usepackage{mathtools}
\usepackage{colortbl}
\usepackage{here}
\usepackage{hyperref}
\usepackage{mathrsfs}
\usepackage{algpseudocode}
\usepackage{xcolor}
\usepackage[normalem]{ulem}
\usepackage{comment}
\usepackage{amsthm}
\usepackage{booktabs}
\usepackage{bbold}
\usepackage{comment}
\usepackage{subcaption}
\usepackage{relsize}
\usepackage{tikz}
\usepackage{pgfplots}
\usepackage{float}
\usetikzlibrary{arrows.meta,decorations.pathmorphing,decorations.markings}

\definecolor{fieldgreen}{RGB}{20,100,40}
\definecolor{photonred}{RGB}{220,60,40}
\definecolor{polorange}{RGB}{230,120,40}
\definecolor{polyellow}{RGB}{245,175,60}
\definecolor{photonfill}{RGB}{240,225,235}
\definecolor{sourcefill}{RGB}{225,240,230}
\definecolor{vacred}{RGB}{210,20,20}

\theoremstyle{plain}        % italic body, bold head — for core results
\newtheorem{theorem}{Theorem}
\theoremstyle{definition}   % upright body, bold head — for definitions, examples
\newtheorem{definition}{Definition}[section]

\theoremstyle{remark}       % upright body, italic head — for remarks, notes
\newcommand{\cs}{c_{s}}
\newcommand{\om}{\omega}
\newcommand{\tR}{\tilde{R}}

\title{\boldmath Microcausality Bounds on Dispersion Laws} 

\author[a, b, c]{Paolo Creminelli,}
\author[c, d]{Michael Mancini,}
\author[c, d]{Warin Patrick McBlain}

\affiliation[a]{ICTP, International Centre for Theoretical Physics, Strada Costiera 11, 34151, Trieste, Italy}
\affiliation[b]{IFPU - Institute for Fundamental Physics of the Universe, Via Beirut 2, 34014, Trieste, Italy}
\affiliation[c]{INFN, Sezione di Trieste, via Valerio 2, 34127 Trieste, Italy}
\affiliation[d]{SISSA, International School for Advanced Studies, via Bonomea, 265, 34136 Trieste, Italy}

\emailAdd{creminel@ictp.it}
\emailAdd{mmancini@sissa.it}
\emailAdd{wmcblain@sissa.it}

\abstract{Microcausality holds in any state, including those that spontaneously break Lorentz invariance, and requires the poles (and other singularities)  of the retarded Green's function to satisfy $\mathrm{Im}\,\omega \le |\mathrm{Im}\,k|$ throughout the complex plane. Building on the work of Heller, Serantes, Spali\'nski and Withers, we study its consequences for dispersion relations that are linear at small $k$, like Goldstone bosons, assuming tree-level, non-dissipative dynamics. We show that microcausality is equivalent to requiring that $k \pm \omega(k)$ be typically real functions. This yields two-sided bounds on all higher-order coefficients of $\omega(k)$ in units of its radius of convergence $R$, the scale at which the linear Goldstone description breaks down. The optimal constraints are characterised by an infinite hierarchy of Toeplitz determinants. We derive analogous bounds on $\omega^2(k)$, whose radius of convergence $\tilde R$ marks the scale at which new states must appear. The bounds shrink to zero as the sound speed $c_s$ approaches 1: for a fixed higher-derivative correction, both $R$ and $\tilde R$ become parametrically smaller than naive dimensional analysis suggests. In particular, in the EFT of Inflation all higher-derivative quadratic operators must vanish as $c_s \to 1$. We test the bounds in several models: a $U(1)$ superfluid, light in a dielectric medium, photon propagation in a strong magnetic field, colour-flavour-locked quark matter, and a homogeneous solid.  Finally, we extend the bounds to gapped modes.}

\begin{document}
\maketitle
\flushbottom

\section{Introduction} \label{sec:intro}

Our understanding of quantum field theory (QFT) relies to a large extent on Lorentz invariance. In most physical systems, however, Lorentz invariance is spontaneously broken by the state. In cosmology there is always a preferred frame, the one in which matter is at rest; in condensed matter, and more generally in any QFT at finite temperature or chemical potential, the medium defines a preferred frame. We stress that we are concerned with the \emph{spontaneous} breaking of Lorentz invariance, and not in a fundamental breaking of this symmetry.

Without Lorentz invariance QFT is considerably richer, and correspondingly less understood. In this paper we address what is arguably the simplest question one can ask in this setting. We assume that boosts are broken, i.e.~there is a preferred frame, and we study the dispersion relation $\omega(k)$ of a weakly coupled excitation, i.e.~a pole of the retarded Green's function. If Lorentz invariance were linearly realised, the answer would be unique: $\omega^2 = k^2 + m^2$. Without it, $\omega(k)$ is a priori an arbitrary function. Microcausality---the vanishing of commutators of local operators at spacelike separation---is a property of the operator algebra and holds whatever the state, including states that break Lorentz invariance~\cite{Dubovsky:2007ac, Cintia:2025fzn}. Our goal is to understand what constraints microcausality imposes on $\omega(k)$.

Microcausality is mapped to analyticity in Fourier space. In particular, as we will review, the poles of the Green's functions, whose location is described by the dispersion relation $\omega(k)$ must satisfy
\begin{equation}
    \Im\omega(k)\le|\Im k| \;. \label{eq:intro_mc}
\end{equation}
Heller, Serantes, Spali\'nski and Withers realised that this condition, imposed throughout the complex $k$-plane, is very constraining~\cite{Heller:2022ejw,Heller:2023jtd}. Applying it to hydrodynamic modes, they derived rigorous two-sided bounds on all the coefficients of their small-$k$ expansion, in units of its radius of convergence. 

In this paper we study the consequences of~\eqref{eq:intro_mc} for dispersion relations that are linear at small momentum,
\begin{equation}
    \omega(k)=\pm\Big(c_s k+\sum_{n=1}^\infty\alpha_{2n+1}k^{2n+1}\Big)\;,\qquad |k|<R \;, \label{eq:intro_series}
\end{equation}
with $R$ the radius of convergence. This applies to Goldstone bosons\footnote{When Lorentz is broken Goldstone bosons can also have dispersion relations that start quadratic in the momentum $\omega \propto k^2$, known as Type II Goldstone in the nomenclature of \cite{Nielsen:1975hm}. Here we focus on Type I, although the $\omega^2$-bounds can also be used for type II.} (phonons in superfluids and solids, the inflaton fluctuation in the effective field theory (EFT) of Inflation~\cite{Cheung:2007st}, the scalar mode of dark energy \cite{Creminelli:2008wc,Gubitosi:2012hu,Gleyzes:2014dya} \ldots) but also, for instance, to photons travelling in a medium. The coefficients $\alpha_{2n+1}$ are encoded in the higher-derivative operators of the EFT.  As in \cite{Heller:2022ejw,Heller:2023jtd}, we neglect loops in the EFT, so that at low energy the singularities of the Green's function are only poles and $\omega(k)$ is analytic in a disc around $k=0$. This is the case, for instance, when the UV completion is weakly coupled. Unlike \cite{Heller:2022ejw,Heller:2023jtd}, we restrict to non-dissipative dynamics and assume that $\omega(k)$ is real for real $k$.

In Section \ref{subsec:u1_superfluid}, to guide the discussion, we present a simple UV complete model of a $U(1)$ superfluid.  In Section~\ref{subsec:omega_bounds} we show that condition~\eqref{eq:intro_mc} is equivalent to the requirement that the two functions $k\pm\omega(k)$ be typically real~\cite{Rogosinski1932} in the disc $|k|<R$, i.e.~that their imaginary part has the same sign as $\Im k$. The Taylor coefficients of typically real functions obey universal bounds: each coefficient satisfies the Rogosinski bounds, while the complete set of constraints on the sequence of coefficients is characterised by an infinite hierarchy of Toeplitz determinants (see Appendix~\ref{sec:tr_series}). (Related tools from geometric function theory have recently found applications in the $S$-matrix bootstrap~\cite{Haldar:2021rri,Raman:2021pkf}.) The allowed region for the dimensionless coefficients $\alpha_{2n+1}R^{2n}$ is obtained by intersecting the constraints of the two functions $k\pm\omega(k)$, eq.~\eqref{eq:true_bounds}, and its projections, computed numerically, are shown in Figure~\ref{fig:rogo_toeplitz_bounds_a}.

Rather than as bounds on the coefficients at fixed $R$, these results should be seen as {\em upper bounds on $R$} once some coefficients of the series are known: beyond this bound the series cannot satisfy microcausality \eqref{eq:intro_mc}, so something must happen for $|k| \le R$ to save the day. (This is analogous to how unitarity bounds are used to estimate the scale of new physics.) In all the examples we study, $\omega(k)$ is microcausal up to the radius of convergence of the series. However, one can also consider situations in which microcausality, eq.~\eqref{eq:intro_mc}, holds only in a smaller disc. For instance, some approximation may break down, so that the system apparently loses microcausality before the radius of convergence is reached; or a pole may ``hide'' on a second sheet of the Green function and become unphysical (see the discussion in \cite{Gavassino:2026fil, Brants:2026qlw}). In these cases our results apply with $R$ the maximum radius within which microcausality holds. For simplicity we will keep calling $R$ the radius of convergence, but it should always be understood as the maximum radius for which the series \eqref{eq:intro_series} is microcausal.

The bounds become particularly interesting as $c_s\to1$ (Section~\ref{sec:cs_to_1}), where the allowed region shrinks to zero linearly in $1-c_s$:
\begin{equation}
    -(2n+1)(1-c_s)\le\alpha_{2n+1}R^{2n}\le\kappa_{2n+1}(1-c_s)\;, \label{eq:intro_csto1}
\end{equation}
with $\kappa_{2n+1}>0$ defined in eq.~\eqref{eq:kappa_maintext}.  That microcausality forbids any higher-order correction for $c_s=1$ was shown in~\cite{Heller:2022ejw}, and in~\cite{Hui:2025aja} it was shown that a dispersion relation at order $k^3$ is microcausal only in a window of momenta that shrinks as $c_s\to1$.

The scale $R$ has a transparent physical meaning: it is the momentum at which the description in terms of a linear dispersion relation plus small corrections breaks down. This does not necessarily require new states. In the $U(1)$ superfluid of Section~\ref{subsec:u1_superfluid}, for $c_s<1/\sqrt 2$, $R$ is set by the collision of the two Goldstone branches $\pm\omega(k)$, a branch point of $\sqrt{\omega^2}$. In physical terms, $1/R$ is the so-called healing length of the superfluid, above which the Bogoliubov spectrum crosses over to the non-relativistic one, $\omega\simeq k^2/2M$. New degrees of freedom appear instead at the radius of convergence of $\omega^2(k)$ that we denote $\tilde R$.
%\MM{Here we can discuss on what we mean with new dof. From what I understood $\tR$ is the scale where $\om^2$ mixes with other  modes, which however can also be light modes already encoded in the EFT. So from the EFT point of view there is no new physics entering, but it's just that different poles of the effective G collide.} 
This shows that the scales $R$ and $\tilde R$ have a different physical meaning.

Therefore, in Section~\ref{sec:omega2_bounds} we derive analogous bounds on the coefficients $\tilde\alpha_{2n}$ of $\omega^2(k)$, in units of $\tilde R$. In this case the microcausality condition is not linear in the function, and we only obtain necessary conditions: a $c_s$-independent bound, $|\tilde\alpha_{2n}|\tilde R^{2(n-1)}\le1$ (Section~\ref{sec:cs independent omega2}), and a $c_s$-dependent one, which follows from the typical-realness of $(k^2-\omega^2)/k$ and again shrinks to zero as $c_s\to1$ (Section~\ref{sec:cs dependent omega2}). Together, our bounds, once a higher-derivative term is known, allow to set rigorous upper bounds on the two scales $R$ and $\tilde R$.
%\MM{After writing the paragraph about $R_M$ remember to say that if $\min[R_M, \tR] = R_M$ our bounds put an upper constraint on the radius of microcausality.}

In Section~\ref{sec:tests} we test the bounds in several models with spontaneously broken Lorentz invariance: light in a medium described by a Drude-Lorentz model; a photon in a strong homogeneous magnetic field, whose radius of convergence is set by the pair-production threshold; the octet and singlet Goldstones of colour-flavour-locked quark matter, for which the radius of convergence is determined numerically with the Domb-Sykes method (Appendix~\ref{app:domb_sykes}). A model of a homogeneous solid is studied in Appendix~\ref{app:solid}.

The implications for cosmology are briefly discussed in Section~\ref{sec:cosmo}. In particular, the bounds \eqref{eq:intro_csto1} imply that, at the level of the quadratic action, the EFT of Inflation is forced to become Lorentz invariant as $c_s\to1$, as conjectured in~\cite{Baumann:2015nta}. The same conclusions apply to the EFT of Dark Energy. In Section~\ref{sec:mass_terms} we extend the bounds on $\omega^2(k)$ to gapped modes. We conclude in Section~\ref{sec:conclusions} with many open directions.

\section{Microcausality bounds}
\label{sec:microcausality bounds}
Our starting point is the retarded two-point function: given a bosonic, local, and gauge-invariant operator $\mathcal{O}$, the retarded Green's function is defined as 
\begin{equation}
    G(t, \vec{x}; t', \vec{x}') = i\theta(t-t') \langle[\mathcal{O}(t, \vec{x}), \mathcal{O}(t', \vec{x}')]\rangle \label{eq:greenfuncspace}.
\end{equation}
As evident from \eqref{eq:greenfuncspace}, the retarded two-point function has support in the future light cone i.e.~$t \geq |\vec{x}|$. Although Lorentz symmetry is spontaneously broken in the theories that we will consider in this paper, the notion of microcausality still exists. Indeed, microcausality is an operatorial statement and thus is valid for any state (or density matrix), including Lorentz-breaking ones (see, for example, \cite{Dubovsky:2007ac}).

For states with time-translational invariance\footnote{The unbroken time-translations can be a diagonal subgroup of the original time-translations and some internal symmetry: this will happen in the superfluid example below.} and spatial homogeneity, the retarded two-point function will depend only on the differences $t - t'$ and $\vec{x} - \vec{x}'$, and we can work with the retarded two-point function in Fourier space:
\begin{equation}
    \Tilde{G}(\omega, \vec{k}) = \int \mathrm{d}^d x ~G(0;t, \vec{x})e^{i(\omega t - \vec{k}\cdot \vec{x})} \label{eq:greenfuncmom}.
\end{equation}
Under the assumption that the operator $\mathcal{O}$ is Hermitian, $G(t,\vec{x})$ will be real, and its Fourier transform $\Tilde{G}(\omega, \vec{k})$ satisfies the reality condition: 
\begin{equation}
    \Tilde{G}^*(\omega, \vec{k}) = \Tilde{G}(-\omega^*, -\vec{k}^*). \label{eq:green_real}
\end{equation}
Taking $\omega = \omega_R + i \omega_I$ and $\vec{k} = \vec{k}_R + i \vec{k}_I$, \eqref{eq:greenfuncmom} becomes
\begin{equation}
    \Tilde{G}(\omega, \vec{k}) = \int \mathrm{d}^d x ~G(0; t, \vec{x})e^{i(\omega_R t - \vec{k}_R\cdot \vec{x})}e^{-(\omega_I t - \vec{k}_I \cdot \vec{x})} \label{eq:greenfuncisitanal}
\end{equation}
and, given the support of $G(t,\vec{x})$ and the assumption that $G(t,\vec{x})$ is polynomially bounded\footnote{This excludes unstable systems whose $G(t,\vec x)$ grows exponentially in time.}, \eqref{eq:greenfuncisitanal} converges in the region
\begin{equation}
    \Im \omega > |\Im \vec{k}|. \label{eq:analyticity_vector}
\end{equation}
As a consequence, we can conclude that $\Tilde{G}(\omega, \vec{k})$ must be analytic within this region.

For simplicity, we further restrict to systems with isotropy, although many of our results can be straightforwardly generalised to the anisotropic case. Due to rotational symmetry, $G(t,\vec{x})$ will only depend on $\vec{x}$ through $x^2$ and, accordingly, $\Tilde{G}(\omega, \vec{k})$ will only depend on $\vec{k}$ through $k^2$. Thus one can reduce to study the retarded two-point function in terms of the two complex variables $\omega$ and\footnote{Astute readers might raise concerns that in general $\vec{k}_R$ and $\vec{k}_I$ need not align with each other. However, given a point $(\omega, \vec{k})$ in the region \eqref{eq:analyticity_vector}, one can always find another point in \eqref{eq:analyticity_vector} with the same $\vec{k}^2$ such that $\vec{k}_R$ and $\vec{k}_I$ align. See Footnote 13 of \cite{Creminelli:2025rxj} and Appendix B of \cite{Hui:2025aja} for details.} $k$ with an additional symmetry:
\begin{equation}
    \Tilde{G}(\omega, -k) = \Tilde{G}(\omega, k). \label{eq:green_rot_inv}
\end{equation}
Following \eqref{eq:analyticity_vector}, $\Tilde{G}(\omega, k)$ will be analytic within the region
\begin{equation}
    \Im \omega > |\Im k|. \label{eq:analyticity}
\end{equation}
A more general and rigorous discussion on the properties of the retarded two-point function when boosts are broken can be found in \cite{Creminelli:2025rxj}.

\subsection{Goldstone dispersion relation} \label{subsec:dispersion_rel}
In the remainder of this paper we will make the assumption that in the EFT loops of the Goldstone can be neglected\footnote{Including loops, the coefficients of the operators that we will constrain below start running. Notice that, in a theory with derivative interactions, an operator of the form $(\partial^2\pi)^2$ does not run: only operators with more derivatives do. See the explicit calculation in \cite{Hui:2025aja}.}. In this limit, at low energy the singularities of $\Tilde{G}(\omega, k)$ are simply poles whose location is given by the function $\omega(k)$. This assumption is valid when the UV theory is weakly coupled (or when there is some large $N$ expansion). This assumption is analogous to the limit in which the (Lorentz-invariant) $S$-matrix reduces to a polynomial: see for example \cite{Bellazzini:2020cot}. The reality condition \eqref{eq:green_real}, together with rotational invariance, implies that if one has a pole at $\omega(k)$ with $k$ real, one has also a pole at $-\omega^*(k)$.

We assume $\Im\,\omega (k)= 0$ for real $k$, i.e.~no dissipation. Dissipation would describe the presence of additional degrees of freedom, into which the Goldstone can decay, like in the case of hydrodynamics modes studied in \cite{Heller:2022ejw,Heller:2023jtd}. Without dissipation, for a given pole at $\omega(k)$ with $k$ real, one has also a pole at $-\omega(k)$.

Within the tree-level EFT, the dispersion relation $\omega(k)$ will be analytic in an open disc of radius\footnote{We should remind the reader that $R$, despite its name, has units of momentum.} $R$—the radius of convergence of $\omega(k)$—centred at $k = 0$. One has
\begin{equation}
    \omega(k) = \pm \left(c_s k + \sum_{n = 1}^{\infty} \alpha_{2n+1} k^{2n+1}\right), \qquad  |k|< R \label{eq:omega_series_odd}
\end{equation}
with $c_s$ being the sound speed of the Goldstone.   Each coefficient $\alpha_{2n+1}$ is dimensionful and can be expressed in units of $R$.

Given that $\Tilde{G}(\omega, k)$ is analytic within the region in \eqref{eq:analyticity}, poles of $\Tilde{G}(\omega, k)$ must lie within
\begin{equation}
    \Im \omega(k) \leq |\Im k|. \label{eq:bound_omega}
\end{equation}
In the rest of the paper we are going to study the consequences of this microcausality condition on the dispersion relation $\omega(k)$. One might suspect that this boils down to requiring the subluminality of the group velocity\footnote{Notice that there is a sign ambiguity in $v_g$ given that the dispersion relation is of the form $\pm \omega(k)$.}
\begin{equation}
    |v_g(k)| \equiv \left|\frac{\mathrm{d}\omega}{\mathrm{d}k} \right| \leq 1 \;. \label{eq:vgdef}
\end{equation}
As nicely discussed in \cite{Hui:2025aja}, this is, of course, a consequence of microcausality, but it does not contain all the information of \eqref{eq:bound_omega}. For a non-dissipative system, the dispersion relation $\omega(k)$ is real for real $k$. When $k$ is close to the real axis
\begin{equation}
    \Im \omega(k) \approx \frac{\mathrm{d}\omega}{\mathrm{d}k} \Im k.
\end{equation}
Using \eqref{eq:bound_omega} we recover \eqref{eq:vgdef}.
This implies in particular that the speed of sound $c_s$ must be subluminal, $c_s \le 1$. The subluminality of the group velocity is just a consequence of \eqref{eq:bound_omega} close to the real axis and does not guarantee the validity of microcausality in the complex plane. As we will see, microcausality is more than subluminal group velocity.

\subsection{A case study: the \texorpdfstring{$U(1)$}{U(1)} superfluid}\label{subsec:u1_superfluid}
Before proceeding it is useful to present an explicit model; several others will be discussed below. This will also shed light on the nature of the singularities of the function $\omega(k)$. 

Let us consider a renormalizable\footnote{The theory has a Landau pole at exponentially high energies, but this will not play a role since we will discuss this theory at tree level.} theory of a complex scalar $\Phi$ with global $U(1)$ symmetry (see, for example, Refs.~\cite{Creminelli:2023kze, Hui:2023pxc}):
\begin{equation}
    S[\Phi] = \int \mathrm{d}^4 x \left[-|\partial \Phi|^2 - M^2 |\Phi|^2 - \lambda |\Phi|^4 \right]. \label{eq:u1_action}
\end{equation}
The complex scalar $\Phi$ can be written in polar form as
\begin{equation}
    \Phi(x) = \frac{\rho(x)}{\sqrt{2}} e^{i\theta(x)/v},
\end{equation}
in which $\rho(x)$ is the radial gapped mode, $\theta(x)$ is the gapless Goldstone, and $v$ is the vacuum expectation value of $\rho(x)$. Next, we introduce a chemical potential $\mu$ through a time-dependent background $\theta(x) = v \mu t + \pi(x)$. We are interested in a state where the $U(1)$ is spontaneously broken.  For $M^2$ positive one needs a chemical potential, with $\mu^2 > M^2$, to have spontaneous symmetry breaking and give rise to the Goldstone. This actually describes a realistic weakly coupled superfluid, see for instance \cite{Esposito:2026diu}. One can also consider the case $M^2<0$ and in this case the $U(1)$ is spontaneously broken also for $\mu =0$.\footnote{This is the case studied in \cite{Creminelli:2023kze, Hui:2023pxc} to avoid discontinuities for $\mu \to 0$. For $M^2 < 0$, $c_s$ ranges from 1 ($\mu = 0$) to $1/\sqrt{3}$ ($\mu \to \infty$). For $M^2 > 0$, $c_s$ ranges from $1/\sqrt{3}$ to 0.  We thank A. Esposito for discussions on this point.} 

\begin{figure}
    \centering
    \begin{tikzpicture}
  \begin{axis}[
      width=10cm, height=8cm,
      domain=0:5, samples=200, smooth,
      xmin=0, xmax=3.4, ymin=0, ymax=3.4,
      axis lines=left,
      axis line style={-{Triangle[length=2.5mm]}, thick},
      xtick={0,1,2,3},
      ytick={0,1,2,3},
      xlabel={$k/\Delta$},
      ylabel={$\omega/\Delta$},
      xlabel style={at={(ticklabel* cs:1)}, anchor=west},
      ylabel style={at={(ticklabel* cs:1)}, anchor=south, rotate=-90},
      every axis plot/.append style={very thick},
    ]
    % omega_+ (red)
    \addplot[red] {sqrt(x^2 + 0.5*(1 + sqrt(1 + 4*(1-(0.5)^2)*x^2)))};
    % omega_- (blue)
    \addplot[blue] {sqrt(x^2 + 0.5*(1 - sqrt(1 + 4*(1-(0.5)^2)*x^2)))};
 
    \node[red,  above left] at (axis cs:2.7,3.2) {$\omega_+$};
    \node[blue, below right] at (axis cs:2.9,2.3) {$\omega_-$};
  \end{axis}
\end{tikzpicture}
    \caption{Plot of \eqref{eq:dispersion_u1} for gapped ($+$, in red) and gapless ($-$, in blue) modes. Here, $c_s = 0.5$.}
    \label{fig:dispersion_u1}
\end{figure}
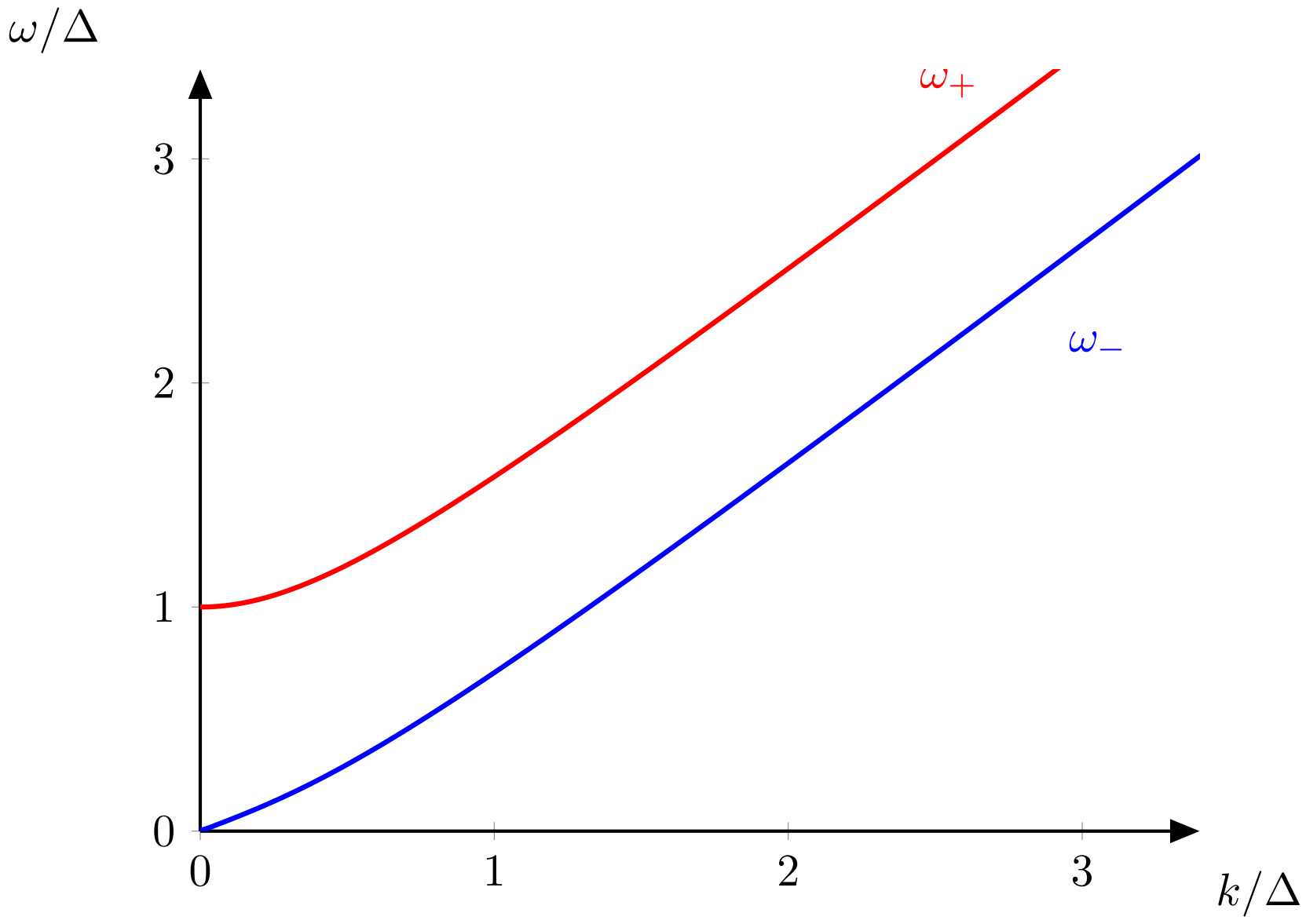

In both cases, the dispersion relations for the Goldstone and the radial mode are given by\footnote{Here and below, to avoid confusion we suppress the $\pm$ sign in front of the external square root.}

\begin{equation}
    \omega_\pm(k) = \sqrt{k^2 + \frac{1}{2}\Delta^2 \left(1 \pm \sqrt{1 + \frac{4(1 - c_s^2)k^2}{\Delta^2}}\right)} \label{eq:dispersion_u1}
\end{equation} 
in which $\Delta$ is the gap of the radial excitation ($+$) and $c_s$ is the sound speed of the Goldstone ($-$), with $0 \leq c_s \leq 1$  (see Figure \ref{fig:dispersion_u1}). These parameters are given by
\begin{equation}
    \Delta^2 = 2\left(-M^2 + 3\mu^2\right), \qquad c_s^2 = \frac{\mu^2 - M^2}{3 \mu^2 - M^2}.
    \label{eq:parameters_U1}
\end{equation}

Let us study the radius of convergence $R$ of the Goldstone dispersion relation $\omega_-(k)$. There are two sources of singularities. The first one are the branch points from the inner square root, located at
\begin{equation}
    1 + \frac{4(1 - c_s^2)k^2}{\Delta^2} = 0 \quad \implies \quad k_* = \pm i \frac{\Delta}{2\sqrt{1 - c_s^2}}~.\label{eq:bp_u1_om2}
\end{equation}
These branch points exist for all values of $c_s$, receding towards infinity as $c_s\to 1$. When $|k| < |k_*|$, the dispersion relations of the two modes $\omega^2_\pm$ are separately analytic. However, at $k_*$ the two modes $\omega^2_\pm$ `collide' with each other (see the discussion in \cite{Hui:2023pxc}). At $k_*$ the description in terms of a single mode breaks down and new states must be included.
The second source of singularities is the outer square root of \eqref{eq:dispersion_u1}
\begin{equation}
    k^2 + \frac{1}{2}\Delta^2 \left(1 - \sqrt{1 + \frac{4(1 - c_s^2)k^2}{\Delta^2}}\right) = 0 \quad \implies \quad  k_* = \pm i c_s \Delta~. \label{eq:Deltacsdef}
\end{equation}
One can check that these branch points exist only when $0 \leq c_s \leq 1/\sqrt{2}$. In this case, the two Goldstone poles `collide'.

To better understand the physical meaning of this singularity, it is useful to expand the Goldstone dispersion relation at small $k$. 
\begin{equation}
    \omega^2_-(k) \approx c_s^2 k^2 \left[1 + \frac{(1-c_s^2)^2 k^2}{c_s^2 \Delta^2} - \frac{2 (1-c_s^2)^3 k^4}{c_s^2 \Delta^4} + \dots \right].
    \label{eq:non_linear_corrections}
\end{equation}
Let us focus on the non-relativistic case $c_s \ll 1$. The first two terms become comparable at $k = c_s \Delta$, while the rest of the series is still negligible. This scale corresponds to $R$, since a branch point will arise from the external square root. At this scale one cannot treat the Goldstone as a linear dispersion relation plus small correction. Indeed the first two terms of the series give the celebrated Bogoliubov spectrum of superfluids: for $k \ll R$ one has the Goldstone, while for $k \gg R$ one has the non-relativistic spectrum $\omega = k^2/2M$ (notice that $\Delta = 2 M$ in the non-relativistic regime, $\mu - M \ll M$, see eq.~\eqref{eq:parameters_U1}). The scale $1/R$ is called healing length (or correlation length) for a superfluid, see e.g.~\cite{Esposito:2026diu}. Notice that both below and above $R$ one has a single degree of freedom: no new states appear, but the linear description of the Goldstone breaks down. If we go to higher momenta then the $\omega^2$ series \eqref{eq:non_linear_corrections} breaks down, because of the singularities \eqref{eq:bp_u1_om2}. At this scale one has to consider additional degrees of freedom. 

It is clear that the radius of convergence of $\omega(k)$ and $\omega^2(k)$ have in general a different physical meaning. We will call $\tilde R$ the radius of convergence of the latter.

\begin{figure}[h]
    \centering
    \begin{tikzpicture}[x=5cm, y=3.5cm, >=latex]
 
\def\xmax{1.15}
\def\ymax{1.65}
\def\xcut{0.95}   % where the top curve is cut off (near-vertical rise)
 
% axes
\draw[->, thick] (0,0) -- (\xmax,0) node[right] {$c_s$};
\draw[->, thick] (0,0) -- (0,\ymax) node[above left] {$R/\Delta, \Tilde{R}/\Delta$};
 
% dashed guide lines to the crossing point (1/sqrt2, 1/sqrt2)
\draw[gray, dashed] (0,{1/sqrt(2)}) -- ({1/sqrt(2)},{1/sqrt(2)}) -- ({1/sqrt(2)},0);
 
% bottom line: R/Delta = c_s, drawn only for 0 <= c_s <= 1/sqrt2
\draw[very thick] (0,0) -- ({1/sqrt(2)},{1/sqrt(2)});
 
% top curve: R/Delta = 1/(2 sqrt(1-c_s^2))
% solid for 1/sqrt2 <= c_s <= xcut
\draw[very thick] plot[domain={1/sqrt(2)}:\xcut, samples=60] (\x,{1/(2*sqrt(1-\x*\x))});
% long-dashed for 0 <= c_s <= 1/sqrt2
\draw[very thick, red ,dash pattern=on 9pt off 6pt] plot[domain=0:\xcut, samples=60] (\x,{1/(2*sqrt(1-\x*\x))});
 
% axis labels
\node[below left] at (0,0) {$0$};
\node[below] at (1,0) {$1$};
\node[below] at ({1/sqrt(2)},0) {$\frac{1}{\sqrt{2}}$};
\node[left] at (0,{1/sqrt(2)}) {$\frac{1}{\sqrt{2}}$};
\node[left] at (0,{1/2}) {$\frac{1}{2}$};
 
\end{tikzpicture}
    \caption{Radius of convergence of the gapless dispersion relation $\omega_-(k)$ (solid black) and its square $\omega_-^2(k)$ (dashed red), as a function of the sound speed $c_s$.}
    \label{fig:radius_u1}
\end{figure}
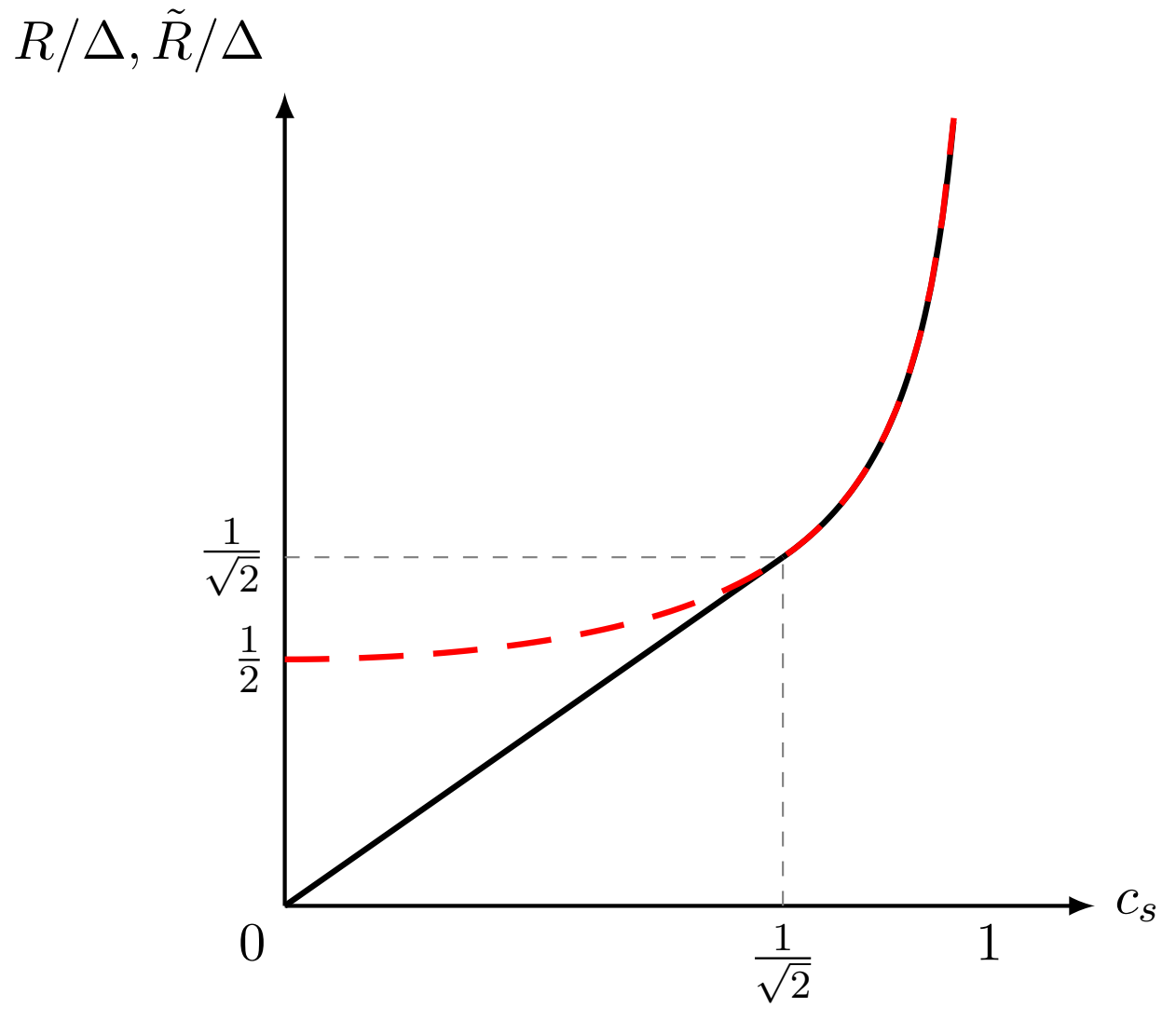

For the system at hand the radii of convergence are (see Figure \ref{fig:radius_u1}):
\begin{equation}
    R = \begin{cases}
        c_s \Delta, & 0 \le c_s < \frac{1}{\sqrt{2}}\\
        \frac{ \Delta }{2\sqrt{1-c_s^2}}, & \frac{1}{\sqrt{2}} \leq c_s \le 1
    \end{cases}; \qquad \tilde R = \frac{ \Delta }{2\sqrt{1-c_s^2}}, \quad 0 \le c_s \le 1.\label{eq:radius_u1}
\end{equation}
To summarise, there are two kind of singularities and associated radii of convergence:
\begin{enumerate}
    \item $\omega(k)$-singularities. These include the branch point singularities associated to taking $\sqrt{\omega^2}$. The associated radius of convergence is $R$. At $|k| > R$ the description in terms of a linear Goldstone breaks down.
    \item $\omega^2(k)$-singularities.  The radius of convergence dictated by this type of singularities will be denoted $\Tilde{R}$. They are associated with new modes appearing: at $|k| > \tilde R$, one has more degrees of freedom. 
\end{enumerate}
Notice that we describe singularities as `collisions' with other poles, either of the Goldstone itself or of additional modes.
%\MM{This sentence confuses me a bit, because we have to distinguish between singularities of G and of the dispersion laws describing the Goldstone mode or additional modes. $\tR$ for example is fixed by the collision of the Goldstone with another singularity of G, which can be either a branch point or a pole.} 
The reason is that it was proven in \cite{Heller:2022ejw}, with the additional assumption that $\tilde{G}(\omega,k)$ is meromorphic, that the only singularities compatible with microcausality are branch points (no poles or essential singularities are allowed): at the point the dispersion relation develops additional solutions. We prove an analogous statement for $\omega^2(k)$ in Appendix \ref{app:omega2}.

In the rest of the paper we will study the constraints imposed by microcausality of the series expansion of $\omega(k)$ and $\omega^2(k)$.

\subsection{Bounds on \texorpdfstring{$\omega(k)$}{omega(k)}}\label{subsec:omega_bounds}
\label{sec:bounds_omega}
\begin{figure}[H]
    \centering
\begin{tikzpicture}[x=2cm, y=2cm, >=latex]
 
\definecolor{myred}{RGB}{219,59,38}
\definecolor{myblue}{RGB}{50,116,181}
 
% filled half-disks
\fill[myred] (0,0) ++(0:1) arc[start angle=0, end angle=180, radius=1] -- cycle;
\fill[myblue] (0,0) ++(180:1) arc[start angle=180, end angle=360, radius=1] -- cycle;
 
% circle outline (dashed)
\draw[line width=1.2pt, dashed] (0,0) circle (1);
 
% axes (default thickness)
\draw[->] (-1.6,0) -- (1.6,0) node[right] {Re $z$};
\draw[->] (0,-1.6) -- (0,1.6) node[above] {Im $z$};
 
% |z|<1 label
\node at (1,1) {$|z|<1$};
 
\end{tikzpicture}
    \caption{A diagram showing $\Im f(z)$ within the unit disc, which is positive (in red) in the upper-half plane and negative (in blue) in the lower-half plane.}
    \label{fig:typireal}
\end{figure}
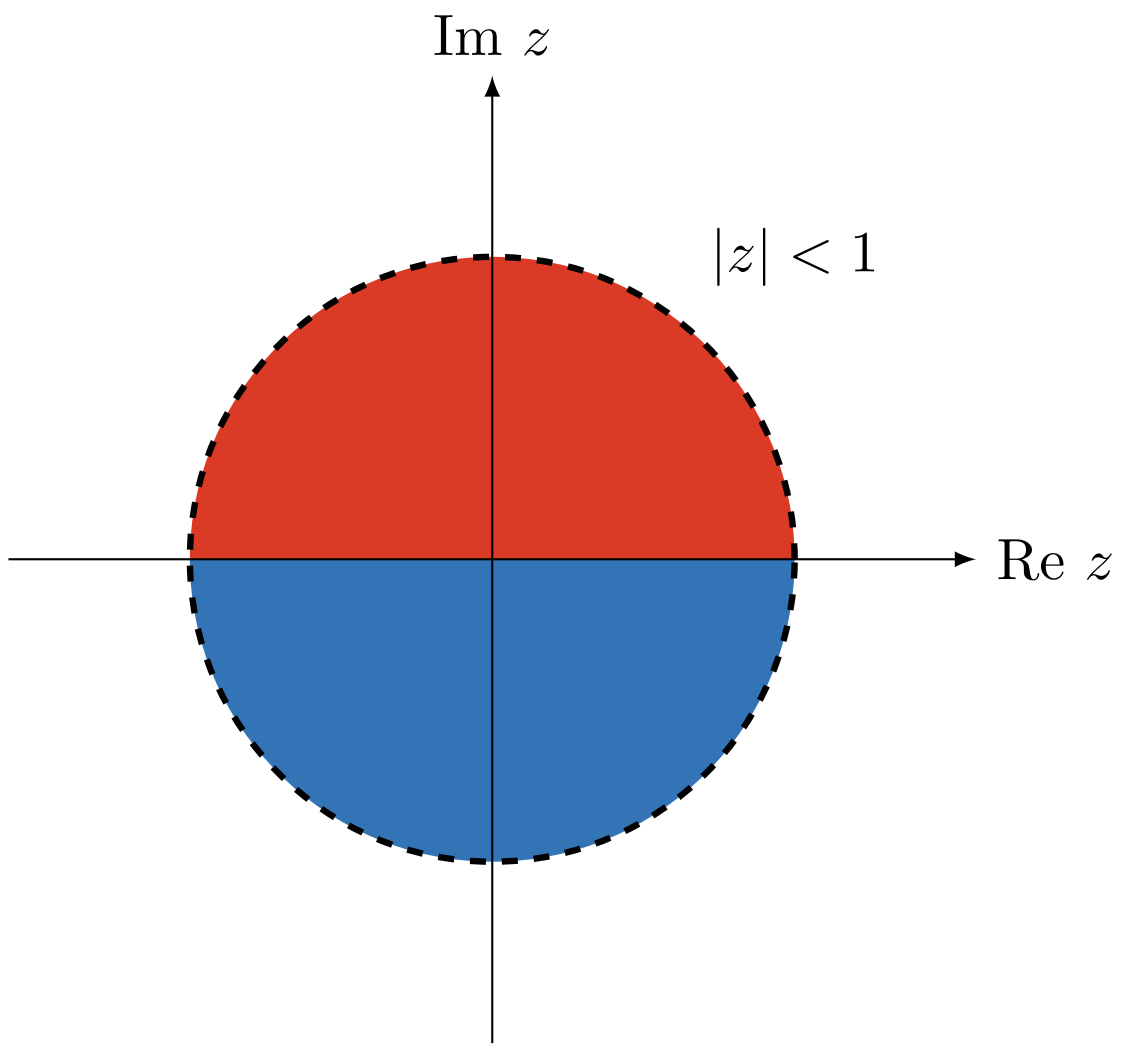

We will use microcausality to put bounds on $\omega(k)$ in units of $R$—the radius of convergence of $\omega(k)$.  Consider \eqref{eq:bound_omega}. When $\Im k > 0$ one has
\begin{equation}
    \pm\Im \omega \leq \Im k \quad \implies \quad \Im [k \mp \omega] \geq 0 \label{eq:dispersion_tr_-}
\end{equation}
while when $\Im k < 0$ one has
\begin{equation}
    \pm \Im \omega \leq -\Im k \quad \implies \quad \Im [k \pm \omega] \leq 0. \label{eq:dispersion_tr_+}
\end{equation}
From Eqs.\eqref{eq:dispersion_tr_-} and \eqref{eq:dispersion_tr_+} we can conclude that $\omega(k)$ satisfies microcausality if and only if both
\begin{equation}
    k + \omega(k) \qquad \text{and} \qquad k - \omega(k) 
    \label{eq:omega_pm}
\end{equation}
have imaginary part which is positive in the upper-half plane and negative in the lower-half plane. Functions with this property are known as typically real functions, first introduced by W.~Rogosinski in \cite{Rogosinski1932}. See Figure \ref{fig:typireal}.
\begin{definition}[Typically real functions]
    A function $f(z)$ of the complex variable $z$, satisfying $f(0) = 0, f'(0) \neq 0$ and analytic in the open unit disc $|z|< 1$ is said to be typically real if it is real valued along the real axis and has a positive imaginary part in the upper-half plane and a negative imaginary part in the lower-half plane:
\begin{equation}
    \Im f(z) \Im z > 0\quad\text{when}\quad\Im z \neq 0. \label{eq:tr_definition_maintext}
\end{equation}
\end{definition}
An important property of this class of functions is that the Taylor coefficients must satisfy universal bounds. We refer the reader to Appendix \ref{sec:tr_series} for all the relevant mathematical details, which we will summarise briefly in the main text when relevant. 
\subsubsection{Rogosinski bounds}
Consider a function $f(z)$ of the complex variable $z$ which is analytic in the open disc $|z|<1$ and odd in $z$. Its power series\footnote{Without loss of generality, we may normalise the leading coefficient in the power series i.e.~$b_1 = 1$.} is
\begin{equation}
    f(z) = z + \sum_{n = 1}^{\infty} b_{2n+1} z^{2n+1}. \label{eq:complex_series_maintext}
\end{equation}

\begin{theorem}[Rogosinski bounds]
    Let $f(z)$ be defined as in Eq.~\eqref{eq:complex_series_maintext}. If the function $f(z)$ is typically real, then the coefficients $b_{2n+1}$ are real and satisfy
    \begin{equation}
        -\kappa_{2n+1} \leq b_{2n+1} \leq 2n+1, \label{eq:rogosinski_bound_maintext}
    \end{equation}
    where
    \begin{equation}
    \kappa_{2n+1} \equiv -\min_{0 < \theta \leq \frac{\pi}{2}} \left[\frac{\sin (2n+1)\theta}{\sin\theta}\right] > 0. \label{eq:kappa_maintext}
\end{equation}
The interval \eqref{eq:rogosinski_bound_maintext} is optimal: if one chooses a particular $b_{2n+1}$ in the interval, there exists an odd typically real function with that coefficient. 
\end{theorem}
Going back to our dispersion relation: we insert \eqref{eq:omega_series_odd} into \eqref{eq:omega_pm} and normalise the coefficient of the first term to obtain
\begin{equation}
    \omega_+(k) \equiv k + \sum_{n=1}^{\infty}  \frac{\alpha_{2n+1}}{1 + c_s} k^{2n+1} \qquad \text{and} \qquad \omega_-(k) \equiv k - \sum_{n=1}^{\infty}  \frac{\alpha_{2n+1}}{1 - c_s} k^{2n+1}. \label{eq:omega_cs_pm}
\end{equation}
%\MM{Since here $\om_{\pm}(k)$ are the normalised dispersion relations, should we call differently the roots of \eqref{eq:dispersion_u1}? Maybe like $\pm\om(k)$.}
Since $\omega(k)$ is analytic in a disc of radius $R$ around $k=0$, also the typically real functions $\omega_+(k)$ and $\omega_-(k)$ are analytic in the same
disc. To apply the theorem above, we have to rescale everything in units of $R$. Introducing the normalised
variable $z \equiv k/R$ and the rescaled quantities $f_\pm \equiv \omega_\pm/R$ and
$a_{2n+1} \equiv \alpha_{2n+1}R^{2n}$, \eqref{eq:omega_cs_pm} becomes
\begin{equation}
	f_+(z) = z + \sum_{n=1}^{\infty} b_{2n+1}^{(+)} z^{2n+1}
	\qquad \text{and} \qquad
	f_-(z) = z + \sum_{n=1}^{\infty} b_{2n+1}^{(-)} z^{2n+1},
	\label{eq:f_cs_pm}
\end{equation}
with
\begin{equation}
	b_{2n+1}^{(\pm)} = \pm\, \frac{a_{2n+1}}{1 \pm c_s}.
	\label{eq:b_pm}
\end{equation}
 Applying \eqref{eq:rogosinski_bound_maintext} onto \eqref{eq:b_pm}, and intersecting the two bounds, we get (notice that $\kappa_{2n+1} < (2n +1)$ for any $n$):
\begin{equation}
    -\min\left[\kappa_{2n+1}(1 + c_s), (2n+1)(1 - c_s)\right] \leq \alpha_{2n+1}R^{2n} \leq \kappa_{2n+1} (1 - c_s) \label{eq:bounds_simple_intersect}
\end{equation}

It is important to notice that the intervals shrink to zero in the limit $c_s \to 1^-$. This is a qualitative difference with respect to the bounds obtained\footnote{We defer a detailed comparison with \cite{Heller:2022ejw,Heller:2023jtd} once dissipation is included: work in progress together with B.~Salehian and B.~Withers.} in \cite{Heller:2022ejw,Heller:2023jtd}. We will come back to this property below.

Notice that, while the Rogosinski bounds \eqref{eq:rogosinski_bound_maintext} are optimal, the intervals \eqref{eq:bounds_simple_intersect} are not: if one chooses a value for a particular $\alpha_{2n+1}$ in the interval, it is not guaranteed that a microcausal dispersion relation with that coefficient exists. The reason is that being typically real is a joint property of the entire sequence of coefficients \eqref{eq:complex_series_maintext}. The Rogosinski intervals \eqref{eq:rogosinski_bound_maintext} are obtained marginalising over all the other coefficients of the series. In our case we have to impose that both functions in \eqref{eq:f_cs_pm} are typically real. The optimal procedure is to first intersect the allowed sequence of coefficients and then to project on the coefficient, or set of coefficients, we are interested in. This is what we are doing next.

\subsubsection{Rogosinski-Toeplitz bounds}
Returning to the power series in Eq.~\eqref{eq:complex_series_maintext}, the full set of constraints implied by typically realness is outlined in Ref.~\cite{Rogosinski1932} through the algebraic characterisation of Toeplitz \cite{schmudgen2017moment}. Firstly, we will define the \textit{representative} of $f(z)$.
\begin{definition}[Representative of a function] 
    Let $f(z)$ be defined as in Eq.~\eqref{eq:complex_series_maintext}. For an integer $N \geq 1$, we call
    \begin{equation}
        p_N \equiv \left(b_{2n+1}\right)_{n =1}^{N} \in \mathbb{R}^N
    \end{equation}
    the $N$-th representative of $f(z)$.
\end{definition}
As $f(z)$ ranges over all possible typically real functions, its $N$-th representatives sweep out a well-defined closed, convex region $K_N \subset \mathbb{R}^N$. For $p_N$ to lie in this region $K_N$, an associated Toeplitz matrix built from the coefficients $b_{2n + 1}$ must be positive-semidefinite, as made precise in Theorem \ref{thm:toeplitz_maintext}.
\begin{theorem}[Toeplitz conditions] \label{thm:toeplitz_maintext}
    Let $f(z)$ be defined as in Eq.~\eqref{eq:complex_series_maintext}. The complex function $f(z)$ will be typically real if and only if all of its $N$-th representatives lie within their corresponding closed convex hull $K_N$ characterised by the $N$ Toeplitz determinants
    \begin{equation}
        D_{k+1} = \begin{vmatrix}
        2 & b_3 - b_1 & b_5 - b_3 & \dots & b_{2k+1}-b_{2k-1}\\
        b_3 - b_1 & 2 & b_3 - b_1 & \dots & b_{2k-1}-b_{2k-3}\\
        b_5 - b_3 & b_3 - b_1 & 2 & \dots & \vdots\\
        \vdots & \vdots & \vdots & 2 & \vdots\\
        b_{2k+1}-b_{2k-1} & b_{2k-1}-b_{2k-3} & \dots & \dots & 2
        \end{vmatrix}, \label{eq:toeplitz_matrix_mainText}
    \end{equation}
    in which $1 \leq k \leq N$ and $b_1 = 1$. These determinants are either all strictly positive, or positive up to some index and vanishing from that index onward, the latter case corresponding to the boundary of $K_N$.
\end{theorem}
Each $D_{k+1}$ results in a polynomial inequality in the coefficients $\left\{b_{2n+1}\right\}_{n = 1}^{N}$; together, these $N$ inequalities describe the closed convex region $K_N$—the \emph{Rogosinski-Toeplitz spectrahedron}. An example of the two-dimensional spectrahedron $K_2$ is shown in Figure \ref{fig:k2_maintext}, while the full details of the spectrahedron $K_N$ can be found in Appendix~\ref{subsec:rogo_toeplitz_poly}. 
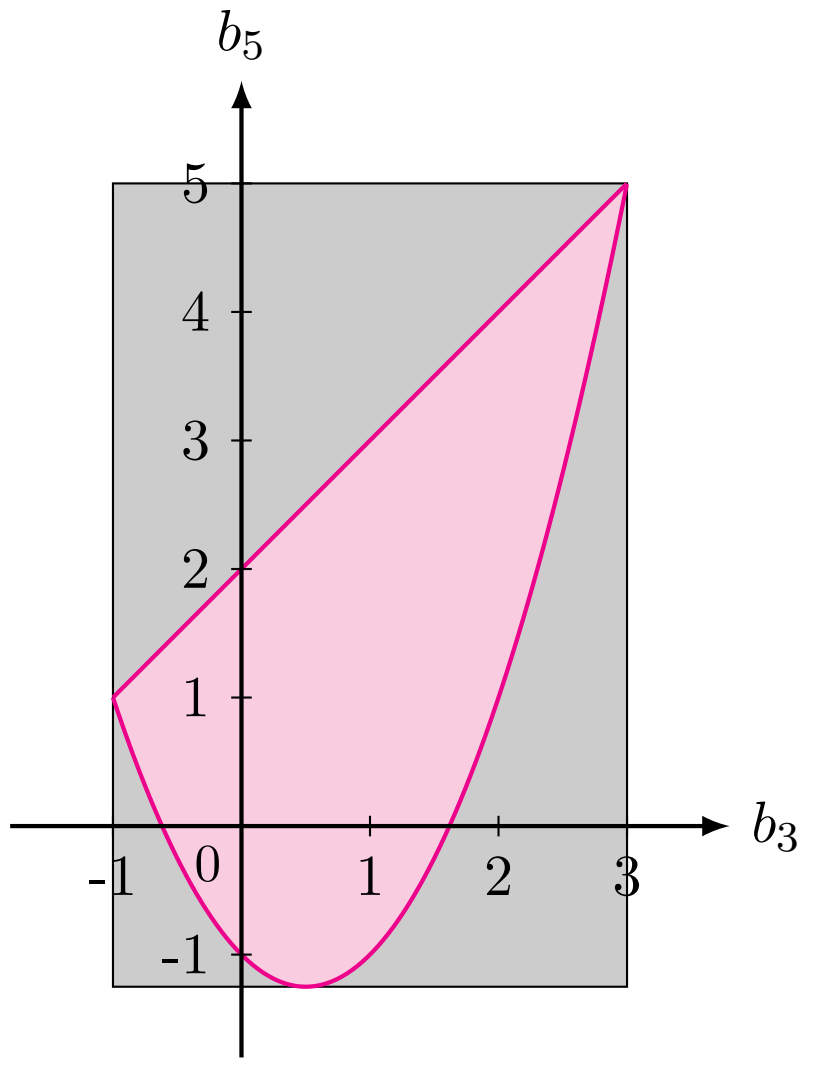
\begin{figure}[H]
    \centering
    \begin{tikzpicture}[>=latex, scale=0.85]
 
%\definecolor{magenta}{RGB}{149,82,197}
 
% axis limits
\def\xmin{-1.8}
\def\xmax{3.8}
\def\ymin{-1.8}
\def\ymax{5.8}
 
% bounding rectangle -1 <= b3 <= 3, -5/4 <= b5 <= 5 (drawn behind the hull)
\draw[black, fill=black, fill opacity=0.2] (-1,-1.25) rectangle (3,5);
 
% shaded region: b3^2 - b3 - 1 <= b5 <= b3 + 2, for b3 in [-1,3]
\fill[magenta!25] plot[domain=-1:3, samples=100] (\x,{\x+2})
   -- plot[domain=3:-1, samples=100] (\x,{\x*\x-\x-1}) -- cycle;
 
% boundary curves
\draw[magenta, thick] plot[domain=-1:3, samples=100] (\x,{\x+2});
\draw[magenta, thick] plot[domain=-1:3, samples=100] (\x,{\x*\x-\x-1});
 
%\node[left] at (-0.15,-1.25) {\scriptsize $-\frac{5}{4}$};
 
% axes (drawn last, on top of the shaded regions)
\draw[->, thick] (\xmin,0) -- (\xmax,0) node[right] {$b_3$};
\draw[->, thick] (0,\ymin) -- (0,\ymax) node[above] {$b_5$};
 
% ticks
\foreach \x in {-1,1,2,3}
  \draw (\x,-0.08) -- (\x,0.08) node[below=4pt] {\x};
\foreach \y in {-1,1,2,3,4,5}
  \draw (-0.08,\y) -- (0.08,\y) node[left=4pt] {\y};
\node[below left] at (0,0) {\small $0$};
 
\end{tikzpicture}
    \caption{Plot of the two-dimensional Rogosinski-Toeplitz spectrahedron $K_2$ (magenta) as calculated in Appendix \ref{subsec:rogo_toeplitz_poly}, compared to the Rogosinski bounds \eqref{eq:rogosinski_bound_maintext} (black). The Rogosinski bounds can be viewed as the projection of $K_2$, which is itself a projection of $K_\infty$.}
    \label{fig:k2_maintext}
\end{figure}
As $N\to\infty$, an infinite hierarchy of determinant inequalities must be imposed, one for each newly introduced coefficient. The result is $K_\infty$: an infinite dimensional spectrahedron\footnote{Technically speaking, the term `spectrahedron' is only used for a finite-dimensional convex hull (see, e.g.~\cite{schmudgen2017moment}). Here, we will relax its definition slightly to describe both finite and infinite-dimensional cases.} such that $f(z)$, as given in Eq.~\eqref{eq:complex_series_maintext}, is typically real if and only if its coefficient sequence satisfies
\begin{equation}
    \left\{b_{2n+1}\right\}_{n = 1}^{\infty} \equiv \left\{b_{2n+1}\right\} \in K_\infty.
\end{equation}

Each finite $K_N$ is recovered as the projection of $K_\infty$ onto its first $N$ coordinates $\left(b_{2n+1}\right)_{n = 1}^{N}$. In this sense, one can treat $K_\infty$ as the fundamental object and $K_N$ as its projection. The Rogosinski bound on a single coefficient $b_{2n+1}$ \eqref{eq:rogosinski_bound_maintext}, is the projection of $K_N$—itself the projection of $K_\infty$—onto each individual coordinate (see Figure \ref{fig:k2_maintext}). 

\begin{figure}[H]
    \centering
    \begin{subfigure}[b]{0.3\textwidth}
        \centering
        \resizebox{\textwidth}{!}{%
\begin{tikzpicture}[x=0.685700cm, y=0.446200cm, >=latex]
 
\definecolor{myred}{RGB}{143,54,37}
\definecolor{myredfill}{RGB}{238,200,193}
\definecolor{myblue}{RGB}{45,73,110}
\definecolor{mybluefill}{RGB}{199,216,233}
\definecolor{mypurple}{RGB}{100,78,145}
\definecolor{mypurplefill}{RGB}{199,183,224}
\definecolor{mycross}{RGB}{130,73,130}
 
\def\xmin{-3.400000}
\def\xmax{3.600000}
\def\ymin{-6.500000}
\def\ymax{6.500000}
\def\sred{1.000000}
\def\sblue{-1.000000}
 
% --- red region: (1+cs) K2 ---
\fill[myredfill] plot[domain=-1:3, samples=150] ({\sred*\x},{\sred*(\x+2)})
   -- plot[domain=3:-1, samples=150] ({\sred*\x},{\sred*(\x*\x-\x-1)}) -- cycle;
\draw[myred, thick] plot[domain=-1:3, samples=150] ({\sred*\x},{\sred*(\x+2)});
\draw[myred, thick] plot[domain=-1:3, samples=150] ({\sred*\x},{\sred*(\x*\x-\x-1)});
 
% --- blue region: -(1-cs) K2 ---
\fill[mybluefill] plot[domain=-1:3, samples=150] ({\sblue*\x},{\sblue*(\x+2)})
   -- plot[domain=3:-1, samples=150] ({\sblue*\x},{\sblue*(\x*\x-\x-1)}) -- cycle;
\draw[myblue, thick] plot[domain=-1:3, samples=150] ({\sblue*\x},{\sblue*(\x+2)});
\draw[myblue, thick] plot[domain=-1:3, samples=150] ({\sblue*\x},{\sblue*(\x*\x-\x-1)});
 
% --- intersected region, filled and outlined in purple ---
\fill[mypurplefill] plot[smooth cycle, tension=0.15] coordinates {(1.0000,-1.0000) (0.9374,-1.0587) (0.8723,-1.1114) (0.8073,-1.1556) (0.7472,-1.1889) (0.6871,-1.2150) (0.6220,-1.2351) (0.5620,-1.2462) (0.4969,-1.2500) (0.4368,-1.2460) (0.3717,-1.2335) (0.3116,-1.2145) (0.2466,-1.1858) (0.1865,-1.1517) (0.1214,-1.1067) (0.0613,-1.0576) (-0.0038,-0.9962) (-0.0688,-0.9264) (-0.1289,-0.8545) (-0.1940,-0.7684) (-0.2541,-0.6814) (-0.3141,-0.5872) (-0.3792,-0.4770) (-0.4443,-0.3583) (-0.5044,-0.2412) (-0.5695,-0.1063) (-0.6295,0.0259) (-0.6946,0.1771) (-0.7597,0.3368) (-0.8298,0.5183) (-0.8999,0.7096) (-1.0000,1.0000) (-0.9975,1.0025) (-0.9374,1.0587) (-0.8723,1.1114) (-0.8073,1.1556) (-0.7472,1.1889) (-0.6871,1.2150) (-0.6220,1.2351) (-0.5620,1.2462) (-0.4969,1.2500) (-0.4368,1.2460) (-0.3717,1.2335) (-0.3116,1.2145) (-0.2466,1.1858) (-0.1865,1.1517) (-0.1214,1.1067) (-0.0613,1.0576) (0.0038,0.9962) (0.0688,0.9264) (0.1289,0.8545) (0.1940,0.7684) (0.2541,0.6814) (0.3141,0.5872) (0.3792,0.4770) (0.4443,0.3583) (0.5044,0.2412) (0.5695,0.1063) (0.6295,-0.0259) (0.6946,-0.1771) (0.7597,-0.3368) (0.8298,-0.5183) (0.8999,-0.7096) (1.0000,-1.0000)};
\draw[mypurple, thick] plot[smooth cycle, tension=0.15] coordinates {(1.0000,-1.0000) (0.9374,-1.0587) (0.8723,-1.1114) (0.8073,-1.1556) (0.7472,-1.1889) (0.6871,-1.2150) (0.6220,-1.2351) (0.5620,-1.2462) (0.4969,-1.2500) (0.4368,-1.2460) (0.3717,-1.2335) (0.3116,-1.2145) (0.2466,-1.1858) (0.1865,-1.1517) (0.1214,-1.1067) (0.0613,-1.0576) (-0.0038,-0.9962) (-0.0688,-0.9264) (-0.1289,-0.8545) (-0.1940,-0.7684) (-0.2541,-0.6814) (-0.3141,-0.5872) (-0.3792,-0.4770) (-0.4443,-0.3583) (-0.5044,-0.2412) (-0.5695,-0.1063) (-0.6295,0.0259) (-0.6946,0.1771) (-0.7597,0.3368) (-0.8298,0.5183) (-0.8999,0.7096) (-1.0000,1.0000) (-0.9975,1.0025) (-0.9374,1.0587) (-0.8723,1.1114) (-0.8073,1.1556) (-0.7472,1.1889) (-0.6871,1.2150) (-0.6220,1.2351) (-0.5620,1.2462) (-0.4969,1.2500) (-0.4368,1.2460) (-0.3717,1.2335) (-0.3116,1.2145) (-0.2466,1.1858) (-0.1865,1.1517) (-0.1214,1.1067) (-0.0613,1.0576) (0.0038,0.9962) (0.0688,0.9264) (0.1289,0.8545) (0.1940,0.7684) (0.2541,0.6814) (0.3141,0.5872) (0.3792,0.4770) (0.4443,0.3583) (0.5044,0.2412) (0.5695,0.1063) (0.6295,-0.0259) (0.6946,-0.1771) (0.7597,-0.3368) (0.8298,-0.5183) (0.8999,-0.7096) (1.0000,-1.0000)};
 
% --- axes, physically broken around the cross span so no black can ever show through it ---
\draw[thick] (\xmin,0) -- (-1.000000,0);
\draw[->, thick] (1.000000,0) -- (\xmax,0) node[right] {$a_3$};
\draw[thick] (0,\ymin) -- (0,-1.250000);
\draw[->, thick] (0,1.250000) -- (0,\ymax) node[above] {$a_5$};
\foreach \t in {-3,-2,-1,1,2,3}
  \draw (\t,-0.12) -- (\t,0.12) node[below=3pt] {\t};
\foreach \t in {-6,-4,-2,2,4,6}
  \draw (-0.12,\t) -- (0.12,\t) node[left=3pt] {\t};
 
% --- projection of the intersection onto each axis (drawn last) ---
\draw[mycross, line width=1.6pt] (-1.000000,0) -- (1.000000,0);
\draw[mycross, line width=1.6pt] (0,-1.250000) -- (0,1.250000);
\filldraw (-1.000000,0) circle (1.6pt);
\filldraw (1.000000,0) circle (1.6pt);
\filldraw (0,-1.250000) circle (1.6pt);
\filldraw (0,1.250000) circle (1.6pt);
 
\end{tikzpicture}
        }
        \caption{$c_s = 0$}
        \label{fig:k2_cs=0}
    \end{subfigure}
    \hfill
    \begin{subfigure}[b]{0.3\textwidth}
        \centering
        \resizebox{\textwidth}{!}{%
\begin{tikzpicture}[x=0.631600cm, y=0.532100cm, >=latex]
 
\definecolor{myred}{RGB}{143,54,37}
\definecolor{myredfill}{RGB}{238,200,193}
\definecolor{myblue}{RGB}{45,73,110}
\definecolor{mybluefill}{RGB}{199,216,233}
\definecolor{mypurple}{RGB}{100,78,145}
\definecolor{mypurplefill}{RGB}{199,183,224}
\definecolor{mycross}{RGB}{130,73,130}
 
\def\xmin{-2.300000}
\def\xmax{5.300000}
\def\ymin{-2.300000}
\def\ymax{8.600000}
\def\sred{1.500000}
\def\sblue{-0.500000}
 
% --- red region: (1+cs) K2 ---
\fill[myredfill] plot[domain=-1:3, samples=150] ({\sred*\x},{\sred*(\x+2)})
   -- plot[domain=3:-1, samples=150] ({\sred*\x},{\sred*(\x*\x-\x-1)}) -- cycle;
\draw[myred, thick] plot[domain=-1:3, samples=150] ({\sred*\x},{\sred*(\x+2)});
\draw[myred, thick] plot[domain=-1:3, samples=150] ({\sred*\x},{\sred*(\x*\x-\x-1)});
 
% --- blue region: -(1-cs) K2 ---
\fill[mybluefill] plot[domain=-1:3, samples=150] ({\sblue*\x},{\sblue*(\x+2)})
   -- plot[domain=3:-1, samples=150] ({\sblue*\x},{\sblue*(\x*\x-\x-1)}) -- cycle;
\draw[myblue, thick] plot[domain=-1:3, samples=150] ({\sblue*\x},{\sblue*(\x+2)});
\draw[myblue, thick] plot[domain=-1:3, samples=150] ({\sblue*\x},{\sblue*(\x*\x-\x-1)});
 
% --- intersected region, filled and outlined in purple ---
\fill[mypurplefill] plot[smooth cycle, tension=0.15] coordinates {(-0.2320,-1.2320) (-0.3135,-1.1210) (-0.3961,-0.9993) (-0.4787,-0.8685) (-0.5538,-0.7417) (-0.6289,-0.6074) (-0.7115,-0.4510) (-0.7866,-0.3009) (-0.8660,-0.1340) (-0.7891,0.0437) (-0.7115,0.1990) (-0.6364,0.3264) (-0.5989,0.3816) (-0.5588,0.4343) (-0.5188,0.4805) (-0.4812,0.5181) (-0.4437,0.5500) (-0.4061,0.5762) (-0.3661,0.5980) (-0.3285,0.6127) (-0.2910,0.6216) (-0.2509,0.6250) (-0.2034,0.6207) (-0.1558,0.6073) (-0.1083,0.5848) (-0.0632,0.5552) (-0.0156,0.5152) (0.0319,0.4660) (0.0795,0.4079) (0.1245,0.3445) (0.1721,0.2687) (0.2171,0.1885) (0.2647,0.0952) (0.3123,-0.0073) (0.3598,-0.1188) (0.4074,-0.2393) (0.5000,-0.5000) (-0.2320,-1.2320)};
\draw[mypurple, thick] plot[smooth cycle, tension=0.15] coordinates {(-0.2320,-1.2320) (-0.3135,-1.1210) (-0.3961,-0.9993) (-0.4787,-0.8685) (-0.5538,-0.7417) (-0.6289,-0.6074) (-0.7115,-0.4510) (-0.7866,-0.3009) (-0.8660,-0.1340) (-0.7891,0.0437) (-0.7115,0.1990) (-0.6364,0.3264) (-0.5989,0.3816) (-0.5588,0.4343) (-0.5188,0.4805) (-0.4812,0.5181) (-0.4437,0.5500) (-0.4061,0.5762) (-0.3661,0.5980) (-0.3285,0.6127) (-0.2910,0.6216) (-0.2509,0.6250) (-0.2034,0.6207) (-0.1558,0.6073) (-0.1083,0.5848) (-0.0632,0.5552) (-0.0156,0.5152) (0.0319,0.4660) (0.0795,0.4079) (0.1245,0.3445) (0.1721,0.2687) (0.2171,0.1885) (0.2647,0.0952) (0.3123,-0.0073) (0.3598,-0.1188) (0.4074,-0.2393) (0.5000,-0.5000) (-0.2320,-1.2320)};
 
% --- axes, physically broken around the cross span so no black can ever show through it ---
\draw[thick] (\xmin,0) -- (-0.866000,0);
\draw[->, thick] (0.500000,0) -- (\xmax,0) node[right] {$a_3$};
\draw[thick] (0,\ymin) -- (0,-1.232000);
\draw[->, thick] (0,0.625000) -- (0,\ymax) node[above] {$a_5$};
\foreach \t in {-2,-1,1,2,3,4,5}
  \draw (\t,-0.12) -- (\t,0.12) node[below=3pt] {\t};
\foreach \t in {-2,2,4,6,8}
  \draw (-0.12,\t) -- (0.12,\t) node[left=3pt] {\t};
 
% --- projection of the intersection onto each axis (drawn last) ---
\draw[mycross, line width=1.6pt] (-0.866000,0) -- (0.500000,0);
\draw[mycross, line width=1.6pt] (0,-1.232000) -- (0,0.625000);
\filldraw (-0.866000,0) circle (1.6pt);
\filldraw (0.500000,0) circle (1.6pt);
\filldraw (0,-1.232000) circle (1.6pt);
\filldraw (0,0.625000) circle (1.6pt);
 
\end{tikzpicture}
        }
        \caption{$c_s = 0.5$}
        \label{fig:k2_cs=0.5}
    \end{subfigure}
    \hfill
    \begin{subfigure}[b]{0.3\textwidth}
        \centering
        \resizebox{\textwidth}{!}{%
\begin{tikzpicture}[x=0.558100cm, y=0.449600cm, >=latex]
 
\definecolor{myred}{RGB}{143,54,37}
\definecolor{myredfill}{RGB}{238,200,193}
\definecolor{myblue}{RGB}{45,73,110}
\definecolor{mybluefill}{RGB}{199,216,233}
\definecolor{mypurple}{RGB}{100,78,145}
\definecolor{mypurplefill}{RGB}{199,183,224}
\definecolor{mycross}{RGB}{130,73,130}
 
\def\xmin{-2.300000}
\def\xmax{6.300000}
\def\ymin{-2.300000}
\def\ymax{10.600000}
\def\sred{1.850000}
\def\sblue{-0.150000}
 
% --- red region: (1+cs) K2 ---
\fill[myredfill] plot[domain=-1:3, samples=150] ({\sred*\x},{\sred*(\x+2)})
   -- plot[domain=3:-1, samples=150] ({\sred*\x},{\sred*(\x*\x-\x-1)}) -- cycle;
\draw[myred, thick] plot[domain=-1:3, samples=150] ({\sred*\x},{\sred*(\x+2)});
\draw[myred, thick] plot[domain=-1:3, samples=150] ({\sred*\x},{\sred*(\x*\x-\x-1)});
 
% --- blue region: -(1-cs) K2 ---
\fill[mybluefill] plot[domain=-1:3, samples=150] ({\sblue*\x},{\sblue*(\x+2)})
   -- plot[domain=3:-1, samples=150] ({\sblue*\x},{\sblue*(\x*\x-\x-1)}) -- cycle;
\draw[myblue, thick] plot[domain=-1:3, samples=150] ({\sblue*\x},{\sblue*(\x+2)});
\draw[myblue, thick] plot[domain=-1:3, samples=150] ({\sblue*\x},{\sblue*(\x*\x-\x-1)});
 
% --- intersected region, filled and outlined in purple ---
\fill[mypurplefill] plot[smooth cycle, tension=0.15] coordinates {(0.1492,-0.1508) (-0.3148,-0.6148) (-0.4500,-0.7500) (-0.4125,-0.5717) (-0.3749,-0.4121) (-0.3374,-0.2714) (-0.2998,-0.1494) (-0.2623,-0.0463) (-0.2247,0.0381) (-0.1872,0.1036) (-0.1496,0.1504) (-0.1309,0.1667) (-0.1121,0.1783) (-0.0933,0.1853) (-0.0745,0.1875) (-0.0558,0.1850) (-0.0370,0.1779) (-0.0182,0.1660) (0.0006,0.1494) (0.0381,0.1022) (0.0764,0.0347) (0.1162,-0.0562) (0.1492,-0.1508)};
\draw[mypurple, thick] plot[smooth cycle, tension=0.15] coordinates {(0.1492,-0.1508) (-0.3148,-0.6148) (-0.4500,-0.7500) (-0.4125,-0.5717) (-0.3749,-0.4121) (-0.3374,-0.2714) (-0.2998,-0.1494) (-0.2623,-0.0463) (-0.2247,0.0381) (-0.1872,0.1036) (-0.1496,0.1504) (-0.1309,0.1667) (-0.1121,0.1783) (-0.0933,0.1853) (-0.0745,0.1875) (-0.0558,0.1850) (-0.0370,0.1779) (-0.0182,0.1660) (0.0006,0.1494) (0.0381,0.1022) (0.0764,0.0347) (0.1162,-0.0562) (0.1492,-0.1508)};
 
% --- axes, physically broken around the cross span so no black can ever show through it ---
\draw[thick] (\xmin,0) -- (-0.450000,0);
\draw[->, thick] (0.150000,0) -- (\xmax,0) node[right] {$a_3$};
\draw[thick] (0,\ymin) -- (0,-0.750000);
\draw[->, thick] (0,0.187500) -- (0,\ymax) node[above] {$a_5$};
\foreach \t in {-2,2,4,6}
  \draw (\t,-0.12) -- (\t,0.12) node[below=3pt] {\t};
\foreach \t in {-2,2,4,6,8,10}
  \draw (-0.12,\t) -- (0.12,\t) node[left=3pt] {\t};
 
% --- projection of the intersection onto each axis (drawn last) ---
\draw[mycross, line width=1.6pt] (-0.450000,0) -- (0.150000,0);
\draw[mycross, line width=1.6pt] (0,-0.750000) -- (0,0.187500);
\filldraw (-0.450000,0) circle (1.6pt);
\filldraw (0.150000,0) circle (1.6pt);
\filldraw (0,-0.750000) circle (1.6pt);
\filldraw (0,0.187500) circle (1.6pt);
 
\end{tikzpicture}
        }
        \caption{$c_s = 0.85$}
        \label{fig:k2_cs=0.85}
    \end{subfigure}
    \caption{The intersection between $-(1 - c_s)K_2$ (blue) and $(1 + c_s)K_2$ (red) for different values of $c_s$. The intersected region (purple) describes the necessary but not sufficient bounds on $(a_3, a_5)$ which are consistent with microcausality.}%\MM{Maybe the coefficients in the axes have to be written a little bit bigger based on what Paolo said?}}
    \label{fig:k2_cs}
\end{figure}
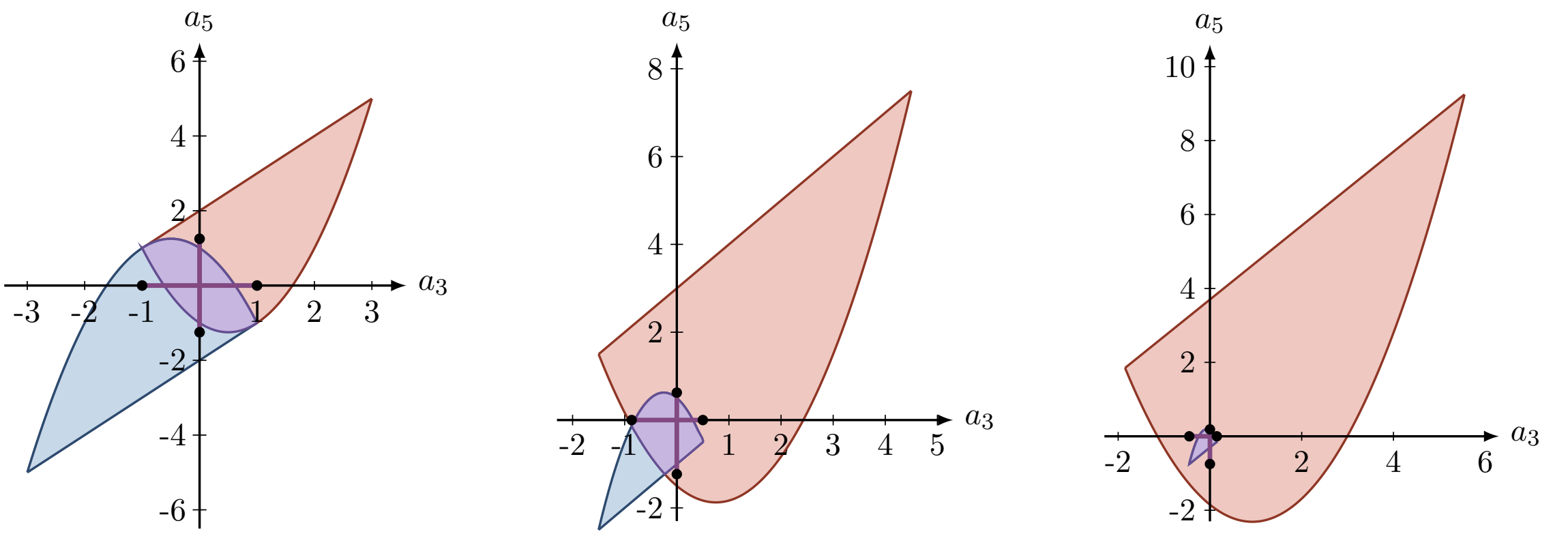

Going back to the functions in Eq.~\eqref{eq:omega_cs_pm}, for $\omega_{\pm}(k)$ to both be typically real, the sequence of coefficients $\{a_{2n+1}\} = \{\alpha_{2n+1}R^{2n}\}$ must satisfy
\begin{equation}
    \{a_{2n + 1}\} = \{\alpha_{2n+1} R^{2n}\} \in -(1 - c_s)K_{\infty} \cap (1 + c_s) K_\infty~. \label{eq:true_bounds}
\end{equation}
The bound in Eq.~\eqref{eq:true_bounds} is optimal, in the sense that it is exactly equivalent to the microcausality bound in Eq.~\eqref{eq:bound_omega}. In practice, however, one can only work with $K_N$ i.e. 
\begin{equation}
    \{a_{2n + 1}\}_{n = 1}^{N} \in -(1 - c_s)K_{N} \cap (1 + c_s) K_N. \label{eq:bounds_project_intersect}
\end{equation}
Similarly to what happened for the intervals for $N=1$, the intersection in Eq.~\eqref{eq:bounds_project_intersect}, as seen in Figure \ref{fig:k2_cs} for $N=2$, yields necessary but not sufficient bounds on $\{a_{2n + 1}\}_{n = 1}^{N}$. This is because $-(1 - c_s)K_{N}$ and $(1 + c_s) K_N$ are both projections of $-(1 - c_s)K_{\infty}$ and $(1 + c_s) K_\infty$, respectively. Intersecting the projections of two sets will always give a set larger than or equal to the projection of the intersection of the two sets. Thus, any bounds derived from $K_N$ is automatically satisfied by the true constraint in Eq.~\eqref{eq:true_bounds}, but the converse need not hold.

\begin{figure}[]
    \centering
    \includegraphics[width=\linewidth]{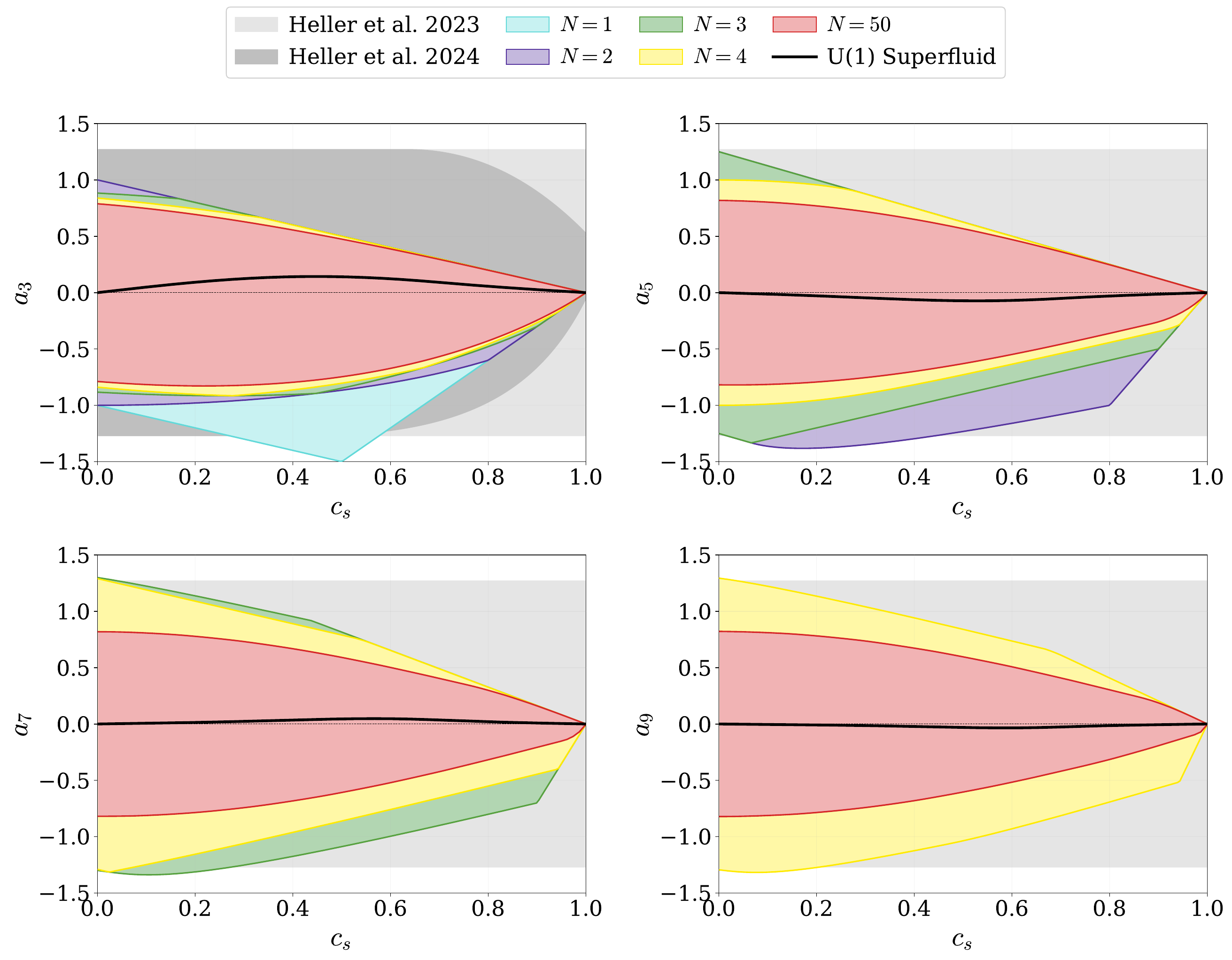}
    \caption{The Rogosinski-Toeplitz bounds on $a_3$, $a_5$, $a_7$ and $a_9$. Each coloured region corresponds to the $N+1$ truncation of the Toeplitz determinant. For $N=2$, the purple region describes $-(1 - c_s)K_2 \cap (1 + c_s)K_2$, shown (also in purple) in Figure \ref{fig:k2_cs}. As $N \to \infty$, the bounds converge towards the optimal region \eqref{eq:true_bounds} shown in red. The grey regions are the bounds given by \cite{Heller:2022ejw} (light grey) and \cite{Heller:2023jtd} (dark grey). The values of $a_3$, $a_5$, $a_7$ and $a_9$ for the $U(1)$ superfluid studied in Section \ref{subsec:u1_superfluid} are also shown as the black curve from $c_s = 0$ to 1.}
    \label{fig:rogo_toeplitz_bounds_a}
\end{figure}

Computing the intersection $-(1-c_s)K_N \cap (1+c_s)K_N$ can be formulated as a semi-definite programming (SDP) problem (see, e.g.~Ref.~\cite{schmudgen2017moment}). We solve this numerically for each increasing $N$. Figure \ref{fig:rogo_toeplitz_bounds_a} illustrates the resulting bounds on $a_3, a_5, a_7$ and $a_9$ for each $N$ as a function of $c_s$, alongside with bounds given by Refs.~\cite{Heller:2022ejw} and~\cite{Heller:2023jtd} for comparison. As expected, the bounds tighten monotonically with increasing $N$. By $N=50$, the bounds visually converge over the entire range of $c_s$. We also include the dispersion relation coefficients calculated from the $U(1)$ model in Section \ref{subsec:u1_superfluid}, which of course lie inside the allowed region for any value of $c_s$.
%\MM{This is related to the comment I have already made. Since the U(1) superfluid model is in another section, different from this one, it's not clear that you're referring to that model here, in my opinion. So either we somehow put everything in one big section or we make it clear that here we are analyzing the U(1) model.}
These bounds give the range of possible values of the higher-order coefficients in the dispersion relation, in units of the radius of convergence $R$. Perhaps it is more physical to reverse the logic. From the measurement of one of the higher-order coefficients (together with $c_s$) one has an upper bound on the scale at which the Goldstone description (linear dispersion relation plus corrections) must break down. 

One can ask whether it is possible to write bounds that do not depend on the radius of convergence $R$: for instance a combination of the $\alpha$'s with the same units, say $\alpha_5^2 - 3\alpha_3 \alpha_7$, could be constrained to be positive. It is easy to realise that this is not possible. Suppose the coefficients admit a bound independent of $R$. Then these bounds would hold even when one restricts the analysis to a disk of smaller radius $r < R$, no matter how small $r$ is.  But microcausality cannot give any bound in the limit $r \to 0$, because for sufficiently small $r$ all higher-derivative corrections are negigible: one can linearise the dispersion relation, $\omega = c_s k$, and the only constraint is $c_s \leq 1$.

\subsubsection{\texorpdfstring{$c_s \to 1$}{cs to 1} limit}
\label{sec:cs_to_1}
As mentioned above, the bounds become particularly interesting when $c_s \to 1$: in this limit the bounds do not only make precise what dimensional analysis would suggest, but they introduce a non-trivial parametric dependence on $1-c_s$. In \cite{Heller:2022ejw}, it was already proven that for $c_s =1$ all the higher-order coefficients must vanish.

In the $c_s \to 1$ limit, the intersection in \eqref{eq:true_bounds} is dominated by $-(1 - c_s)K_\infty$, so that the bounds reduce to
\begin{equation}
    \{a_{2n + 1}\} = \{\alpha_{2n+1} R^{2n}\} \in -(1 - c_s)K_{\infty}.
\end{equation}
The optimal constraints on each individual $a_{2n+1} = \alpha_{2n+1}R^{2n}$ are the Rogosinski intervals \eqref{eq:rogosinski_bound_maintext}:
\begin{equation}
  -(2n+1) (1 - c_s) \leq \alpha_{2n+1} R^{2n} \leq \kappa_{2n+1} (1 - c_s)~. \label{eq:bounds_csto1}
\end{equation}
The bounds become tighter as $c_s \to 1$ and the window for the $\alpha_{2n+1}$ shrinks to zero. More physically, for a fixed given higher-order coefficient, in the limit $c_s \to 1$ the scale $R$, at which the linear Goldstone description breaks, goes to zero. 

This conclusion is not new. In \cite{Hui:2025aja}, the authors considered the leading correction\footnote{The definition of $\alpha_3$ used in \cite{Hui:2025aja} differs from ours by a factor of $c_s$. We stick to our notation.} to the linear Goldstone dispersion relation:
\begin{equation}
    \omega(k) \simeq \pm (c_s k + \alpha_3 k^3)
\end{equation}
and showed that $\omega(k)$ remains microcausal when $k$ lies in the interval
\begin{equation}
    -(1 - c_s) \leq \alpha_{3} |k|^2 \leq \frac{(1 - c_s)}{3}. \label{eq:bound_hui}
\end{equation}
If one goes to higher $|k|$, higher-order corrections in the dispersion relation must become relevant. Our bounds have a stronger content because they say something about the scale at which the series stops converging. Notice that the interval \eqref{eq:bound_hui} is a factor of 3 smaller than the Rogosinski's \eqref{eq:bounds_csto1}. The reason is that in the case of  \eqref{eq:bound_hui}, one is setting to zero all the higher order terms $\alpha_5, \alpha_7,\dots = 0$, while \eqref{eq:bounds_csto1} holds for any continuation of the series.
As emphasized in \cite{Hui:2025aja}, it is here clear that microcausality is much stronger than requiring the subluminality of the group velocity: a negative $\alpha_3$ would push towards slower velocity, but is still constrained by \eqref{eq:bound_omega}. Moreover all higher-order corrections in the dispersion relation are constrained as well.

Finally, we identify the dispersion relation which saturates the Rogosinski bounds \eqref{eq:bounds_csto1} in this limit. Every odd typically real function admits the Robertson representation (see Appendix \ref{sec:tr_series}): an integral representation with a unit positive measure $\mathrm{d}\mu(t)$ on $[0,1]$ of the elementary
function
\begin{equation}
  f(z;t) = \frac{z(1+z^2)}{(1+z^2)^2-4z^2t^2}
         = z + \sum_{n = 1}^{\infty} U_{2n}(t)\, z^{2n+1}\,,
\end{equation}
in which $U_{2n}(t)$ are Chebyshev polynomials of the second kind. Each Taylor
coefficient is therefore linear in the measure,
$b_{2n+1}=\int_0^1 U_{2n}(t)\,\mathrm{d}\mu(t)$,
and lies between the minimum and maximum of $U_{2n}(t)$ on $t \in [0,1]$. A bound is saturated only if
$\mu(t)$ is supported where $U_{2n}(t)$ attains the corresponding extremum,
i.e.\ $\mu(t)=\delta(t-t_*)$ and $f(z)=f(z;t_*)$. The extremising dispersion
relation is thus
\begin{equation}
  \omega(k) = \pm k\left[1-(1-c_s)\,
  \frac{1+k^2/R^2}{(1+k^2/R^2)^2-4t_*^2k^2/R^2}\right].
\end{equation}
The lower end of~\eqref{eq:bounds_csto1} is attained for every $a_{2n+1}$ at
$t_*=1$, where $U_{2n}(1)=2n+1$. The upper end is attained at the
$n$-dependent $t_*$ minimising $U_{2n}$ on $[0,1]$, with
$\kappa_{2n+1}=-U_{2n}(t_*)$ as in~\eqref{eq:kappa_maintext}. Since the poles of $f(z;t_*)$ lie on $|z|=1$, the extremal $\omega(k)$ has poles at $|k|=R$ rather than the branch points. Therefore these bounds cannot be saturated if one assumes a meromorphic $\tilde{G}(\omega,k)$.
%\MM{Even though I think we can go really close to saturate them. Should we write it? For example in the example spitted by Claude $\om$ had only square root branch points but still went really close to maximise the $a_3$ coefficient.}

\subsection{Bounds on \texorpdfstring{$\om^2(k)$}{omega2(k)}}
\label{sec:omega2_bounds}
The $U(1)$ example studied above makes clear the different physical meaning of the radius of convergence of $\omega(k)$ and $\omega^2(k)$. 
We now turn to the study of $\omega^2(k)$, whose radius of analyticity around $k=0$ is $\tilde R$. At $\tilde R$ one encounters genuinely new states.

As discussed in Section \ref{subsec:dispersion_rel}, due to rotational symmetry, $\omega^2(k)$ must be an even function of $k$. Working around $k = 0$, we can write $\omega^2(k)$ as a series in $k^2$ as
\begin{equation}
    \omega^2(k) = \sum_{n = 1}^{\infty} \Tilde{\alpha}_{2n} k^{2n} \label{eq:omega2_series}
\end{equation}
and the coefficients $\Tilde{\alpha}_{2n}$ are real (no dissipation). Also, $\Tilde{\alpha}_{2n}$ are dimensionful coefficients that can be expressed in terms of the radius of convergence $\Tilde{R}$ as
\begin{equation}
    \Tilde{\alpha}_{2n} = \frac{\Tilde{a}_{2n}}{\Tilde{R}^{2(n-1)}}
\end{equation}
where $\Tilde{a}_{2n}$, instead, are dimensionless. 

Our goal is to derive bounds on the coefficients $\tilde{\alpha}_{2n}$ using microcausality. However, the inequality \eqref{eq:bound_omega} is written in terms of $\omega(k)$. If we take the square-root of $\omega^2(k)$ we go back to the previous analysis, introducing branch points at $\omega^2(k) = 0$. Thus, we need to re-express the microcausality condition \eqref{eq:bound_omega} in terms of $\omega^2(k)$, and use this new form to impose bounds on $\Tilde{\alpha}_{2n}$. As discussed in Section~\ref{subsec:dispersion_rel}, one must impose microcausality on both signs of $\om = \pm\sqrt{\om^2(k)}$. Therefore, we can rewrite \eqref{eq:bound_omega} as
\begin{equation}
    \pm \Im \sqrt{\om^2(k)} \leq |\Im k| \quad \Longleftrightarrow \quad (\Im\om)^2 \leq (\Im k)^2.
    \label{eq:square_bound_omega}
\end{equation}
Then, using $\om = \Re \om + i\Im \om$, we can rewrite \eqref{eq:square_bound_omega} in terms of $\omega^2$: first,
\begin{align}
    |\omega^2| = |\omega|^2 = (\Re\omega)^2 + (\Im \omega)^2  \label{eq:replace1},
\end{align}
and second,
\begin{align}
    \Re (\omega^2) = (\Re\omega)^2 - (\Im \omega)^2 \label{eq:replace2}.
\end{align}
Equations \eqref{eq:replace1} and \eqref{eq:replace2} allow us to rewrite \eqref{eq:square_bound_omega} in terms of $\omega^2(k)$ as
\begin{equation}
    \frac{|\om^2| - \Re(\om^2)}{2} \leq (\Im k)^2.
    \label{eq:bounds_omega2}
\end{equation}
In the remainder of this section, we will derive bounds on the coefficients $\Tilde{\alpha}_{2n}$ of \eqref{eq:omega2_series}, using the microcausality condition \eqref{eq:bounds_omega2}, via two independent approaches.

\subsubsection{\texorpdfstring{$\cs$}{cs}-independent bounds}
\label{sec:cs independent omega2}
The microcausality condition \eqref{eq:bounds_omega2} together with the inequality
\begin{equation}
    |\om^2| -\Re(\om^2) \geq -2\Re(\om^2)
\end{equation}
imply the simpler bound:
\begin{equation}
  \Re(-\om^2) \le (\Im k)^2.
  \label{eq:simpler_bound_omega2}
\end{equation}
Notice that while \eqref{eq:bounds_omega2} is equivalent to microcausality, the inequality \eqref{eq:simpler_bound_omega2} is necessary but not sufficient: the bounds that we are going to get cannot be ``optimal''. The advantage is that the new condition is linear in $\omega^2(k)$, while the original inequality contained the absolute value. Following an approach similar to \cite{Heller:2022ejw}, we let $k = re^{i\theta}$, with $r\in[0,\tR)$, and decompose each coefficient $\Tilde{\alpha}_{2n}$ as
\begin{equation}
  \Tilde{\alpha}_{2n} = |\Tilde{\alpha}_{2n}| e^{i\theta_{2n}}.
  \label{eq:a2n-phase}
\end{equation}
The coefficients $\Tilde{\alpha}_{2n}$ are real, which means that
\begin{equation}
    \Tilde{\alpha}_{2n} = |\Tilde{\alpha}_{2n}|\cos(\theta_{2n}), \quad \theta_{2n} = \begin{cases}
        0\quad \text{when}\quad  \Tilde{\alpha}_{2n} > 0\\
        \pi \quad \text{when}\quad  \Tilde{\alpha}_{2n} < 0.
    \end{cases}\label{eq:alpha_2n_real}
\end{equation} 
From \eqref{eq:omega2_series}, we have
\begin{align}
  \Re(-\om^2) &= -\sum_{n=1}^{\infty} \Tilde{\alpha}_{2n} r^{2n} \cos(2n\theta) \equiv U(r,\theta) \label{eq:U-def} \\
  (\Im k)^2 &= r^2\sin^2\theta.
\end{align}
The bound in \eqref{eq:simpler_bound_omega2} now becomes 
\begin{equation}
    U(r,\theta) \leq r^2\sin^2\theta. \label{eq:simpler_bound_omega2_U}
\end{equation}
To extract the coefficient $\tilde\alpha_{2n}$ from \eqref{eq:U-def}, we multiply both
sides by $\cos(2n\theta)$ and integrate over $\theta \in [0, 2\pi)$:
\begin{equation}
  |\Tilde{\alpha}_{2n}| r^{2n} = \Tilde{\alpha}_{2n}\cos(\theta_{2n}) r^{2n} = -\frac{1}{\pi}\int_0^{2\pi} \mathrm{d}\theta\, U(r,\theta)\cos(2n\theta)\cos(\theta_{2n})~,
  \label{eq:a2n-extract}
\end{equation}
where we have exploited the orthogonality of cosines. 
In order to use \eqref{eq:simpler_bound_omega2_U} to put bounds on $\Tilde{\alpha}_{2n}$, we must ensure that whatever multiplies $U(r,\theta)$ inside the integral in \eqref{eq:a2n-extract} is positive. To do so, notice that since
\begin{equation}
    \int_0^{2\pi} \mathrm{d}\theta\, U(r,\theta) \sim \int_0^{2\pi}\cos(2n\theta)\,\mathrm{d}\theta = 0
    \label{eq:U_vanishing}
\end{equation}
for every integer $n\ge1$, then
\begin{align}
    |\Tilde{\alpha}_{2n}| r^{2n}
    &= \frac{1}{\pi}\int_0^{2\pi} \mathrm{d}\theta\, U(r,\theta)\Big[1 - \cos(2n\theta)\cos(\theta_{2n})\Big]\\
    &\leq \frac{r^2}{\pi} \int_0^{2\pi} \mathrm{d}\theta\, \sin^2\theta \Big[1 - \cos(2n\theta)\cos(\theta_{2n})\Big],
    \label{eq:a2n-minus-a0}
\end{align}
where we have used \eqref{eq:simpler_bound_omega2_U} for the last inequality. The integral on the right-hand side of \eqref{eq:a2n-minus-a0} can be evaluated:
\begin{equation}
  \int_0^{2\pi} \mathrm{d}\theta\, \sin^2\theta\Big[1-\cos(2n\theta)\cos(\theta_{2n})\Big] = \pi + \frac{\cos(\theta_{2n})\sin(4n\pi)}{4n(n^2-1)},
  \label{eq:integral-result}
\end{equation}
which for $n\ne1$ reduces to $\pi$, since $\sin(4n\pi)=0$ for all integer $n$. Thus, \eqref{eq:a2n-minus-a0} yields
\begin{equation}
  |\Tilde a_{2n}| = |\Tilde{\alpha}_{2n}|\tR^{2(n-1)} \le 1, \qquad n \ge 1~,
  \label{eq:cs_independent_bound}
\end{equation}
where we have sent $r\to\tR$ i.e.~the radius of convergence of $\omega^2$. To obtain the above inequality for $n=1$ one should not follow the steps above (which gives $|\Tilde{\alpha}_2| = \cs^2 \leq \frac32$) but return to \eqref{eq:bounds_omega2} and take $k \to 0$. One finds $\cs^2 \leq 1$ consistent with group-velocity arguments.

\subsubsection{\texorpdfstring{$c_s$}{cs}-dependent bounds}
\label{sec:cs dependent omega2}
In this subsection, we show how to improve the bound in \eqref{eq:cs_independent_bound} by introducing an explicit dependence on the sound speed $c_s$.

We return to the microcausality condition on $\omega^2(k)$ in \eqref{eq:bounds_omega2}
and rewrite $\Im k$ in terms of $k^2$ in analogy with
Eqs.~\eqref{eq:replace1} and \eqref{eq:replace2}. This yields, after some rearrangements,
\begin{equation}
    \Re(k^2-\om^2) \le |k|^2 - |\om^2|.
  \label{eq:rt-start}
\end{equation}
We will proceed by substituting \eqref{eq:omega2_series} into \eqref{eq:rt-start} to find
\begin{equation}
    (1-\cs^2)\Re(k^2)-\sum_{n = 2}^{\infty}\tilde{\alpha}_{2n}\Re(k^{2n}) \leq |k|^2\left[1-\left|\cs^2+\sum_{n = 2}^{\infty}\tilde{\alpha}_{2n}k^{2n-2}\right|\right]. \label{eq:linear_rt_constraints_old}
\end{equation}
The following inequality 
\begin{equation}
    \left|\cs^2+\sum_{n = 2}^{\infty}\tilde{\alpha}_{2n} k^{2n-2}\right| \geq \cs^2+\sum_{n = 2}^{\infty} \tilde{\alpha}_{2n}\Re(k^{2n-2}) \label{eq:inequality}
\end{equation}
allows us to rewrite \eqref{eq:linear_rt_constraints_old} as
\begin{equation}
      (1-\cs^2)\Re(k^2) - \sum_{n=2}^{\infty} \tilde{\alpha}_{2n}\Re(k^{2n}) \leq |k|^2 \left[1-\cs^2 - \sum_{n=2}^{\infty}\tilde{\alpha}_{2n}\Re(k^{2n-2})\right].
      \label{eq:linear_rt_constraints}
\end{equation}
Due to \eqref{eq:inequality}, we now have a necessary inequality, but not equivalent to microcausality: again we are going to derive bounds which are not ``optimal''. Writing $k = re^{i\theta}, \, r\in [0,\tR)$, \eqref{eq:linear_rt_constraints} becomes
\begin{equation}
  \sum_{n=2}^{\infty} \tilde{\alpha}_{2n} r^{2n}\Big\{\cos[2(n-1)\theta] - \cos(2n\theta)\Big\} \le (1-\cs^2) r^2 [1-\cos(2\theta)].
  \label{eq:rt-polar}
\end{equation}
We can use the difference-to-product of cosines identity and the half-angle identity to rewrite \eqref{eq:rt-polar} as
\begin{equation}
  \sum_{n=2}^{\infty} \tilde{\alpha}_{2n} r^{2n} \sin(\theta)\sin[(2n-1)\theta] \le (1-\cs^2) r^2 \sin(\theta)^2.
  \label{eq:rt-with-sin}
\end{equation}
Assuming that $\sin(\theta)>0$ i.e.~$\Im k>0$, we can divide both sides of \eqref{eq:rt-with-sin} by $r\sin(\theta)$ and rewriting in terms of $k$ to obtain
\begin{equation}
  \sum_{n=2}^{\infty} \tilde{\alpha}_{2n}\Im(k^{2n-1}) \le (1-\cs^2)\Im k.
  \label{eq:rt-divided}
\end{equation}
Defining 
\begin{equation}
    \Tilde{\omega}^2(k) \equiv (1-\cs^2)k - \sum_{n=2}^{\infty} \tilde{\alpha}_{2n} k^{2n-1} = \frac{k^2-\om^2}{k}, \label{eq:F}
\end{equation}
\eqref{eq:rt-divided} is equivalent to
\begin{equation}
  \Im \Tilde{\omega}^2 \ge 0, \qquad \text{if } \Im(k)>0.
  \label{eq:F-positive}
\end{equation}
Similarly, if $\sin(\theta)<0$ i.e.~$\Im k<0$, we can divide both sides of \eqref{eq:rt-with-sin} by $r\sin(\theta)$ to obtain
\begin{equation}
  \Im \Tilde{\omega}^2 \le 0, \qquad \text{if } \Im(k)<0.
  \label{eq:F-negative}
\end{equation}
Thus, comparing \eqref{eq:F-positive} and \eqref{eq:F-negative} to \eqref{eq:tr_definition_maintext} and also taking in mind that all the coefficients $\tilde{\alpha}_{2n}$ are real, we find that the function in \eqref{eq:F} must be typically real. To compare directly with \eqref{eq:complex_series_maintext}, we normalise \eqref{eq:F} by
$(1-c_s^2)\tilde R$, introducing $f(z) \equiv \tilde\omega^2/[(1-\cs^2)\tilde R]$,
$z \equiv k/\tilde R$, and $\tilde a_{2n} \equiv \tilde\alpha_{2n}\tilde R^{2(n-1)}$. This
gives
\begin{equation}
    f(z) = z - \sum_{n=2}^{\infty} \frac{\tilde a_{2n}}{1-c_s^2} z^{2n-1} \, .
\end{equation}
Contrary to Section \ref{subsec:omega_bounds}, here we have a single typically real function. Hence, we can immediately use the Rogosinski bound in \eqref{eq:rogosinski_bound_maintext}, yielding
\begin{equation}
    -(2n-1)(1-\cs^2) \le \Tilde{a}_{2n} = \Tilde{\alpha}_{2n}\tR^{2(n-1)}  \le \kappa_{2n-1}(1-\cs^2) \label{eq:omega2boundscs}
\end{equation}
in which $\kappa_{2n-1}$ is defined as in \eqref{eq:kappa_maintext}.

\begin{figure}[H]
    \centering
    \includegraphics[width=\linewidth]{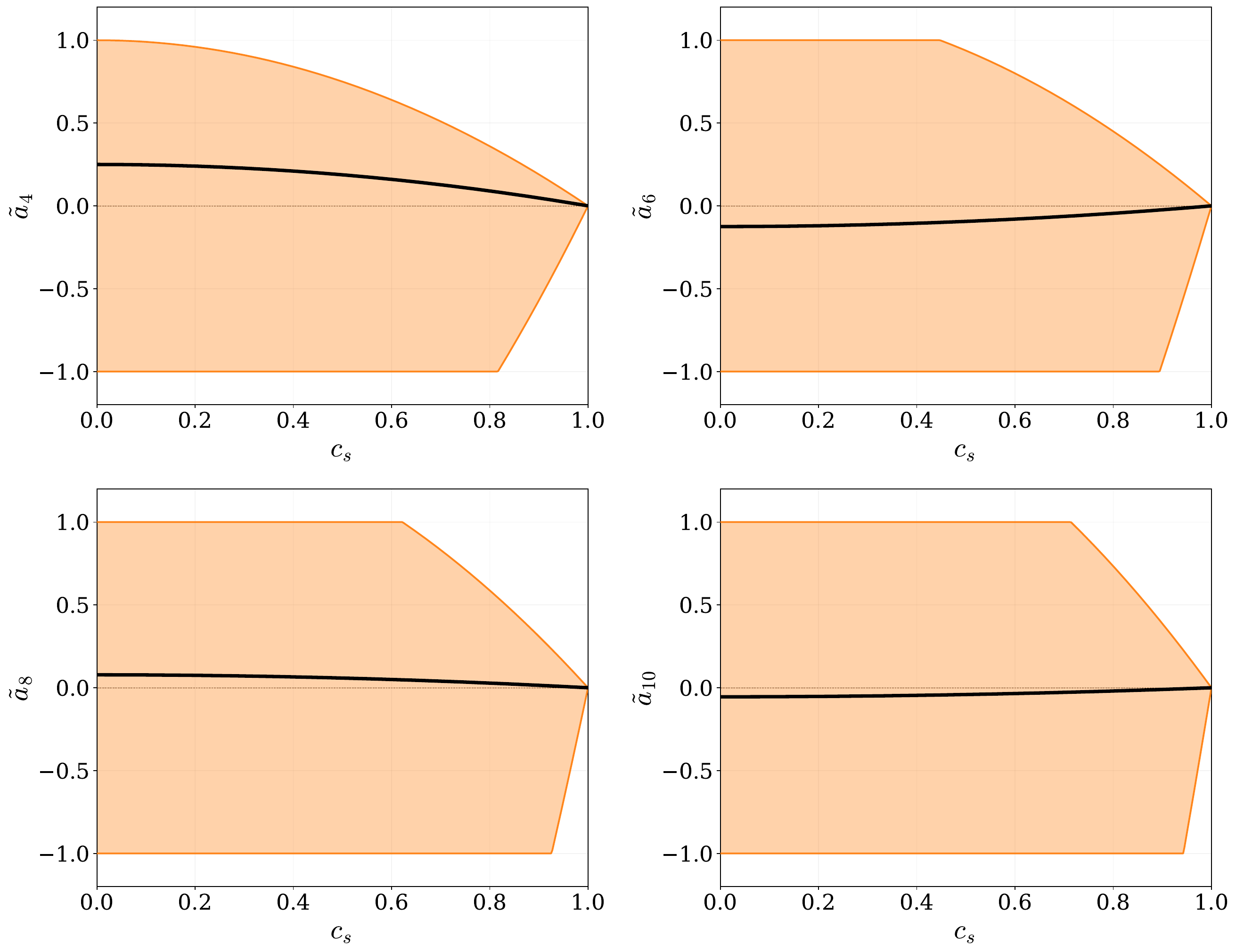}
    \caption{The bound \eqref{eq:mixed_bound} on $\Tilde{a}_4, \Tilde{a}_6, \Tilde{a}_8, \Tilde{a}_{10}$. The corresponding values for the $U(1)$ superfluid studied in Section \ref{subsec:u1_superfluid} are also shown (black line) for comparison.}
    \label{fig:tilde_a2n}
\end{figure}

\vspace{5 mm}
The two approaches of Sections \ref{sec:cs independent omega2} and \ref{sec:cs dependent omega2} are complementary. On one hand, the $\cs$-dependent bounds \eqref{eq:omega2boundscs} go to zero as $\cs\to 1$. However, moving away from this limit, they are worse than the bounds \eqref{eq:cs_independent_bound} especially for large $n$.  This motivates the combination of the bounds to obtain 
\begin{equation}
  -\min\big[(2n-1)(1-\cs^2),\,1\big] \le \Tilde{a}_{2n} \le \min\big[\kappa_{2n-1}(1-\cs^2),\,1\big]~.
  \label{eq:mixed_bound}
\end{equation}
These bounds are represented in Figure~\ref{fig:tilde_a2n} together with the coefficients of the $U(1)$ superfluid model.\footnote{It is also possible to prove analytically, for any $n$, the bounds on the gapless dispersion relation $\om^2(k)$ for the $U(1)$ superfluid model. The proof proceeds very similarly to the one in appendix \ref{app:checkmassive} for gapped dispersion relations.}

\begin{figure}[h]
    \centering
    \includegraphics[width=0.8\linewidth]{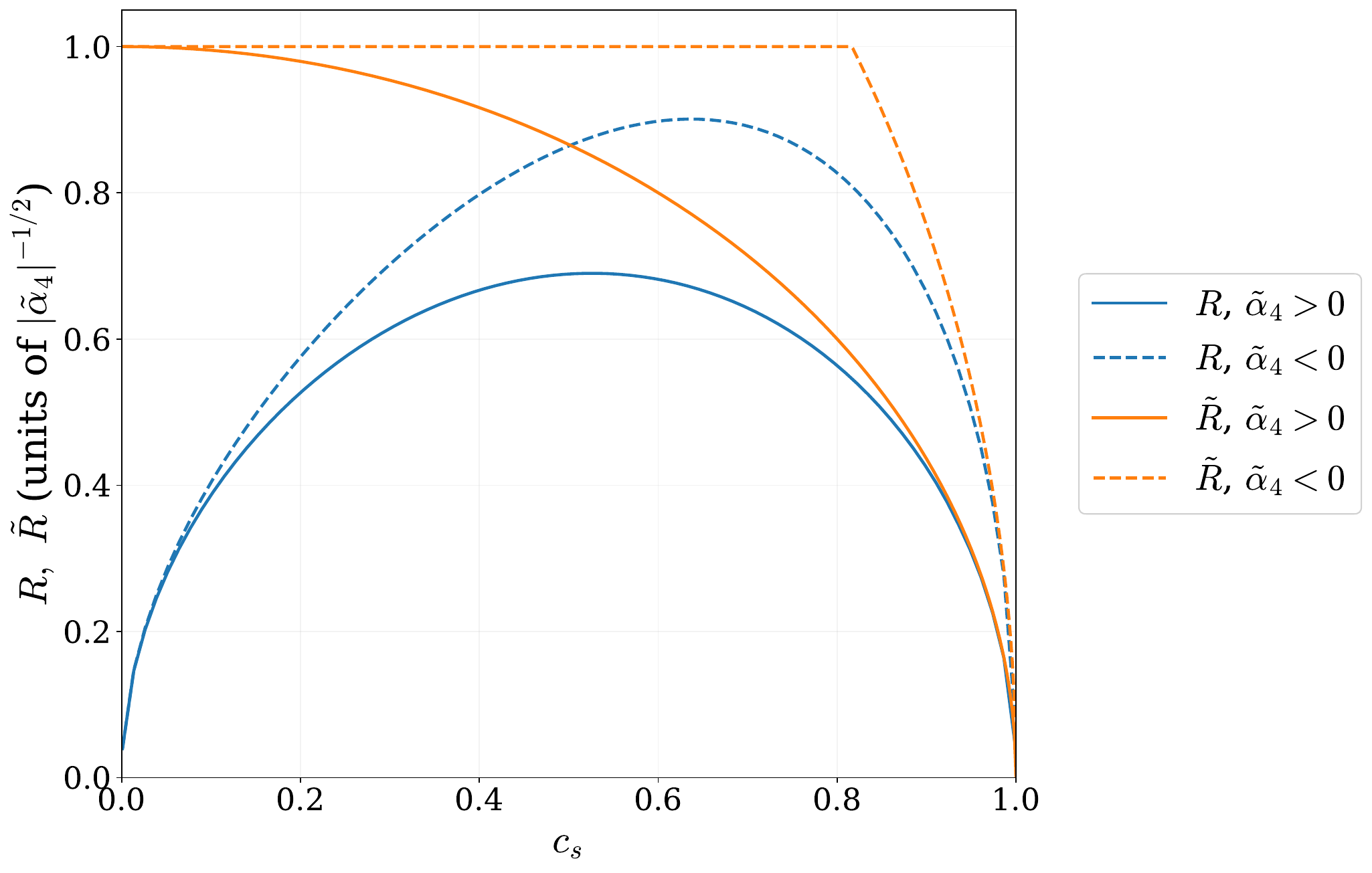}
    \caption{The maximum possible values of $R$ (in blue) and $\Tilde{R}$ (in orange) as function of the sound speed $c_s$ in units of $|\Tilde{\alpha}|_4^{-1/2}$.
    %\MM{I think here should be in units of $|\tilde{\alpha_4}|^{-1/2}$ both in the caption and also in the plot}. 
    Solid lines correspond to $\Tilde{\alpha}_4 > 0$ while dashed lines correspond to $\Tilde{\alpha}_4 < 0$.}
    \label{fig:R_compare}
\end{figure}

Summarizing, we have obtained two-sided bounds on the coefficients $\alpha_{2n+1}$ and $\Tilde{\alpha}_{2n}$ in units of the radii of convergence $R$ and $\Tilde{R}$. Again, one can revert the logic: if one measures the sound speed $c_s$ and the first higher-derivative correction $\Tilde{\alpha}_4$ (notice that $\alpha_3 = \Tilde{\alpha}_4/(2 c_s)$), one has upper bounds on $R$, the scale at which the linear Goldstone dispersion relation breaks down, and $\tilde R$, the scale at which new states must enter the game. See Figure \ref{fig:R_compare}.

Finally, we observe that in the $\cs\to 1$ limit one always has $R\to \tilde{R}$. Recall that $R$ is determined by the singularities of $\om(k)$ associated with taking the square root of $\om^2(k)$, while $\tilde{R}$ is determined by the singularities of $\om^2(k)$ itself. As the sound speed approaches unity, however, the higher-order coefficients in the expansion of $\om^2(k)$ vanish as $\mathcal{O}(1-\cs^2)$, as can be easily seen from \eqref{eq:mixed_bound}:
\begin{equation}
    \om^2(k)=k^2 + \mathcal{O}(1-\cs^2).
\end{equation}
Since the only zero of $\om^2(k)$ in this limit is at $k=0$, and this zero has even multiplicity, it does not give rise to a branch point of $\om(k)$. Therefore, the singularities associated with taking the square root recede to infinity
%\MM{Maybe here it's better "the singularities associated with taking the square root are pushed to the boundary $|k| = \tR$?} 
as $\cs\to 1$, leaving only the ones inherited from $\om^2(k)$. It follows that $R=\tilde{R}$ in this limit.

\section{Models and checks of the bounds}
\label{sec:tests}
In this section we will test our bounds on several theories with spontaneously Lorentz-breaking. Another model, related to the EFT of solids, is studied in appendix \ref{app:solid}.
\begin{figure}[p]   % 'p' = float page: only floats, no text
  \centering
  \begin{subfigure}[t]{\textwidth}
    \centering
    \caption{Bounds on $\omega(k)$}
    \label{fig:om_checks}
    \includegraphics[width=0.85\textwidth]{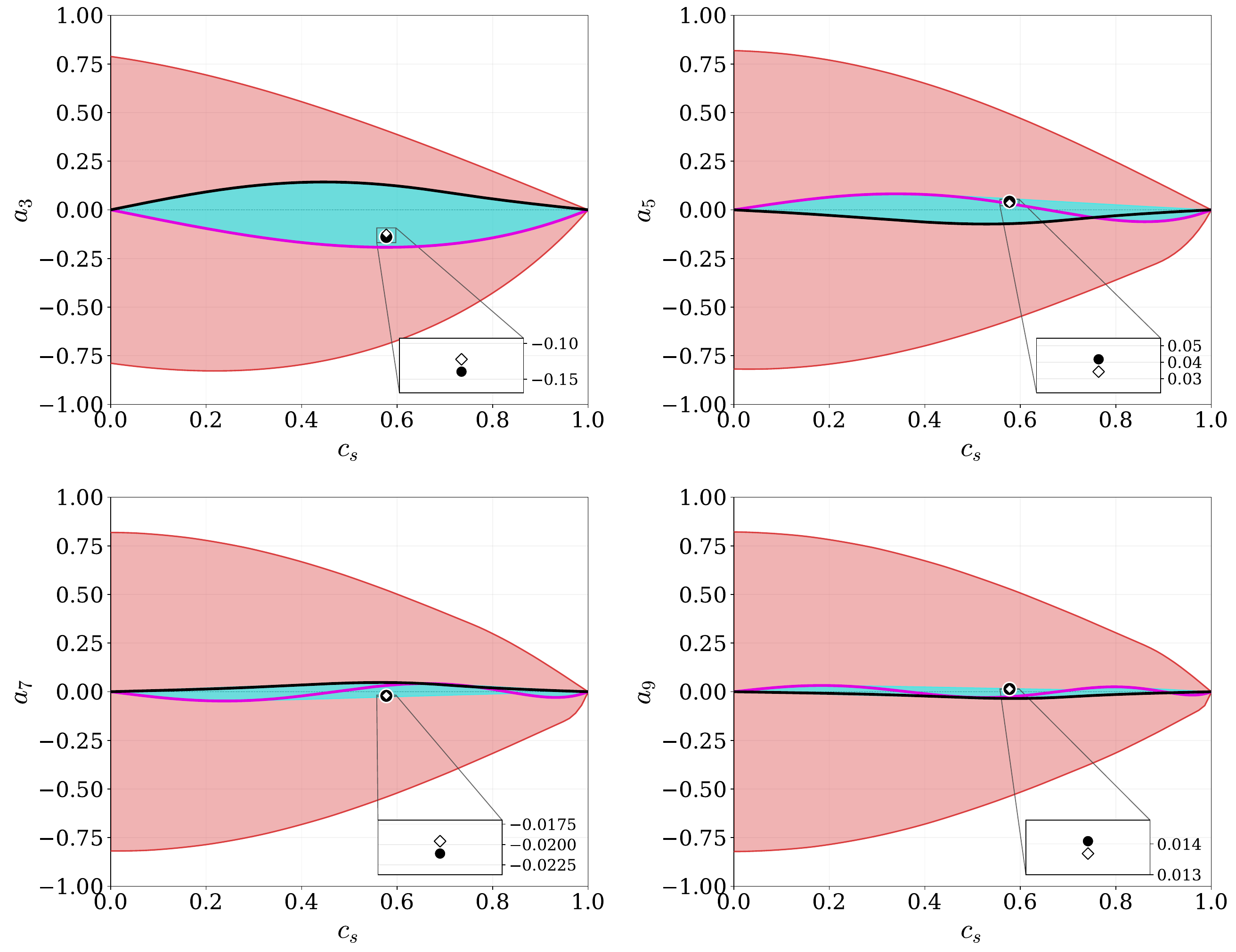}
  \end{subfigure}
  \begin{subfigure}[t]{\textwidth}
    \centering
       \caption{Bounds on $\omega^2(k)$}
    \label{fig:om2_checks}
    \includegraphics[width=0.85\textwidth]{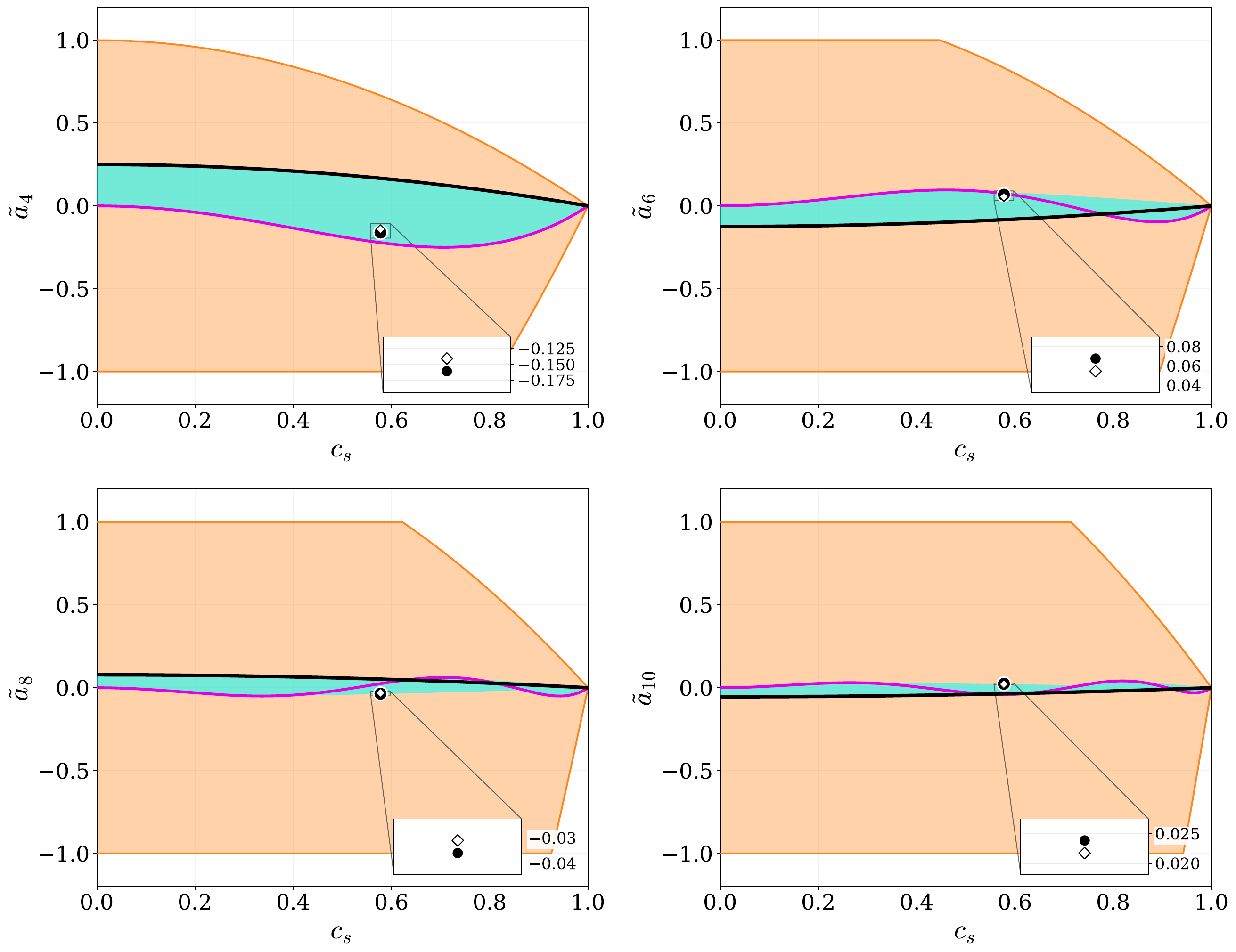}
  \end{subfigure}
  \caption{Drude-Lorentz model: $c_r = 0$ (magenta curve), $0 \leq c_r \leq 1$ (cyan region), $c_r = 1$ (black curve—equivalent to the $U(1)$ superfluid). Colour superconductor: octet Goldstones (black circle) and singlet Goldstones (white diamond).}
  \label{fig:tested_bounds_all}
\end{figure}

\subsection{Light in a medium: Drude-Lorentz Model}
When an electromagnetic field propagates in a medium, the bound charges are displaced from equilibrium and oscillate, resulting in a frequency-dependent refractive index, see e.g.~\cite{Jackson:1998nia}.  One can model this with an action of the form (see also \cite{Dubovsky:2007ac})
\begin{equation}
    S = \int \mathrm{d}^4x \left[\frac{1}{2}\dot{h}^2 - \frac{1}{2} c_r^2 (\vec{\nabla} h)^2 - \frac{1}{2}M^2 h^2 - \frac{1}{2}(\partial \pi)^2 - g h \dot{\pi} \right] \label{eq:action_drude}
\end{equation}
in which $h$ is a field with mass $M$ playing the role of the dipole oscillators in the medium (it should be a 3-vector, but we suppress its indices for simplicity). $\pi$ is a massless field and plays the role of the photon (again one should talk about the 3-vector $\vec A$ but we reduced it to a scalar toy-model). The mixing $g h \dot\pi $ describes the coupling between the photon and the dipoles, since $\dot\pi$ is analogous to the electric field $\dot {\vec A}$.  Localised dipoles are usually treated as independent one from the others, but here we also included a spatial kinetic term, which introduces a non-zero speed of propagation of the dipole mode. Note that when $c_r = 1$, the action \eqref{eq:action_drude} takes the same form as the $U(1)$ superfluid discussed above, with $g = 2\mu$
(the derivation of its action can be found for instance in \cite{Creminelli:2023kze}), with the angular variable taking the role of the photon and the radial mode the one of the dipoles. The action above is microcausal for $c_r \le 1$.

From \eqref{eq:action_drude}, we work in Fourier space and rewrite the action as
\begin{equation}
    S = \frac{1}{2} \int \frac{\mathrm{d}^4k}{(2\pi)^4}  \begin{bmatrix}
        \Tilde{h}(-\omega, -k)&\Tilde{\pi}(-\omega, -k)
    \end{bmatrix}
    \begin{bmatrix}
        (\omega^2 - c_r^2 k^2) - M^2 & -ig\omega  \\
        ig\omega  & (\omega^2 - k^2)
    \end{bmatrix}
    \begin{bmatrix}
        \Tilde{h}(\omega,k)\\
        \Tilde{\pi}(\omega, k)
    \end{bmatrix}.
\end{equation}
The dispersion relations for both modes correspond to the zero of the determinant of the kinetic matrix:
\begin{equation}
    (\omega^2 - c_r^2 k^2 - M^2)(\omega^2 - k^2) - g^2\omega^2 = 0. \label{eq:detK_drude}
\end{equation}
There are two branches of solutions: the gapped mode $(+)$ and the gapless mode $(-)$, with dispersion relations given by
\begin{equation}
     \omega_\pm^2 = \frac{1}{2}\left\{[\Delta^2 + (1 + c_r^2)k^2] \pm \sqrt{[\Delta^2 - (1 - c_r^2)k^2]^2 + 4\Delta^2 (1-c_s^2) k^2}\right\}
\end{equation}
in which $\Delta^2 = M^2 + g^2$ and $c_s^2 = 1 - g^2/\Delta^2$ (it would be perhaps more appropriate to call it $c_\gamma$). Notice that when $c_r = 1$, the dispersion relation reduces to that of the $U(1)$ superfluid \eqref{eq:dispersion_u1}.

Similar to the $U(1)$ case, the gapless dispersion relation $\omega_-(k)$ will have two possible sources of singularities. The first source are the branch points from the inner square root, which occur when
\begin{equation}
    [\Delta^2 - (1 - c_r^2)k^2]^2 + 4\Delta^2 (1-c_s^2) k^2 = 0.
\end{equation}
Solving the quadratic, we find
\begin{equation}
        \frac{k_*^2}{\Delta^2} = \frac{-(1 + c_r^2 - 2c_s^2)\pm 2\sqrt{(1-c_s^2)(c_r^2 - c_s^2)}}{(1 - c_r^2)^2}. \label{eq:dispersion_drude}
\end{equation}
The expression inside the square root will be positive when $c_s < c_r$ and negative when $c_s > c_r$. Therefore, we have two cases to consider.
\begin{enumerate}
    \item \emph{$c_s \leq c_r$}: In this case, $k_*^2$ will always be negative, meaning that $k_*$ is purely imaginary. Thus, we have
    \begin{equation}
        k_* = \pm i\Delta \left[\frac{\sqrt{1 - c_s^2}\pm \sqrt{c_r^2 - c_s^2}}{1 - c_r^2}\right].
    \end{equation}
    Notice that when $c_r \to 1$, the inner plus branch grows to infinity while the inner minus branch approaches the $U(1)$ value \eqref{eq:bp_u1_om2}.
    \item \emph{$c_s > c_r$}: In this case, we get
    \begin{equation}
        \frac{k_*^2}{\Delta^2} = \frac{-(1 + c_r^2 - 2c_s^2)\pm 2i\sqrt{(1-c_s^2)(c_s^2 - c_r^2)}}{(1 - c_r^2)^2}.
    \end{equation}
    Notice that when $c_s = c_r$, $k_*$ is purely imaginary, while when $c_s = 1$, $k_*$ is purely real. Taking the square root and rewrite in polar form, we find
    \begin{equation}
        k_* = \pm \frac{\Delta}{\sqrt{1 - c_r^2}} \exp \left[\pm i\frac{1}{2}\arg\left(\frac{k_*^2}{\Delta^2}\right)\right].
    \end{equation}
\end{enumerate}
The second source are the branch points from the outer square root, which occur when $\omega^2_- = 0$. Returning to the quadratic \eqref{eq:detK_drude}, setting $\omega = 0$ and solve for $k$, we find
\begin{equation}
    k_* = \pm i \frac{c_s}{c_r} \Delta,\qquad 0 \leq c_s \leq \frac{c_r}{\sqrt{1 + c_r^2}}.
\end{equation}
As such, the radius of convergence of the dispersion relation is given by
\begin{equation}
    R = \begin{cases}
        \frac{c_s}{c_r} \Delta, & 0 < c_s \leq \frac{c_r}{\sqrt{1 + c_r^2}},\\
        \left[\frac{\sqrt{1 - c_s^2} - \sqrt{c_r^2 - c_s^2}}{1 - c_r^2}\right]\Delta , & \frac{c_r}{\sqrt{1 + c_r^2}} < c_s \leq c_r,\\
         \frac{\Delta}{\sqrt{1 - c_r^2}}, & c_r < c_s < 1,
    \end{cases}; \quad \tilde R = \begin{cases}
        \left[\frac{\sqrt{1 - c_s^2} - \sqrt{c_r^2 - c_s^2}}{1 - c_r^2}\right]\Delta , & 0 < c_s \leq c_r,\\
         \frac{\Delta}{\sqrt{1 - c_r^2}}, & c_r < c_s < 1.
    \end{cases}
\end{equation}
The bounds are checked in Figure \ref{fig:tested_bounds_all}.

For non-relativistic media, the response is approximately local in space: $c_r \ll 1$. In this limit, the radius of convergence for $\omega(k)$ and $\omega^2(k)$ is the same. For all values of $c_s$ one has
\begin{equation}
    R = \Tilde{R} = \Delta.
\end{equation}

One can also consider the limit when the effect of the medium is small so that $c_s$ is close to 1.
In this limit, the radius of convergence for $\omega(k)$ and $\omega^2(k)$ is the same, as discussed. For all values of $c_r$
\begin{equation}
    R = \Tilde{R} = \frac{\Delta}{\sqrt{1 - c_r^2}}.
\end{equation}
In this case, the coefficients of the dispersion relation can be explicitly written for all values of $n$, allowing a direct check of the bounds:

$\omega(k)$:
In this limit, the series expansion becomes
\begin{equation}
    \omega(k) = c_s k - (1 - c_s) \sum_{n = 1}^{\infty} \frac{(1 - c_r^2)^n}{\Delta^{2n}} k^{2n+1}.
\end{equation}
Thus, we have $\alpha_{2n+1} R^{2n} = -(1 - c_s)$, which satisfies the bounds \eqref{eq:bounds_csto1} for all values of $n$.

$\omega^2(k)$:
In this limit, the series expansion becomes
\begin{equation}
    \omega^2(k) = c_s^2 k^2 - (1 - c_s^2)\sum_{n = 2}^{\infty} \frac{(1 - c_r^2)^{n - 1}}{\Delta^{2n-2}} k^{2n}.
\end{equation}
Thus, we have $\Tilde{\alpha}_{2n+2} \tilde R^{2n} = -(1 - c_s^2)$, which satisfies the bounds \eqref{eq:mixed_bound} for all values of $n$.

\subsection{Light propagation in a homogeneous magnetic field}
Loops of electrons make QED effectively non-linear. Therefore, in an external homogeneous electric or magnetic field, a photon develops a non-trivial dispersion relation.\footnote{Contrary to our general discussion, here also rotations are broken. Our bounds apply straightforwardly once the direction of propagation is chosen.} The study of this setup started with the Euler-Heisenberg effective Lagrangian~\cite{Heisenberg:1936nmg} and was later refined to apply in a variety of kinematic regimes: here we will follow Ref.~\cite{Karbstein:2013ufa}.
\begin{figure}[H]
\centering
\def\panelheight{5cm}% common height of both panels
%
% ---------- (a) Photon entering a magnetic field ----------
\begin{subfigure}[b]{0.48\linewidth}
\centering
\resizebox{!}{\panelheight}{%
\begin{tikzpicture}[
    tri/.style={-{Triangle[length=3mm,width=3mm]}},
    fieldline/.style={line width=1.6pt, fieldgreen, tri},
]
% magnetic field lines
\foreach \x in {3,4.2,5.4,6.6,7.8}{
    \draw[fieldline] (\x,-1.4) -- (\x,5.4);
}
\node[fieldgreen, anchor=south west, font=\huge] at (8.0,5.1) {$\vec{B}$};
 
% two photon wave packets (wave packet + momentum), heads at y = 3.6 and y = 0.4
\foreach \y in {3.6,0.4}{
    \filldraw[photonred, fill=photonfill, line width=1.2pt]
        (-0.8,\y) ellipse [x radius=0.8cm, y radius=0.34cm];
    \draw[photonred, line width=1.2pt, smooth, samples=80, domain=-1.45:-0.15]
        plot (\x, {\y + 0.24*sin(deg(2*pi*(\x+1.45)/0.43))});
    \draw[photonred, line width=1.6pt, tri] (0,\y) -- (1.6,\y)
        node[above=1mm, pos=0.85, font=\huge] {$\vec{k}$};
}
 
% upper photon: parallel polarisation (along B)
\draw[polorange, line width=1.6pt, tri] (0,3.6) -- (0,4.8)
    node[above=1mm, font=\huge] {$\vec{\epsilon}_\parallel$};
 
% lower photon: perpendicular polarisation (out of the page)
\filldraw[polorange, fill=polyellow, line width=1.6pt] (0,0.4) circle (1.7mm);
\fill[polorange] (0,0.4) circle (0.7mm);
\node[polorange, below=3mm, font=\huge] at (0,0.4) {$\vec{\epsilon}_\perp$};
\end{tikzpicture}%
}
\caption{Photons entering a constant homogeneous external magnetic field.}
\label{fig:photon-B}
\end{subfigure}
\hfill
%
% ---------- (b) Vacuum polarisation in the external field ----------
\begin{subfigure}[b]{0.48\linewidth}
\centering
\resizebox{!}{\panelheight}{%
\begin{tikzpicture}[
    photon/.style={decorate,
        decoration={snake, amplitude=1.3mm, segment length=3.5mm,
                    pre length=1pt, post length=1pt}},
    extphoton/.style={photon, fieldgreen, line width=1.2pt},
    vacphoton/.style={photon, vacred, line width=1.4pt,
        decoration={snake, amplitude=2.6mm, segment length=6mm}},
    loop/.style={line width=1.6pt, black, postaction={decorate},
        decoration={markings,
            mark=at position 0.355 with {\arrow{Triangle[length=3mm,width=3mm]}},
            mark=at position 0.855 with {\arrow{Triangle[length=3mm,width=3mm]}}}},
]
\def\R{1.2}     % fermion loop radius
\def\S{3.0}     % distance of external sources from the centre
\def\r{0.38}    % source radius
 
% external-field insertions
\foreach \a in {90,35,-35,-90,-145,145}{
    \draw[extphoton] (\a:\R) -- (\a:{\S-\r});
    \begin{scope}[shift={(\a:\S)}]
        \filldraw[fieldgreen, fill=sourcefill, line width=1.2pt] (0,0) circle (\r);
        \draw[fieldgreen, line width=1.2pt] (45:\r) -- (225:\r) (135:\r) -- (-45:\r);
    \end{scope}
}
 
% dots: infinitely many insertions
\foreach \a in {62.5,-62.5,-117.5,117.5}{
    \foreach \d in {-10,0,10}{
        \fill[fieldgreen] ({\a+\d}:{\S+0.35}) circle (2.2pt);
    }
}
 
% vacuum photons
\draw[vacphoton] (-4.8,0) -- (180:\R);
\draw[vacphoton] (0:\R) -- (4.8,0);
 
% fermion loop
\draw[loop] (\R,0) arc[start angle=0, end angle=360, radius=\R];
\node[font=\huge] at (0,0) {$\Pi_{\mu\nu}$};
\end{tikzpicture}%
}
\caption{One-loop vacuum polarisation in the presence of external strong magnetic field}
\label{fig:vacpol}
\end{subfigure}
\caption{The setup we consider in this section.}
\label{fig:photon-B-vacpol}
\end{figure}
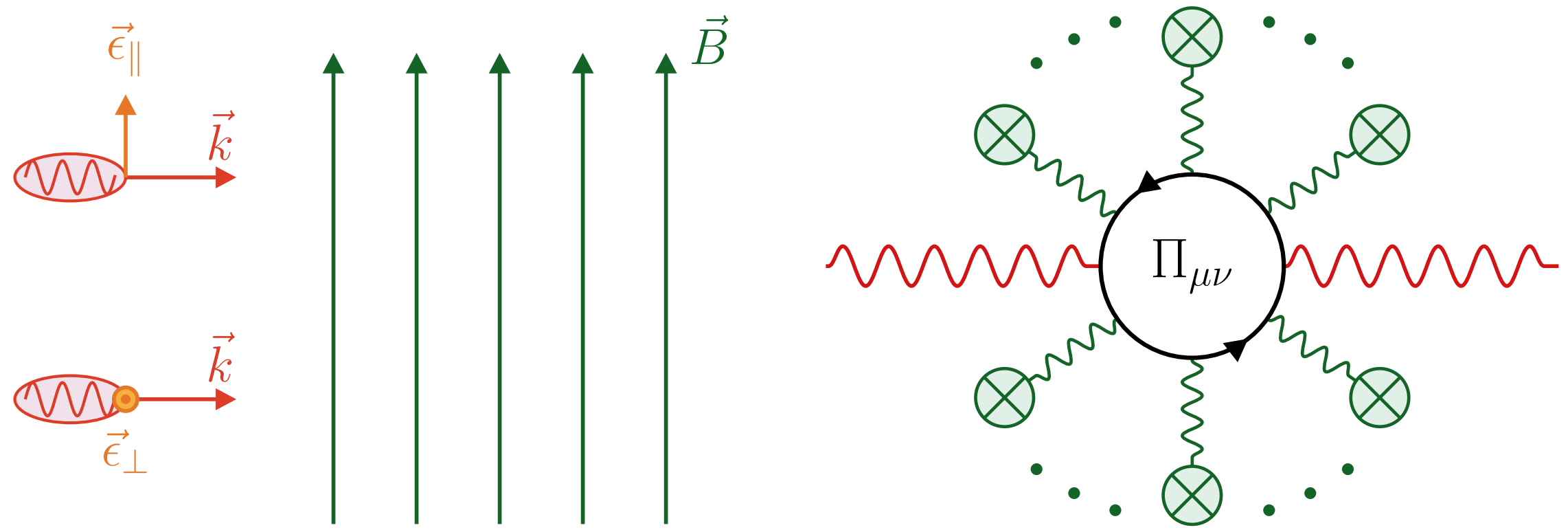
Consider a static and homogeneous magnetic field\footnote{We consider magnetic fields here to avoid the complications of the Schwinger effect, which induces absorption of the photons. For magnetic fields, we will see that pair production is only active above a certain threshold of the photon energy, independently of the magnetic field strength.} $\vec{B}$, and a photon propagating with momentum $\vec{k}\perp\vec{B}$, see figure \ref{fig:photon-B-vacpol}. In this configuration, Lorentz symmetry spontaneously breaks into boosts along the magnetic field and rotations around it. Under this residual symmetry, the photon polarisation tensor $\Pi^{\mu\nu}$ can be decomposed into three components: polarisation parallel to the external field ($\Pi_\parallel$), polarisation perpendicular to the external field ($\Pi_\perp$), and a ``longitudinal'' component ($\Pi_0$) that does not contribute to the on-shell photon propagation.

We work in the strong-field regime: $eB\gg m^2,\,k^2$. (Such extreme magnetic fields are actually possible in magnetars, see for instance \cite{Harding:2006qn}.) In this limit, the components of the photon-polarisation tensor up to one-loop are calculated in \cite{Karbstein:2013ufa} (see his eq.~(188)):
\begin{equation}
\begin{split}
\Re\begin{Bmatrix} \Pi_\parallel \\ \Pi_\perp \\ \Pi_0 \end{Bmatrix}
&= \frac{\alpha}{2\pi}\Bigg[
\left(eB - \frac{k^2}{2}\right)\int_{-1}^{1} \frac{\mathrm{d}\nu}{2}\,
\frac{1-\nu^2}{m^2 - \frac{1 - \nu^2}{4} \omega^2}
\begin{Bmatrix} -\omega^2 \\ 0 \\ 0 \end{Bmatrix} \\[2mm]
&\quad + \frac{2}{3}\left[\ln\!\left(\frac{m^2}{2eB}\right)+\gamma\right](-\omega^2 + k^2)
+ \Sigma\begin{Bmatrix} k^2 \\ -\omega^2 + k^2 \\ -\omega^2 + k^2 \end{Bmatrix}
\\[2mm] &\quad + \left[\frac{2}{3}+\frac{m^2}{eB}\ln\!\left(\frac{m^2}{2eB}\right)\right]
\begin{Bmatrix} k^2 \\ -\omega^2 \\ -\omega^2 + k^2 \end{Bmatrix}\Bigg] + {{\cal O}\left(\frac{1}{eB}\right)}
\label{eq:polarisation_tensors}
\end{split}
\end{equation}
in which $\gamma$ is the Euler-Mascheroni constant and $\Sigma \simeq 0.605$ is another constant defined in \cite{Karbstein:2013ufa}. The $\parallel$ mode also has a non-zero imaginary part given by
\begin{equation}
    \Im\{\Pi_\parallel\} = -\alpha\, eB\, \mathrm{e}^{-\frac{k^2}{2eB}}\,
\frac{\left(\frac{2m}{\omega}\right)^2}{\sqrt{1 - \left(\frac{2m}{\omega}\right)^2}}\,
\theta(\omega^2 - 4m^2) \label{eq:im_eh}
\end{equation}
while $\Im\{\Pi_\perp\} = \Im\{\Pi_0\} = 0$. Thus, in what follows, we restrict our analysis to $\omega < 2m$, where $\Pi_{\parallel}$ has no imaginary part.

The dispersion relation for each polarisation mode can be obtained by solving
\begin{equation}
-\omega^2+k^2+\Pi_p=0 \label{eq:EH_dispersion_relation}
\end{equation}
with $p=\parallel, \perp$ depending on the polarisation of the photon with respect to $\vec{B}$. We work in the regime $\alpha\ll 1$, $eB/m^2\gg 1$ with $\alpha\, eB/m^2\ll1$: the correction to the vacuum dispersion relation is small.

\subsubsection*{\texorpdfstring{$\perp$}{perp}-polarisation}
Solving \eqref{eq:EH_dispersion_relation} for the $\perp$ polarisation mode, we find:
\begin{equation}
\omega = c_s k,\qquad c_s = 1 - \frac{\alpha}{6\pi}.
\end{equation}
This is clearly compatible with microcausality since $c_s <1$ and all higher-order terms vanish. Eventually, for $\omega \to \infty$, photons will come back to the vacuum lightcone, but this happens outside the regime of validity of the strong-field expansion.

\subsubsection*{\texorpdfstring{$\parallel$}{parallel}-polarisation}
Consider \eqref{eq:polarisation_tensors}: in the strong-field limit, the $\mathcal{O}(eB)$ term dominates. Evaluating the integral in the leading term one finds
\begin{equation}
    \Pi_\parallel \approx \frac{\alpha eB}{2\pi} \int_{-1}^{1} \frac{\mathrm{d}\nu}{2}\,
\frac{(1-\nu^2)(-\omega^2)}{m^2 - \frac{1 - \nu^2}{4} \omega^2} = \frac{2\alpha eB}{\pi} \left[1 - \frac{\sin^{-1}\left(\frac{\omega}{2m}\right)}{\left(\frac{\omega}{2m}\right) \sqrt{1 - \left(\frac{\omega}{2m}\right)^2}}\right].
\end{equation}
Thus, using \eqref{eq:EH_dispersion_relation}, the dispersion relation is given by
\begin{equation}
    -\omega^2 + k^2 + \frac{2\alpha eB}{\pi} \left[1 - \frac{\sin^{-1}\left(\frac{k}{2m}\right)}{\left(\frac{k}{2m}\right) \sqrt{1 - \left(\frac{k}{2m}\right)^2}}\right] = 0,\label{eq:EH_dispersion_relation_parallel}
\end{equation}
where we used $\omega = k$ in $\Pi_\parallel$ since we are only keeping terms linear $\alpha eB$. The dispersion relation has inverse-square-root branch points at $k = \pm 2 m$. This implies $R = \tilde R = 2m$. Notice this singularity corresponds to the pair-production threshold.

\paragraph{\texorpdfstring{$\om(k)$}{omega(k)}:} 
The expansion of \eqref{eq:EH_dispersion_relation_parallel} gives
\begin{equation}
    \cs \simeq 1- \frac{\alpha eB}{6\pi m^2},\qquad \alpha_{2n+1} \simeq -\frac{\alpha e B}{\pi m^{2n + 2}} \frac{[(n+1)!]^2}{(2n+3)!} \quad (n \geq 1).
\end{equation}
In the regime $\alpha eB/m^2 \ll 1$ one has $1 - c_s \ll 1$. We can compare our results directly to the $c_s \to 1$ limit bounds \eqref{eq:bounds_csto1}. Rewriting $\alpha_{2n+1}$ in terms of $R$ and $c_s$, we get
\begin{equation}
    \alpha_{2n+1}R^{2n} = -3\cdot 2^{2n+1}\frac{[(n+1)!]^2}{(2n+3)!} (1 - c_s)
\end{equation}
which satisfies \eqref{eq:bounds_csto1} for all $n \geq 1$.
\paragraph{\texorpdfstring{$\om^2(k)$}{omega2(k)}:} The coefficients $\tilde{\alpha}_{2n}$ are given by
\begin{equation}
    \tilde{\alpha}_{2n} \simeq -\frac{2\alpha eB}{\pi m^{2n}}\frac{(n!)^2}{(2n+1)!} \quad (n\geq 2).
    \label{eq:EH_omega2_coefficients}
\end{equation}
We rewrite $\Tilde{\alpha}_{2n}$ in terms of $\Tilde{R}$ and $c_s$:
\begin{equation}
    \tilde{\alpha}_{2n}\Tilde{R}^{2n - 2} \simeq -6\cdot 4^{n-1}\frac{(n!)^2}{(2n+1)!}(1-\cs^2),
\end{equation}
which satisfies \eqref{eq:mixed_bound} for all $n \geq 2$.

\subsection{Colour superconductor}
Another system of interest, both for its rich analytic structure and its potential relevance to neutron star cores, is strongly interacting quark matter at high baryon density and low temperature.\footnote{For a comprehensive review of this subject, we refer readers to \cite{Alford:2007xm}.} In the large density regime, due to asymptotic freedom, quantum chromodynamics (QCD) becomes weakly coupled and perturbative; at low temperatures, the quarks form a diquark condensate which spontaneously breaks the colour gauge symmetry. This diquark condensate is the QCD analogue of the Cooper-pair condensate in BCS superconductivity, and the resulting phase is accordingly termed a \textit{colour} superconductor.

Although there can be multiple pairing possibilities between the two quarks, if the up, down and strange quarks\footnote{Only these three flavours are relevant for typical densities in neutron star cores.} are considered light, the preferred pairing is colour-flavour locking (CFL). That is:
\begin{equation}
    \langle {q_L}^{i}_{a} \, {q_L}^{j}_{b} \rangle
= -\langle {q_R}^{i}_{a} \, {q_R}^{j}_{b} \rangle
= \Delta\, \epsilon^{ijk}\epsilon_{abk}
\end{equation}
where $i, j$ and $a, b$ are the colour and flavour indices, respectively, and $\Delta$ is the gap parameter. In this phase, the condensate breaks the original symmetries of the QCD theory in the following pattern
\begin{equation}
    [SU(3)_c] \times SU(3)_V \times SU(3)_A \times U(1)_B \times U(1)_A \quad \to \quad SU(3)_{c + V} \times \mathbb{Z}_2
\end{equation}
with the brackets signifying a gauge group. This yields multiple Goldstones: the breaking of $SU(3)_A$ yields octet Goldstones\footnote{They are also called `pions' due to their similar spectrum to pions in vacuum QCD.} while the breaking of $U(1)_B$ and $U(1)_A$\footnote{Strictly speaking, the $U(1)_A$ symmetry is strongly broken in QCD due to anomalous effects. However, the anomalous effects become suppressed when baryon density $\mu$ is large.} each yield singlet Goldstones.

\subsubsection*{Octet Goldstones}
The inverse of the two-point function of the octet Goldstones is calculated in \cite{Zarembo:2000pj}:
\begin{multline}
    K_o(\omega,k) = -\frac{\mu^2\Delta^2}{24\pi^2 k} \int_0^1 dx \int_0^k d\alpha\,(\alpha^2-\omega^2) \bigg[\frac{5}{x(1-x)(\alpha^2-\omega^2)+\Delta^2}\\+\frac{4-6x}{x(1-x)(\alpha^2-\omega^2)+(4-3x)\Delta^2}\bigg],
\end{multline}
in which $\mu$ is the baryon density and $\Delta$ is the mass gap of the colour superconductor. To solve for the dispersion relation, we set $\Pi(\omega,k) = 0$ and solve for $\omega$ perturbatively.

\begin{figure}[h]
\captionsetup[subfigure]{margin={0.1\linewidth,0pt}}
    \centering
    \begin{subfigure}[c]{0.49\textwidth}
        \centering
        \includegraphics[width=\linewidth]{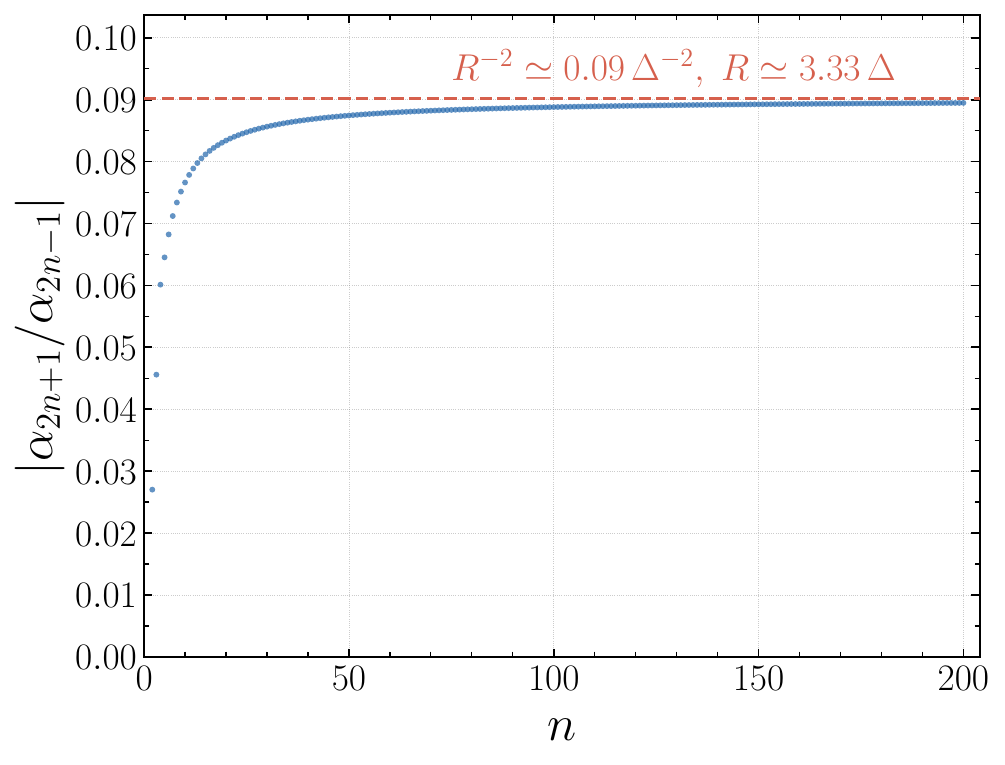}
        \caption{Ratio test}
        \label{fig:pi_ratiotest}
    \end{subfigure}
    \hfill
    \begin{subfigure}[c]{0.49\textwidth}
        \centering
        \includegraphics[width=\linewidth]{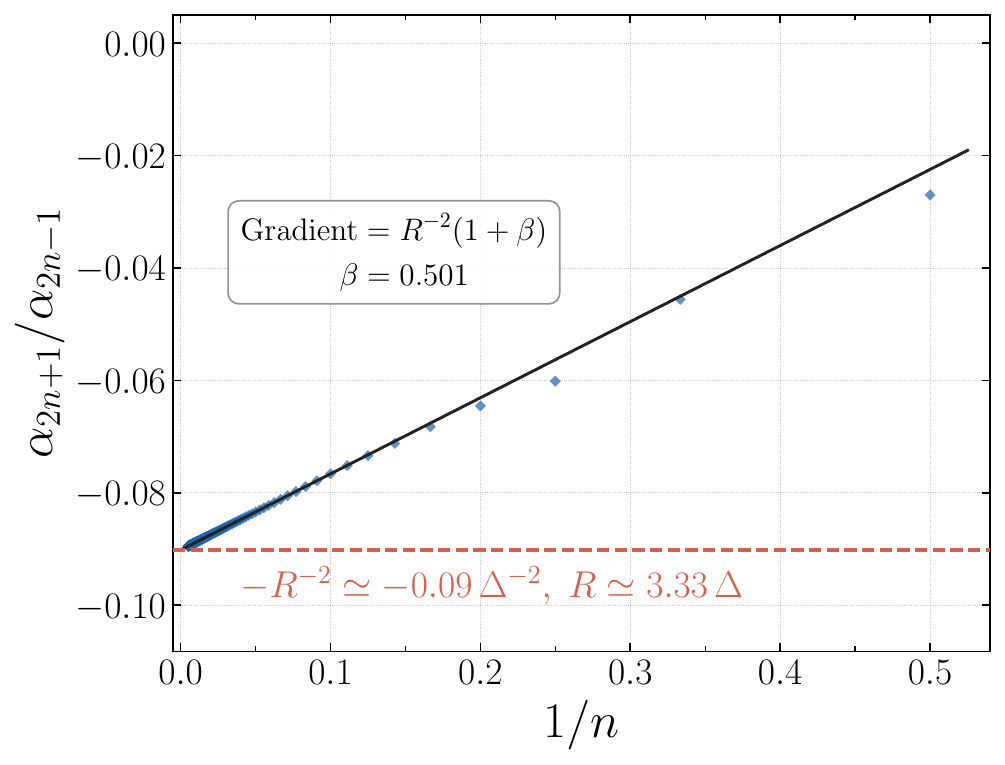}
        \caption{Method of Domb-Sykes}
        \label{fig:pi_dombsykes}
    \end{subfigure}
    \caption{The two methods to determine the radius of convergence applied to the first 200 values of $\alpha_{2n+1}$ for the octet Goldstones. Both methods give a radius of convergence $R \simeq 3.33 \Delta$. Additionally, the Domb-Sykes method shows that $\beta = 0.501$, which indicates a square root branch point.}
    \label{fig:r_pions}
\end{figure}

If $\omega(k)$ has a singularity at real $k^2_*$, as $k \to k_*$, $\omega(k)$ must take the following form:
\begin{equation}
    \omega(k) \sim \left(1 - \frac{k^2}{k_*^2}\right)^\beta
\end{equation}
with $\beta$ being any non-integer. When $\omega(k)$ is solved perturbatively, one obtains a power series of the form \eqref{eq:omega_series_odd}. Given that $\alpha_{2n+1}$ is known numerically, the simplest approach to determine the radius of convergence $R$ is to perform a ratio test. That is, 
\begin{equation}
    R^2 = \lim_{n \to \infty} \left|\frac{\alpha_{2n-1}}{\alpha_{2n+1}}\right|.
\end{equation}
In practice, given that we have access to a finite amount of $\alpha_{2n+1}$, we have to estimate this value by plotting the ratio $|\alpha_{2n-1}/\alpha_{2n+1}|$ against $n$ and `eyeball' the asymptote. A method to reduce this uncertainty is pointed out by C. Domb and M. F. Sykes in \cite{Domb_Sykes_1957}: one could instead plot the inverse ratio $\alpha_{2n+1}/\alpha_{2n-1}$ against $1/n$. This is because for many simple singular functions, the inverse ratio is linear in $1/n$ i.e.
\begin{equation}
    \frac{\alpha_{2n+1}}{\alpha_{2n-1}} \simeq \pm \frac{1}{R^2} \left(1 - \frac{1 + \beta}{n}\right)
\end{equation}
By linear extrapolating the points at small $1/n$, the radius of convergence is given by the $y$-intercept of the straight line. The gradient of the line depends on the type of singularity at $k_*$.
\paragraph{\texorpdfstring{$\om(k)$}{omega(k)}:} Substituting \eqref{eq:omega_series_odd} into $\Pi(\omega,k) = 0$ and solving order by order in $k$. The first few terms of $\omega(k)$ are:
\begin{multline}
    \omega_o(k) = \pm \frac{1}{\sqrt{3}}k\bigg[1 - \frac{1}{135} \frac{123 - 112 \ln 2}{21 - 8 \ln 2} \frac{k^2}{\Delta^2}\\
    + \frac{1}{85050} \frac{152019+32\ln 2(-7641 + 1912 \ln 2)}{(21-8\ln 2)^2} \frac{k^4}{\Delta^4} - \dots\bigg] \label{eq:dispersion_pions}
\end{multline}
which shows that the sound speed is $c_s = 1/\sqrt{3}$. The radius of convergence is determined numerically in Figure \ref{fig:r_pions} using the ratio test (Fig.~\ref{fig:pi_ratiotest}) and the method of Domb-Sykes (Fig. \ref{fig:pi_dombsykes}). Both methods show that $R \simeq 3.33 \Delta$ and that the type of singularities dictating $R$ are square root branch points.
\paragraph{\texorpdfstring{$\om^2(k)$}{omega2(k)}:} We can square \eqref{eq:dispersion_pions} to find, for the first few terms,
\begin{equation}
    \omega_o^2(k) = \frac{1}{3}k^2\bigg[1 - \frac{2}{135} \frac{123 - 112 \ln 2}{21 - 8 \ln 2} \frac{k^2}{\Delta^2} + \frac{8}{25515} \frac{669 - 848 \ln 2}{21-8\ln 2} \frac{k^4}{\Delta^4} - \dots\bigg].
\end{equation}
Following a similar analysis, the radius of convergence is $\Tilde{R} \simeq 3.33 \Delta$. 

\subsubsection*{Singlet Goldstones}
The inverse of the two-point function of the singlet Goldstones is given by
\begin{multline}
 K_s(\omega,k) = -\frac{\mu^2\Delta^2}{\pi^2 k} \int_0^1 dx \int_0^k d\alpha\,(\alpha^2-\omega^2) \bigg[\frac{2}{x(1-x)(\alpha^2-\omega^2)+\Delta^2}\\+\frac{1}{x(1-x)(\alpha^2-\omega^2)+4\Delta^2}\bigg].
\end{multline}
\paragraph{\texorpdfstring{$\om(k)$}{omega(k)}:} Substituting \eqref{eq:omega_series_odd} into $K_s(\omega, k) = 0$ and solving order by order in $k$, the first few terms of $\omega(k)$ are:
\begin{equation}
    \omega_s(k) = \pm \frac{1}{\sqrt{3}} k \left[1 -\frac{11}{540} \frac{k^2}{\Delta^2} + \frac{2249}{4082400} \frac{k^4}{\Delta^4} - \dots \right]. \label{eq:dispersion_singlet}
\end{equation}
which shows that the sound speed is $c_s = 1/\sqrt{3}$. The radius of convergence is determined numerically as before and one gets $R \simeq 3.22 \Delta$, with a square-root branch point. 

\paragraph{\texorpdfstring{$\om^2(k)$}{omega2(k)}:}
Squaring \eqref{eq:dispersion_singlet}, we find, for the first few terms:
\begin{equation}
    \omega_s^2(k) = \frac{1}{3}k^2 \left[1 - \frac{11}{270} \frac{k^2}{\Delta^2} + \frac{43}{28350} \frac{k^4}{\Delta^4} +\cdots\right].
\end{equation}
Following a similar analysis, the radius of convergence is $\Tilde{R} \simeq 3.22\Delta$.

\vspace{3 mm}
\noindent The bounds are checked for both octet and singlet Goldstones in Figure \ref{fig:tested_bounds_all} (\footnote{As discussed in the introduction, $\omega^2(k)$ may not satisfy microcausality until the radius of convergence, but only in a smaller disk. 
If the retarded Green function $\tilde G(\om,k)$ has branch cuts,  due to loops for instance,
as $k$ moves away from the origin, $\om^2(k)$ may reach
one of these branch points or cross one of the cuts, and thereby move onto the second
sheet of $\tilde G(\om,k)$ (see \cite{Gavassino:2026fil, Brants:2026qlw}). This can happen while $\om^2(k)$ remains analytic. When the singularity is on the second sheet, it is not constrained by microcausality. In this situation our bounds should be derived imposing microcausality on a smaller disk: $\tilde R$ will not be the radius of convergence, but the maximal radius for which the pole remains in the first sheet. Notice that in both cases our bounds are upper bounds on the scale of ``new physics'', either in the form of a collision with an extra singularity, or in the presence of a branch cut. The same can be repeated for the radius $R$, associated with the $\omega$-bounds. 

The colour-superconductor model of this section provides an explicit example of this phenomenon.  For the singlet Goldstone, for instance, the retarded two-point function has a square-root branch point
at $\om^2 = k^2 + 4\Delta^2$, associated with the production of a pair of octet
quasiparticles of gap $\Delta$. The root $\om^2(k)$ touches this branch point at
$k \simeq i\,2.95\Delta$. The radius of convergence was on the other hand found to be at $|k| \simeq 3.22\Delta$. The bounds for the colour superconductor model are thus checked using $\tilde R = 2.95\Delta$, since only in this disk the microcausality condition, eq.~\eqref{eq:bounds_omega2}, is guaranteed. (It turns out that, in this specific model, the inequality eq.~\eqref{eq:bounds_omega2} remains true for larger radii, even larger than the radius of convergence, but this is not guaranteed by microcausality.) \label{fnot:radius_of_validity}}).

\section{Cosmological implications}\label{sec:cosmo}
We now comment on the implications of the results in cosmology, focusing on inflation and dark energy.

Inflation is characterised by a light scalar mode whose fluctuations are responsible for the structure we observe in the universe. Our bounds are derived in flat space and therefore apply during inflation only if we restrict to energies and momenta much larger than the Hubble scale, $\omega, k \gg H$. In this regime one can neglect both the curvature of spacetime and the mixing of the scalar mode with gravity (the decoupling limit). Note that, even at these short scales, the time-dependent inflaton background still defines a preferred frame, so Lorentz boosts are spontaneously broken. In general, the time dependence of the background also breaks time-translation invariance. However, one usually assumes that a diagonal combination of time translations and an internal shift symmetry is preserved to a good approximation. This approximate symmetry accounts for the observed near scale invariance of the scalar power spectrum.

Under these approximations, the most general Lagrangian for a single-field model can be written as \cite{Cheung:2007st}
\begin{eqnarray} \label{theeffectiveaction} {\cal {L}}_\pi =  M_{\rm Pl}^2 \dot H
 \, (\partial_\mu \pi)^2+2 M^4_2\left(\dot\pi^2+ \dot\pi^3-\dot\pi \frac{1}{a^2}(\partial_i\pi)^2\right) - \frac{\bar M ^2}{2} \,
\frac{1}{a^4}(\partial_i ^2 \pi)^2 +\ldots
 \end{eqnarray}
The term proportional to $M_2^4$ modifies the kinetic term with two time derivatives: microcausality requires $c_s \le 1$, equivalent to $M_2^4 \ge 0$. It is important to stress that, although our discussion is limited to the quadratic terms in the Lagrangian, namely those that contribute to the dispersion relation of $\pi$, the non-linear realisation of Lorentz invariance relates these quadratic terms to interaction terms. For instance, a reduced speed of sound comes with specific cubic and quartic interactions. Our bounds therefore also constrain interactions, insofar as these are related by symmetry to the dispersion relation. 

The operator proportional to $\bar M^2$ contributes to the dispersion relation of $\pi$ with a term  $k^4$: once one goes to canonical normalization, it contributes to $\tilde\alpha_4$ in the notation above:
 \begin{equation}
    \tilde\alpha_4 = \frac{c_s^2 \bar M^2}{2 M_{\rm Pl}^2 |\dot H|}.
\end{equation}
Our results give rigorous upper bounds on the scales $R$ and $\tilde R$ in terms of $\bar M^2$. Both bounds are relevant for inflation. $\tilde R$ indicates the scale when new modes must be taken into account: the validity of the EFT in inflation requires $\tilde R > H$. The scale $R$ is the maximum scale at which the inflaton can be described as a Goldstone with linear dispersion relations plus small corrections. For $c_s \ll 1$ the scale $R$ becomes small and one has to retain the operator proportional to $\bar M^2$: this is the regime of Ghost Inflation \cite{Arkani-Hamed:2003juy}.

It is remarkable that in the limit $c_s \to 1$, i.e.~when $M_2^4 \ll M_{\rm Pl}^2|\dot H|$, the two cut-offs $R$ and $\tilde R$ become parametrically small compared to the naive scale $\tilde \alpha_4^{-1/2}$. 
During inflation, one has to require $\tilde R \gtrsim H$ (as discussed $R = \tilde R$ for $1-c_s \ll 1$) which gives 
 \begin{equation}
    \frac{|\bar M^2| H^2}{2 M_{\rm Pl}^2 |\dot H|} \lesssim \frac{|\bar M^2| \tilde R^2}{2 M_{\rm Pl}^2 |\dot H|} < 3 (1-c_s^2) \;.
\end{equation}
(For simplicity we have made the interval \eqref{eq:omega2boundscs} symmetric.) The operator proportional to $\bar M^2$ must go to zero as $c_s \to 1$. Even when one evaluates the operator at the maximum scale $\tilde R$, its effect is still suppressed by $1-c_s^2$ compared to the leading kinetic term. The same will apply to all higher-order terms. These arguments rule out a portion of the EFT of Inflation. 
In \cite{Baumann:2015nta} it was conjectured that in the limit $c_s \to 1$ the EFT of Inflation becomes Lorentz invariant. Here we proved rigorously that this is indeed what happens, at least if one concentrates on terms quadratic in $\pi$.

Much of what we said can be repeated verbatim for the EFT of Dark Energy \cite{Creminelli:2008wc,Gubitosi:2012hu,Gleyzes:2014dya}. Most analyses limit the action to terms that give rise to second-order equations of motion \cite{Horndeski:1974wa}. In this case one does not have dispersion, since higher derivative terms are not allowed. Of course higher-derivative terms are always present, although they may be small in some cases, since they are generated by loop corrections. For these operators the bounds discussed above apply.

\section{Including mass terms}
\label{sec:mass_terms}
Up to this point we have focused on gapless dispersion relations, i.e.~those whose $\omega(k)$ vanishes at $k=0$. The microcausality condition, however, is not specific to gapless modes: it applies to every pole of the retarded two-point function, including those of gapped modes. Therefore, in what follows we generalize the microcausality bounds to dispersion relations with a gap. This will apply to pseudo-Goldstone bosons, where the small mass is protected by a small breaking parameter, and genuinely massive states like the radial mode of our $U(1)$ superfluid example. We will only study $\om^2(k)$ and not $\om(k)$. The radius of convergence of $\omega(k) = \sqrt{m^2 + \ldots}$ would set the scale below which the approximation $\omega \simeq m$ plus small corrections is valid and this does not sound an interesting question. The radius of convergence $\tilde R$ of $\omega^2(k)$, on the other hand, sets the momentum scale at which new states become
relevant, similarly to the gapless case.

Let us consider a gapped dispersion relation, written as a power series in $k^2$:
\begin{equation}
    \omega^2(k) = m^2 + c_g^2 k^2 + \sum_{n=2}^{\infty} \tilde{\alpha}_{2n}\, k^{2n},
    \label{eq:gapped_dispersion}
\end{equation}
where $m^2 = \omega^2(0)$, with $m^2 > 0$ the gap. The coefficient of $k^2$ is now denoted with $c_g^2$ (`gapped'), rather than $c_s^2$. In the presence of a gap, microcausality no longer requires $0 \le c_g^2 \le 1$, as it does for $\cs^2$: below we derive the inequality
\begin{equation}
    -\frac{1}{2}-\frac{2m^2}{\tR^2} \leq c_g^2 \leq 1+\frac{m^2}{\tR^2}.
    \label{eq:cg_bound}
\end{equation}
In what follows, we extend the $\cs$-independent and $\cs$-dependent bounds obtained
in Sections~\ref{sec:cs independent omega2} and~\ref{sec:cs dependent omega2} to $m^2 \neq 0$, (now they will be $c_g$ (in)-dependent). We only highlight the points where the gap modifies the argument. 

\subsection{\texorpdfstring{$c_g$}{cg}-independent bounds}
For what concerns the bounds derived in Section \ref{sec:cs independent omega2}, the main difference with respect to the gapless case is that in the presence of a gap $m$, Eq.~\eqref{eq:U_vanishing} does not hold, but rather we have 
\begin{equation}
    \int_0^{2\pi}d\theta \,U(r,\theta) = -2\pi m^2,
    \label{eq:U_gapped}
\end{equation}
where we recall that $U(r,\theta) \equiv \Re(-\om^2)$.
Given this and using Eq.~\eqref{eq:a2n-extract}, we can write for any $n \geq 1$
\begin{equation}
\begin{split}
    |\tilde{\alpha}_{2n}|r^{2n} -2m^2 &= \frac{1}{\pi}\int_0^{2\pi}d\theta \,U(r,\theta)[1-\cos(2n\theta)\cos(\theta_{2n})] \\
    &\leq \frac{r^2}{\pi}\int_0^{2\pi}d\theta\, \sin^2\theta[1-\cos(2n\theta)\cos(\theta_{2n})],
    \end{split}
\end{equation}
where to derive the last inequality we have used the microcausality bound \eqref{eq:simpler_bound_omega2_U}. Also, for brevity, we have written $c_g^2 \equiv \tilde{\alpha}_2$. Proceeding as we did in Section \ref{sec:cs independent omega2} and, in particular, exploiting Eq.~\eqref{eq:integral-result}, we find
\begin{align}
    &|\tilde{\alpha}_{2n}| \leq \frac{1}{\tR^{2n}}(\tR^2 + 2m^2), \quad n \geq 2,
    \label{eq:cg_independent_bounds}\\
    &-\frac{1}{2}-\frac{2m^2}{\tR^2}\leq c_g^2 \leq \frac{3}{2}+\frac{2m^2}{\tR^2}  
\end{align}
where we have sent $r\to \tR$, i.e.~the radius of convergence of $\om^2(k)$. As in the massless case, the upper bound on $c_g^2$ obtained for $n=1$ is not  interesting compared to what we will discuss in the next section. The lower bound, instead, is more informative: although not optimal, it shows that $c_g^2$, while allowed to be negative, cannot be arbitrarily so. A stronger lower bound, at least for $m\gtrsim \tR$, can be obtained, for instance, by using the full microcausality condition~\eqref{eq:bounds_omega2} instead of~\eqref{eq:simpler_bound_omega2_U}, but since it plays no role in what follows we do not pursue it here.

\subsection{\texorpdfstring{$c_g$}{cg}-dependent bounds}
We now go through the computations of Sec.~\ref{sec:cs dependent omega2} introducing a gap in the dispersion relation, as in eq.~\eqref{eq:gapped_dispersion}. Imposing Eq.~\eqref{eq:rt-start}, we find 
\begin{equation}
    -m^2 +(1-c_g^2)\Re(k^2) - \sum_{n = 2}^{\infty}\tilde{\alpha}_{2n}\Re(k^{2n}) \leq |k|^2\left[1-\left|\frac{m^2}{k^2}+c_g^2+\sum_{n = 2}^{\infty}\tilde{\alpha}_{2n}k^{2n-2}\right|\right].
\end{equation}
Using 
\begin{equation}
    \left|\frac{m^2}{k^2}+c_g^2+\sum_{n = 2}^{\infty}\tilde{\alpha}_{2n}k^{2n-2}\right| \geq m^2 \frac{\Re(k^2)}{|k^2|^2}+c_g^2 +\sum_{n = 2}^{\infty}\tilde{\alpha}_{2n}\Re(k^{2n-2})
    \label{eq:source_of_suboptimality}
\end{equation}
and $k = re^{i\theta}, \, r \in [0,\tR)$, it follows that
\begin{equation}
    (1-c_g^2)r^2\sin(\theta)^2 + m^2\sin(\theta)^2 - \sum_{n = 2}^{\infty}\tilde{\alpha}_{2n}r^{2n}\sin(\theta)\sin[(2n-1)\theta] \geq 0,
    \label{eq:gapped_microcausality}
\end{equation}
where we have used the difference-to-product of cosines identity. From this relation we can notice two things. First, dividing both sides of \eqref{eq:gapped_microcausality} by $r^2$ and integrating over $\theta \in [0,\pi]$, we get the upper bound \eqref{eq:cg_bound}, after sending $r \to \tR$. Notice that the resulting constraint on the gapped speed may be suboptimal (contrary to $c_s^2 \le 1$), since the inequalities above give only necessary conditions. Second, restricting to the upper half-disc of analyticity, we can divide both sides of \eqref{eq:gapped_microcausality} by $r\sin(\theta) > 0$. Microcausality implies: 
\begin{equation}
     \Im \tilde{\om}^2 \geq 0
\end{equation}
where 
\begin{equation}
    \tilde{\om}^2(k) \equiv k\left(1-c_g^2 +\frac{m^2}{\tR^2}\right) - \sum_{n = 2}^{\infty}\tilde{\alpha}_{2n}k^{2n-1}.
\end{equation}
In the same way, focusing now to the lower half-disc, we can divide both sides of \eqref{eq:gapped_microcausality} by $r\sin(\theta) < 0$. Microcausality implies 
\begin{equation}
    \Im \tilde{\omega}^2 \leq 0.
\end{equation}
Therefore, we find a typically real function that, introducing the complex variable $z = k/\tR$, can be normalised as 
\begin{equation}
    f(z) \equiv \frac{\tilde{\omega}^2(z\tR)}{\tR(1-c_g^2+m^2/\tR^2)} = z - \sum_{n = 2}^{\infty}\frac{\tilde{\alpha}_{2n}\tR^{2n-2}}{1-c_g^2+m^2/\tR^2}z^{2n-1}.
\end{equation}
The coefficients of this function are thus subject to the Rogosinski bounds~\eqref{eq:rogosinski_bound_maintext}:
\begin{equation}
    -(2n-1)\left(1-c_g^2+\frac{m^2}{\tR^2}\right) \leq \tilde{a}_{2n} \leq k_{2n-1}\left(1-c_g^2+\frac{m^2}{\tR^2}\right),
    \label{eq:cg_dependent_bounds}
\end{equation}
where $k_{2n-1}$ is defined as in Eq.~\eqref{eq:kappa_maintext} and $\tilde{a}_{2n} \equiv \tilde{\alpha}_{2n}\tR^{2n-2}$. Notice that these two-sided bounds do not vanish in the $c_g \to 1$ limit but rather when $c_g^2 \to c_*^2 \equiv 1+m^2/\tR^2$, the upper bound of \eqref{eq:cg_bound}.

Similarly to the bounds in Section \ref{sec:omega2_bounds}, we can find stronger constraints on the coefficients $\tilde{a}_{2n}$ of a gapped dispersion relation $\om^2(k)$ by intersecting Eqs.~\eqref{eq:cg_independent_bounds} and~\eqref{eq:cg_dependent_bounds}. The result is:
\begin{equation}
    -\min\left[(2n-1)\left(c_*^2-c_g^2\right),2c_*^2-1\right] \leq \tilde{a}_{2n} \leq \min\left[k_{2n-1}\left(c_*^2-c_g^2\right),2c_*^2-1\right].
    \label{eq:massive_bounds_inter}
\end{equation}
These bounds are tested for the gapped mode of the $U(1)$ superfluid in appendix \ref{app:checkmassive}.

\section{Conclusions and future directions} \label{sec:conclusions}
In this paper we used microcausality to derive two-sided bounds on the Taylor coefficients, around $k=0$, of $\omega(k)$ and $\omega^2(k)$, the two functions that describe the dispersion relation of gapless modes. The bounds are expressed in units of the respective radii of convergence, $R$ and $\tilde R$, and are shown in figures~\ref{fig:rogo_toeplitz_bounds_a} and~\ref{fig:tilde_a2n}. They become particularly stringent as $c_s \to 1$, see eqs.~\eqref{eq:bounds_csto1} and~\eqref{eq:omega2boundscs}: for a fixed higher-derivative correction, the scale at which the Goldstone description breaks down is parametrically lowered. In the EFT of Inflation, this implies that the theory becomes Lorentz invariant at quadratic order in the limit $c_s \to 1$.

We conclude with some directions in which this work can be extended.
\begin{itemize}
\item The bounds on $\omega(k)$ are {\em optimal}: they are equivalent to imposing microcausality in the disc of convergence. For $\omega^2(k)$ we only obtained necessary, but not sufficient, conditions. Although we do not expect qualitative changes, it would be interesting to reach optimality in this case as well.
\item We did not consider dissipation, so our bounds do not apply to hydrodynamic modes, contrary to those derived in \cite{Heller:2022ejw,Heller:2023jtd}. It is not straightforward to generalise our methods to dissipative systems, but it is plausible that bounds which become stringent as $c_s \to 1$ exist in that case as well.
\item We neglected loops in the EFT. Loops modify the analytic structure of the retarded Green's function by introducing logarithmic branch cuts. Including them would be very valuable, since it would give access to interactions. In particular, one would like to understand the behaviour of all interactions as $c_s \to 1$: is the theory forced to become Lorentz invariant?
\item In the seminal paper \cite{Adams:2006sv}, the positivity of the coefficient of $(\partial\pi)^4$ was derived both from the analyticity of the $S$-matrix and from the requirement of subluminal propagation around a homogeneous background, $\partial\pi = \text{const}$. Since microcausality constrains not only the speed of propagation but all higher-derivative corrections to the dispersion relation, it would be interesting to push the background argument to higher orders and derive constraints on higher-derivative operators in the Goldstone Lagrangian.
\item Our bounds follow from relativistic microcausality and are therefore weak for non-relativistic, condensed-matter systems. Lattice systems obey a different form of causality, the Lieb-Robinson bound \cite{Lieb:1972wy}: an effective, approximate light cone arises from the locality of interactions, which couple only a finite number of neighbouring sites. The implications of this form of causality have recently been studied in \cite{Chowdhury:2025dlx,Chowdhury:2025qyc}, and an analysis similar to ours seems feasible in that context.
\item The observation of the binary neutron star merger GW170817 has established that gravitational waves propagate at the speed of light, up to fractional deviations below $10^{-15}$. This speed may in principle depend on frequency \cite{deRham:2018red}, and one may ask whether LISA, operating at much lower frequencies, could observe a different one. It would be interesting to understand what microcausality allows in this context.
\end{itemize}
Work is in progress in these directions.

\acknowledgments
We are grateful to B.~Bellazzini, R.~Brants, A.~Esposito, M.~Heller, E.~Pajer, A.~Podo, B.~Salehian, G.~Tambalo, B.~Withers and V.~Yingcharoenrat for useful comments. We also would like to thank P.~Zaprawa for directly providing us one of the references used in this project.  WPMcB would like to thank ETH Zürich, IPhT Saclay, APC, and University of Southampton for their hospitality while this work was in progress. 

The authors used Claude (Anthropic) as assistive tools for discussing aspects of the mathematical results, editing the text, and producing the figures. All AI-assisted suggestions were critically assessed: the authors take full responsibility for the final content.

PC is funded by the European Union via the AdG ERC Project ``BestLobe" (Grant Agreement n.~101264087).  Views and opinions expressed are however those of the author(s) only and do not necessarily reflect those of the European Union or the European Research Council. Neither the European Union nor the granting authority can be held responsible for them.

\appendix
\section{Typically real functions} \label{sec:tr_series}
In this section, we collect all the properties of typically real functions, first introduced by Ref.~\cite{Rogosinski1932}, that will be relevant to this paper. A mathematical review can be found in Ref.~\cite{Duren1983}. Another review, with ubiquitous applications in physics, can be found in Ref.~\cite{Raman:2021pkf}.
\subsection{Definitions and theorems}
\begin{definition}[Typically real functions]
    A function $f(z)$ of the complex variable $z$, satisfying $f(0) = 0, f'(0) \neq 0$ and analytic in the open unit disc $|z|< 1$ is said to be typically real if it is real valued along the real axis and has a positive imaginary part in the upper-half plane and a negative imaginary part in the lower-half plane, that is:
\begin{equation}
    \Im f(z) \Im z > 0\quad\text{when}\quad\Im z \neq 0. \label{eq:tr_definition}
\end{equation}
\end{definition}
Of particular interest to us is the sub-class of typically real functions that are odd and analytic in the unit disc $|z| < 1$, with normalised series expansion given by
\begin{equation}
    f(z) = z + \sum_{n=1}^{\infty} b_{2n+1} z^{2n+1}. \label{eq:tr_series}
\end{equation}
\begin{theorem}[Robertson representation] \label{thm:robertson_rep}
    An odd typically real function has a unique Herglotz representation~\textup{\cite{Robertson1935} (see also Ref.~\cite{KoczanZaprawa2006})}
    \begin{equation}
        f(z) = \int_{0}^{1} \frac{z(1+z^2)}{(1 + z^2)^2 - 4z^2 t^2} \mathrm{d}\mu(t) \label{eq:robertson}
    \end{equation}
    in which $\mathrm{d}\mu(t)$ is a unit measure on $[0,1]$.
\end{theorem}
We also define another class of complex functions, interesting for our discussion and first introduced by Ref.~\cite{Caratheodory1911}:
\begin{definition}[Carathéodory functions]
    A complex function $\varphi(z)$ analytic in the unit disc $|z| < 1$ and normalised such that $\varphi(0) = 1$, is said to be of the Carath\'eodory class if $\operatorname{Re}\varphi(z) > 0$ for all $z$ in the unit disc.
\end{definition}
A sub-class of Carathéodory functions of interest to us are even Carathéodory functions, the series expansion of which is given by
\begin{equation}
    \varphi(z) = 1 + \sum_{n = 1}^{\infty} c_{2n}z^{2n}\label{eq:caratheodory}.
\end{equation}
\begin{theorem}[Poisson-Stieljes representation]
    An \emph{even} Carathéodory function has a unique Herglotz representation given by
    \begin{equation}
        \varphi(z) = \int_{0}^{2\pi} \frac{e^{it} + z^2}{e^{it} - z^2} \mathrm{d}\mu(t) \label{eq:poisson_stieljes}
    \end{equation}
    in which $\mathrm{d}\mu(t)$ is a unit measure on $t \in [0,2\pi]$.
\end{theorem}
Odd typically real functions are in one-to-one correspondence with even Carathéodory functions, via the transformation
\begin{equation}
    \varphi (z) = \frac{1 - z^2}{z} f(z). \label{eq:map2cara}
\end{equation}
We will use this result in what follows, as the Herglotz representation of Carathéodory functions provides a solution to the trigonometric moment problem.
\begin{theorem}[Carathéodory-Toeplitz]
    Let $\{c_{2n}\}_{n = 0}^{\infty}$ be a sequence of complex numbers with $c_{-2n} = \bar{c}_{2n}$ and $c_0 = 2$. There exists an even Carathéodory function, as defined in Eq.~\eqref{eq:caratheodory}, \emph{if and only if} the infinite Toeplitz matrix
    \begin{equation}
        T_{\infty} = \left[c_{2(j-i)}\right]_{i,j = 0}^{\infty}
    \end{equation}
    is positive semi-definite.
\end{theorem}
\begin{proof}
    Using the Herglotz representation in Eq.~\eqref{eq:poisson_stieljes} and expanding in $z^2$, we have
    \begin{equation}
        \varphi(z) = 1 + 2\sum_{n = 1}^{\infty} z^{2n}\int_{0}^{2\pi} e^{-int} \mathrm{d} \mu(t)
    \end{equation}
    Comparing to Eq.~\eqref{eq:caratheodory}, we find
    \begin{equation}
        c_{2n} = 2 \int_{0}^{2\pi} e^{-int} \mathrm{d}\mu(t)
    \end{equation}
    This implies that the sequence $\{c_{2n}\}$ is a solution to the Trigonometric moment problem.
\end{proof}
\noindent We refer to Chapter 11 of Ref.~\cite{schmudgen2017moment} for a detailed proof.
\subsection{Rogosinski bounds}
\begin{theorem}[Rogosinski bounds]
    Let $f(z)$ be defined as in Eq.~\eqref{eq:tr_series}. If the function $f(z)$ is typically real, then the coefficients $b_{2n+1}$ are real and satisfy
    \begin{equation}
        -\kappa_{2n+1} \leq b_{2n+1} \leq 2n+1, \label{eq:rogosinski_bound}
    \end{equation}
    where
    \begin{equation}
    \kappa_{2n+1} \equiv -\min_{0 < \theta \leq \frac{\pi}{2}} \left[\frac{\sin (2n+1)\theta}{\sin\theta}\right] > 0. \label{eq:kappa_2n+1}
\end{equation}
The interval \eqref{eq:rogosinski_bound} is optimal: if one chooses a particular $b_{2n+1}$ in the interval, there exists an odd typically real function with that coefficient. 
\end{theorem}
\begin{proof}
Expanding Eq.~\eqref{eq:robertson} in $z$ and comparing to Eq.~\eqref{eq:tr_series} yields
\begin{equation}
    b_{2n+1} = \int_{0}^{1} U_{2n}(t) \mathrm{d}\nu(t) \label{eq:b_int_measure}
\end{equation}
in which $U_n(t)$ are the Chebyshev polynomials of the second kind. Since $\nu(t)$ is a unit positive measure, $b_{2n+1}$ is bounded above and below by the extrema of $U_{2n}(t)$ on $t\in[0,1]$ and Rogosinski bounds are attained by delta measures at the extremising points. To determine these extrema, we make use of the identity
\begin{equation}
    U_{2n}(\cos\theta) = \frac{\sin[(2n+1)\theta]}{\sin\theta} \label{eq:chebyshev_theta}
\end{equation}
and extremise the right-hand side over $\theta \in \left[0, \frac{\pi}{2}\right]$. Differentiating with respect to $\theta$ and setting the result to zero gives the condition for $\theta$ to extremise Eq.~\eqref{eq:chebyshev_theta}:
\begin{equation}
    \tan[(2n+1)\theta] = (2n + 1)\tan\theta.
\end{equation}
The trivial $\theta = 0$ solution (equivalent to $t = 1$, or $\nu(t) = \delta_1$) corresponds to the maximum of Eq.~\eqref{eq:chebyshev_theta} which is
\begin{equation}
    \max \left[\frac{\sin (2n+1)\theta}{\sin\theta}\right] = \lim_{\theta \to 0} \left[\frac{\sin (2n+1)\theta}{\sin\theta}\right] = 2n+1
\end{equation}
while the solution corresponding to the minimum of Eq.~\eqref{eq:chebyshev_theta}, namely Eq.~\eqref{eq:kappa_2n+1}, is non-trivial and has to be solved numerically. Some values for references, including the asymptotic value for $(2n+1) \to \infty$, are included in Table \ref{tab:kappa}. 
\begin{table}[ht]
  \centering
  \renewcommand{\arraystretch}{1.6}
  \begin{tabular}{cccc}
    \toprule
    $2n+1$ & $\theta$ & $t$ & $\kappa_{2n+1}$ \\
    \midrule
    $3$ & $90.00^\circ$ & $0$      & $1.0000$ \\
    $5$ & $52.24^\circ$ & $0.6124$ & $1.2500$ \\
    $7$ & $37.04^\circ$ & $0.7982$ & $1.6311$ \\
    $9$ & $28.72^\circ$ & $0.8769$ & $2.0395$ \\
    \midrule
    $\to\infty$ & $\to 0$ & $\to 1$ &
      ${\sim}\,\dfrac{2\,(2n+1)}{3\pi}$ \\
    \bottomrule
  \end{tabular}
  \caption{Numerical values of the minimising angle $\theta$,
           the corresponding $t = \cos\theta$, and $\kappa_{2n+1}$,
           for odd indices $2n+1 = 3,5,7,9$, together with the
           leading asymptotic behaviour as $2n+1\to\infty$.}
  \label{tab:kappa}
\end{table}\\
Finally, the optimality of Rogosinski bounds is guaranteed by Theorem \ref{thm:robertson_rep} i.e.~the function which saturates the bound is given by 
\begin{equation}
    f(z;t) = \frac{z(1+z^2)}{(1 + z^2)^2 - 4z^2 t^2}.
\end{equation}
The maximising function is given by
\begin{equation}
    f(z;1) = \frac{z(1+z^2)}{(1-z)^2(1+z)^2} = z + 3z^3 + 5z^5 + \dots
\end{equation}
which universally maximises all the coefficients at $t = 1$. The coefficients of the minimising function, however, do not universally minimise i.e.~each coefficient minimises at different values of $t$ given by Table \ref{tab:kappa}.
\end{proof}
\subsection{Rogosinski-Toeplitz bounds} \label{subsec:rogo_toeplitz_poly}
\begin{definition}[Representative of a function] 
    Let $f(z)$ be defined as in Eq.~\eqref{eq:tr_series}. For an integer $N \geq 1$, we call
    \begin{equation}
        p_N \equiv \left(b_{2n+1}\right)_{n =1}^{N} \in \mathbb{R}^N
    \end{equation}
    the $N$-th representative of $f(z)$.
\end{definition}
\begin{theorem}[Toeplitz conditions]
    Let $f(z)$ be defined as in Eq.~\eqref{eq:tr_series}. The complex function $f(z)$ will be typically real if and only if all of its $N$-th representatives lie within their corresponding closed convex hull $K_N$ characterised by the $N$ Toeplitz determinants
    \begin{equation}
        D_{k+1} = \begin{vmatrix}
        2 & b_3 - b_1 & b_5 - b_3 & \dots & b_{2k+1}-b_{2k-1}\\
        b_3 - b_1 & 2 & b_3 - b_1 & \dots & b_{2k-1}-b_{2k-3}\\
        b_5 - b_3 & b_3 - b_1 & 2 & \dots & \vdots\\
        \vdots & \vdots & \vdots & 2 & \vdots\\
        b_{2k+1}-b_{2k-1} & b_{2k-1}-b_{2k-3} & \dots & \dots & 2
        \end{vmatrix}, \label{eq:toeplitz_matrix}
    \end{equation}
    in which $1 \leq k \leq N$ and $b_1 = 1$. These determinants are either all strictly positive, or positive up to some index and vanishing from that index onward, the latter case corresponding to the boundary of $K_N$.
\end{theorem}
\begin{proof}
    The first half of this theorem is a consequence of \eqref{eq:b_int_measure}: given that the integral is integrated over a measure $\mu(t)$, the $b_{2n+1}$ has to belong to the closure of the convex hull as stated. 
    
    As for the second half, one can utilise \eqref{eq:map2cara}.
    Given \eqref{eq:tr_series}, we have
    \begin{equation}
        \varphi (z) = 1 + \sum_{n = 1}^{\infty} (b_{2n + 1} - b_{2n-1})z^{2n}
    \end{equation}
    Compare to \eqref{eq:caratheodory}, we find that for $n\geq 1$
    \begin{equation}
        c_{2n} = b_{2n + 1} - b_{2n-1},\quad c_0 = 2
    \end{equation}
    and thus they must satisfy \eqref{eq:toeplitz_matrix}. Finally, due to the symmetries of the Toeplitz matrix, the conditions on positive semi-definiteness reduces to the non-negativity of the $n$ Toeplitz determinants.
\end{proof}
Here, we provide one ($K_1$), two ($K_2$), and three ($K_3$) dimensional projections of the infinite-dimensional Rogosinski-Toeplitz spectrahedron $K_\infty$.

\vspace{5mm}
\noindent \emph{$N = 1$}: In this case, the Toeplitz determinant is given by
\begin{equation}
    D_2 = \begin{vmatrix}
        2 & b_3 - 1 \\
        b_3 - 1 & 2
    \end{vmatrix}.
\end{equation}
For this to be non-negative i.e. $D_2 \geq 0$, we have
\begin{align}
    4 - (b_3-1)^2 &\geq 0,\\
    (3-b_3)(1+b_3) &\geq 0,
\end{align}
which gives
\begin{equation}
    -1 \leq b_3 \leq 3. \label{eq:toeplitz_n=2}
\end{equation}
This is $K_1$—a one-dimensional projection of the infinite-dimensional spectrahedron $K_\infty$ onto the $b_3$ axis. Notice that this is equivalent to the Rogosinski bound for $b_3$.

\vspace{5mm}
\noindent \emph{$N = 2$}: In this case, the Toeplitz determinants is given by $D_2 \geq 0$, which yields \eqref{eq:toeplitz_n=2}, and
\begin{equation}
    D_3 = \begin{vmatrix}
        2 & b_3 - 1 & b_5 - b_3 \\
        b_3 - 1 & 2 & b_3 - b_1\\
        b_5 - b_3 & b_3 - 1 & 2
    \end{vmatrix}.
\end{equation}
For this to be non-negative i.e. $D_3 \geq 0$, we have 
\begin{equation}
    [4-2(b_3 -1)^2] + (b_5 - b_3)(b_3 -1)^2 - (b_5 - b_3)^2 \geq 0.
\end{equation}
This is a quadratic in $b_5 - b_3$. Solving this yields
\begin{equation}
    (b_3 - 1)^2 - 2 \leq b_5 - b_3 \leq 2
\end{equation}
Expanding and rearranging yields
\begin{equation}
    b_3^2 - b_3 - 1 \leq b_5 \leq b_3 + 2 \label{eq:toeplitz_n=3}
\end{equation}
which describes the projection of the spectrahedron $K_\infty$ onto the $(b_3, b_5)$ plane as seen in Figure \ref{fig:k2_maintext}.

\begin{figure}[h]
    \centering
    \begin{subfigure}[b]{0.4\textwidth}
        \centering
        % same unit length (\linewidth/9.6) and same bounding box in both panels;
        % text is not scaled, so it matches the original tikzpicture
        \begin{tikzpicture}[>=latex, x=0.1041667\linewidth, y=0.1041667\linewidth]
        \useasboundingbox (-2.6,-2.9) rectangle (7.0,8.2);

        % axis limits
        \def\xmin{-2.4}
        \def\xmax{4.4}
        \def\ymin{-2.4}
        \def\ymax{7.6}
        % minimum of b7 on K_3: b7(b3) = b3^3 - 2 b3^2 - b3 + 1 at b3 = (2+sqrt 7)/3
        \def\bmin{-1.6311}

        % bounding rectangle -1 <= b3 <= 3, b7min <= b7 <= 7 (drawn behind the hull)
        \draw[black, fill=black, fill opacity=0.2] (-1,\bmin) rectangle (3,7);

        % shaded region: left arc, upper line, right arc, lower line
        \fill[magenta!25]
           (-1,-1) -- plot[domain=-1:-0.5, samples=50] (\x,{\x*\x*\x-2*\x*\x-\x+1})
           -- (3,7)
           -- plot[domain=3:1.5, samples=100] (\x,{\x*\x*\x-2*\x*\x-\x+1})
           -- cycle;

        % boundary curves and lines
        \draw[magenta, thick] plot[domain=-1:-0.5, samples=50] (\x,{\x*\x*\x-2*\x*\x-\x+1});
        \draw[magenta, thick] (-0.5,0.875) -- (3,7);               % b7 = 7(b3+1)/4
        \draw[magenta, thick] plot[domain=1.5:3, samples=100] (\x,{\x*\x*\x-2*\x*\x-\x+1});
        \draw[magenta, thick] (-1,-1) -- (1.5,-1.625);             % b7 = -(b3+5)/4

        % axes (drawn last, on top of the shaded regions)
        \draw[->, thick] (\xmin,0) -- (\xmax,0) node[right] {$b_3$};
        \draw[->, thick] (0,\ymin) -- (0,\ymax) node[above] {$b_7$};

        % ticks (fixed absolute length, independent of the unit length)
        \foreach \x in {-2,-1,1,2,3,4}
          \draw (\x,0) ++(0pt,-0.08cm) -- ++(0pt,0.16cm) (\x,0) node[below=4pt] {\x};
        \foreach \y in {-2,-1,1,2,3,4,5,6,7}
          \draw (0,\y) ++(-0.08cm,0pt) -- ++(0.16cm,0pt) (0,\y) node[left=4pt] {\y};
        \node[below left] at (0,0) {\small $0$};

        \end{tikzpicture}
        \caption{Projection of $K_3$ onto $(b_3, b_7)$ plane.}
        \label{fig:k3_b3b7}
    \end{subfigure}
    \hfill
    \begin{subfigure}[b]{0.4\textwidth}
        \centering
        % same unit length (\linewidth/9.6) and same bounding box in both panels;
        % text is not scaled, so it matches the original tikzpicture
        \begin{tikzpicture}[>=latex, x=0.1041667\linewidth, y=0.1041667\linewidth]
        \useasboundingbox (-2.6,-2.9) rectangle (7.0,8.2);

        % axis limits
        \def\xmin{-2.4}
        \def\xmax{6.4}
        \def\ymin{-2.4}
        \def\ymax{7.6}
        % minimum of b7 on K_3: b7(b3) = b3^3 - 2 b3^2 - b3 + 1 at b3 = (2+sqrt 7)/3
        \def\bmin{-1.6311}

        % bounding rectangle -5/4 <= b5 <= 5, b7min <= b7 <= 7 (drawn behind the hull)
        \draw[black, fill=black, fill opacity=0.2] (-1.25,\bmin) rectangle (5,7);

        % shaded region: moment curve for s in [1/4,1], closed by the line b7 = b5 + 2
        % (b5,b7) = (16s^2-12s+1, 64s^3-80s^2+24s-1)
        \fill[magenta!25]
           plot[domain=0.25:1, samples=100, variable=\s]
             ({16*\s*\s-12*\s+1},{64*\s*\s*\s-80*\s*\s+24*\s-1})
           -- cycle;

        % boundary curve and line
        \draw[magenta, thick]
           plot[domain=0.25:1, samples=100, variable=\s]
             ({16*\s*\s-12*\s+1},{64*\s*\s*\s-80*\s*\s+24*\s-1});
        \draw[magenta, thick] (-1,1) -- (5,7);                     % b7 = b5 + 2

        % axes (drawn last, on top of the shaded regions)
        \draw[->, thick] (\xmin,0) -- (\xmax,0) node[right] {$b_5$};
        \draw[->, thick] (0,\ymin) -- (0,\ymax) node[above] {$b_7$};

        % ticks (fixed absolute length, independent of the unit length)
        \foreach \x in {-2,-1,1,2,3,4,5,6}
          \draw (\x,0) ++(0pt,-0.08cm) -- ++(0pt,0.16cm) (\x,0) node[below=4pt] {\x};
        \foreach \y in {-2,-1,1,2,3,4,5,6,7}
          \draw (0,\y) ++(-0.08cm,0pt) -- ++(0.16cm,0pt) (0,\y) node[left=4pt] {\y};
        \node[below left] at (0,0) {\small $0$};

        \end{tikzpicture}
        \caption{Projection of $K_3$ onto $(b_5, b_7)$ plane.}
        \label{fig:k3_b5b7}
    \end{subfigure}
    \caption{Projections of $K_3$ onto each plane.}
    \label{fig:k3_proj}
\end{figure}
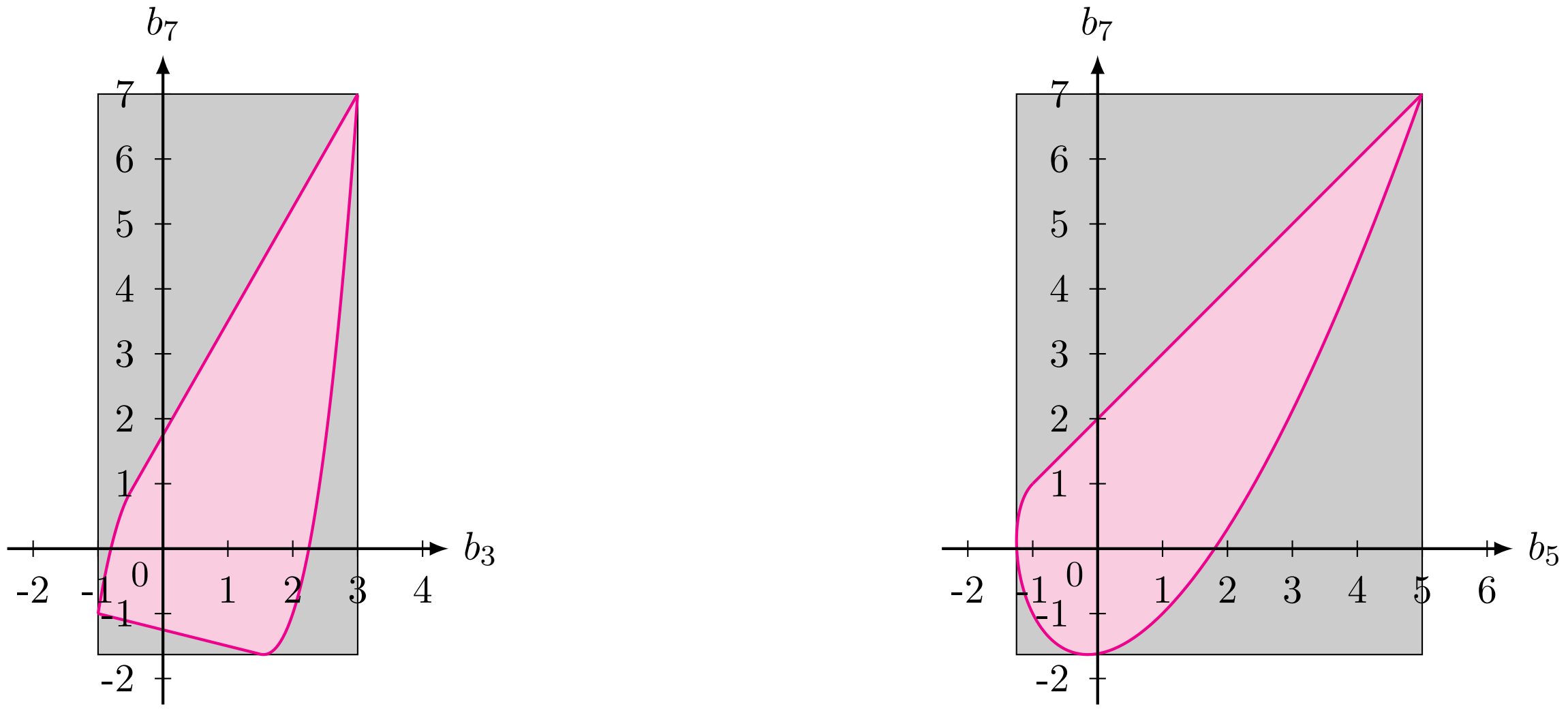

\vspace{5mm}
\noindent \emph{$N = 3$}: In this case, the Toeplitz determinants is given by $D_2 \geq 0$ which yields \eqref{eq:toeplitz_n=2}, $D_3 \geq 0$ which yields \eqref{eq:toeplitz_n=3}, and
\begin{equation}
    D_4 = \begin{vmatrix}
        2 & b_3 - 1 & b_5 - b_3 & b_7 - b_5 \\
        b_3 - 1 & 2 & b_3 - b_1 & b_5 - b_3 \\
        b_5 - b_3 & b_3 - b_1 & 2 & b_3 - 1\\
        b_7 - b_5 & b_5 - b_3 & b_3 - 1 & 2
    \end{vmatrix}.
\end{equation}
For this to be non-negative i.e. $D_4 \geq 0$, we have (defining $x = b_3 - 1$, $y = b_5 - b_3$, and $z = b_7 - b_5$ for brevity)
\begin{equation}
    (x^2-4)z^2 - 2x(x^2 + y^2 - 4y)z + (x^4 -2x^2y^2 +8x^2y-12x^2+           y^4-8y^2 + 16) \geq 0.
\end{equation}
This is a quadratic in $z = b_7 - b_5$. Solving this yields
\begin{equation}
    \frac{(x + y)^2}{x + 2} - 2 \leq z \leq \frac{(x - y)^2}{x - 2} + 2.
\end{equation}
Plugging back $b_3$, $b_5$ and $b_7$, we find
\begin{equation}
    b_5 + \frac{(b_5 - 1)^2}{b_3 + 1} - 2 \leq b_7 \leq b_5 + \frac{(2 b_3 - b_5 - 1)^2}{b_3 - 3} + 2
\end{equation}
which describes the projection of the spectrahedron $K_\infty$ onto the $(b_3, b_5, b_7)$ axes. The projection of $K_3$ onto the $(b_3, b_7)$ plane and $(b_5, b_7)$ plane is shown in Figure \ref{fig:k3_proj}.

\section{Analytic structure of \texorpdfstring{$\om^2(k)$}{omega2(k)}\label{app:omega2}}
In this appendix, we show that, under the assumption that the retarded two-point function is meromorphic\footnote{If $\tilde G(\om,k)$ has branch cuts, the pole $\om^2(k)$ can move onto the
second sheet, either by reaching a branch point or by crossing a cut. Beyond this point
the mode is no longer constrained by \eqref{eq:bound_omega}. In particular, nothing then
prevents $\om^2(k)$ from developing poles, which microcausality would otherwise exclude. For a more detailed discussion, see footnote \ref{fnot:radius_of_validity}.}, $\omega^2(k)$ cannot have poles or essential singularities while satisfying the microcausality condition \eqref{eq:bounds_omega2}. Hence, any isolated singularities of $\omega^2(k)$ outside the open disc of convergence must be a branch point\footnote{We assume that $\omega^2(k)$ does not have non-isolated singularities, such as accumulations of poles or natural boundaries. Thus, in the absence of poles and essential singularities, the only allowed singularities are branch points. This assumption is similar to that made in \cite{Heller:2022ejw}.}. 

Before proceeding with the proof of this claim, we note that it is sufficient to prove that in the presence of poles and essential singularities:
\begin{equation}
    \Re[-\om^2(k)] > 2(\Im k)^2
    \label{eq:absence_of_poles}
\end{equation}
for some complex $k$. From this result, it follows automatically that microcausality in \eqref{eq:bounds_omega2} is violated. Also, the fact that $\om^2(k)$ can only have branch points, is inherited directly from $\om(k)$, which, similarly, cannot have any isolated singularities other than branch points, as proved in Ref.~\cite{Heller:2022ejw}. 

We begin by discussing why a microcausal $\omega^2(k)$ cannot admit essential singularities. The core of the argument relies on the Casorati--Weierstrass theorem (see, for example, \cite{Conway:1978}).

\begin{theorem}[Casorati--Weierstrass]
\label{thm:casorati-weierstrass}
Let $f$ be holomorphic on a punctured neighborhood $0<|k-a|<\varepsilon_0$
of a point $a\in\mathbb{C}$, and suppose $a$ is an essential singularity
of $f$, so that $f$ admits a Laurent expansion about $a$ with infinitely
many terms with negative powers of $(k-a)$. Then, for any
$\varepsilon\in(0,\varepsilon_0]$, $w_0\in\mathbb{C}$, and $\delta>0$,
there exists $\tilde k$ with $0<|\tilde k-a|<\varepsilon$ such that
$|f(\tilde k)-w_0|<\delta$.
\end{theorem}

With this in hand, we can state our theorem and proof.

\begin{theorem}
    If $\omega^2(k)$ is microcausal, then it has no essential singularities.
\end{theorem}
\begin{proof}
    Suppose, for contradiction, that $\omega^2$ has an essential singularity at $k=a$, and let $\varepsilon_0>0$ be such that $\omega^2$ is holomorphic on the punctured disc $0<|k-a|<\varepsilon_0$. Fixing $\varepsilon\in(0,\varepsilon_0]$, for every $k$ in this disc we have
    \begin{equation}
        |\Im k-\Im a|\leq|k-a|<\varepsilon.
    \end{equation}
Using the triangle inequality,
\begin{equation}
|\Im k| = |\Im k- \Im a+\Im a|
\leq|\Im k-\Im a|+|\Im a|<\varepsilon+ |\Im a|,
\end{equation}
so that
\begin{equation}
|\Im k|^{2}<\bigl(\varepsilon+|\Im a|\bigr)^{2}\equiv M_\varepsilon.
\label{eq:Meps-bound}
\end{equation} 
Now choose, for example, $w_0=-2M_\varepsilon-1 \in \mathbb{R}$ and $\delta=1$. By the
Casorati--Weierstrass theorem there exists $\tilde k$ satisfying
$0<|\tilde k-a|<\varepsilon$ such that $|\omega^2(\tilde k)-w_0|<1$. Using
this inequality we can write
\begin{equation}
\Re\,[\omega^2(\tilde k)]-w_0\leq\bigl|\omega^2(\tilde k)-w_0\bigr|<1\implies
\Re\,[\omega^2(\tilde k)]<1+w_0=-2M_\varepsilon,
\end{equation}
and hence, using \eqref{eq:Meps-bound},
\begin{equation}
\Re\,[-\omega^2(\tilde k)]>2M_\varepsilon >2(\Im\tilde k)^{2}.
\end{equation}
This clearly violates the microcausality condition \eqref{eq:bounds_omega2}. Hence, $\omega^2(k)$ cannot have an essential singularity at $k = a$ while respecting microcausality outside the disc of convergence\footnote{The bound \eqref{eq:bound_omega} follows only from microcausality, the polynomial
boundedness of $G(t,\vec x)$ and its reality. Its derivation never refers to a radius
of convergence: it constrains every singularity of $\tilde G(\om,k)$ on the physical sheet. If $\tilde G(\om,k)$ is meromorphic, it has no other sheets, so $\om^2(k)$ can never leave the physical sheet and
\eqref{eq:bound_omega} holds also outside the disc of convergence.}.
\end{proof}

Next, we prove that a microcausal $\omega^2(k)$ cannot have poles using an argument similar to that of \cite{Heller:2022ejw} for $\omega(k)$ itself.

\begin{theorem}
    If $\omega^2(k)$ is microcausal, then it has no poles.
\end{theorem}
\begin{proof}
    Given an isolated pole at $a\in\mathbb{C}$, we can write the Laurent
series, valid in the punctured disc $0<|k-a|<\varepsilon$,
\begin{equation}
\omega^{2}(k)=\sum_{n=n_{\min}}^{\infty}c_n(k-a)^{n},
\end{equation}
for some order of the pole $n_{\min}<0$ with $c_{n_{\min}}\neq0$. Take
$k=a+re^{i\theta}$ with $r\to0^{+}$. Then the leading term dominates, and
\begin{equation}
\Re\,[(\omega^2(k)]=|c_{n_{\min}}|r^{n_{\min}}
\cos[(\theta\,n_{\min}+\arg(c_{n_{\min}})]+\mathcal{O}(r^{n_{\min}+1}).
\label{eq:pole-leading}
\end{equation}
Approaching $k=a$ at the angle
\begin{equation}
\tilde\theta=\frac{\pi-\arg(c_{n_{\min}})}{n_{\min}},
\end{equation}
we have $\cos[\tilde\theta\,n_{\min}+\arg(c_{n_{\min}})]=\cos(\pi)=-1$,
and along this ray \eqref{eq:pole-leading} gives
\begin{equation}
\Re\,[-\omega^2(k)]
=|c_{n_{\min}}|r^{n_{\min}}+ \mathcal{O}(r^{n_{\min}+1})
\;\xrightarrow[r\to0^{+}]{}\;+\infty.
\end{equation}
On the other hand, since $\tilde\theta$ is fixed, $k=a+re^{i\tilde\theta}\to a$
smoothly as $r\to0^{+}$, and thus
\begin{equation}
(\Im k)^{2}=\bigl(\Im a+r\sin\tilde\theta\bigr)^{2}
\;\xrightarrow[r\to0^{+}]{}\;(\Im a)^{2}.
\end{equation}
Therefore, for $r$ small enough,
\begin{equation}
\Re\,[-\omega^2(k)]>2(\Im k)^{2},
\end{equation}
which again violates the microcausality condition in \eqref{eq:simpler_bound_omega2}. Hence $\omega^2(k)$ cannot have a pole at $k = a$ while respecting microcausality outside the disc of convergence.
\end{proof}

\section{Method of Domb-Sykes} \label{app:domb_sykes}
We outline a graphical method used to calculate the radius of convergence of a power series, and the nature of the singularity dictating that radius. We will assume that the function $f(z)$ has only one nearest singularity from the origin, and that singularity lies on the real axis. This section follows closely to the discussion in \cite{vanDyke_1974}, where readers can find a more detailed treatment on this method. 

Consider a power series
\begin{equation}
    f(z) = \sum_{n = 0}^{\infty} c_n z^n. \label{eq:power_series}
\end{equation}
For large values of $n$, the signs of $c_n$ will settle down to a certain pattern. If the signs of $c_n$ are unchanged, then
the nearest singularity lies on the positive real axis. For example:
\begin{equation}
    f(z) = \frac{1}{1 - z} = 1 + z + z^2 + z^3 + \dots.
\end{equation}
However, if the signs of $c_n$ alternate, then the nearest singularity lies on the negative real axis. For example:
\begin{equation}
    f(z) = \frac{1}{1 + z} = 1 - z + z^2 - z^3 + \dots.
\end{equation}
After knowing the position of the singularity, the radius of convergence of \eqref{eq:power_series} can be determined through the ratio test:
\begin{equation}
    R = \lim_{n \to \infty} \left|\frac{c_{n}}{c_{n+1}}\right|.
\end{equation}
In practice, one only has access to a finite number of $c_n$. Thus, one can only estimate the radius of convergence of the finite power series. One way to do so is to plot the ratio $|c_{n}/c_{n+1}|$ against $n$ and determine the horizontal asymptote. However, this can be very unreliable. A more reliable method, first proposed in Ref.~\cite{Domb_Sykes_1957}, is to instead plot the \emph{inverse} of the ratio of coefficients $|c_{n+1}/c_{n}|$ against $1/n$. 

A function $f(z)$ with a singularity at $z_*$ will have the following asymptotic form:
\begin{equation}
    f(z) \sim A\left(1 - \frac{z}{z_*}\right)^{\beta} \label{eq:asymp_form}
\end{equation}
in which $\beta$ dictates the type of singularity of $z_*$: $\beta$ being a negative integer for poles, a non-integer for branch points. $A$ is the amplitude of that singularity: for poles $A$ is the residue, and for branch points $A$ is the discontinuity across the branch cut. Expanding \eqref{eq:asymp_form} around $z = 0$, we have
\begin{align}
    f(z) &= A \sum_{n = 0}^{\infty}z_*^{-n} \frac{\beta(\beta - 1)(\beta - 2)\dots(\beta - n + 1)}{n!} (-1)^n  z^n.\\
    &=  A \sum_{n = 0}^{\infty}z_*^{-n} \frac{(-\beta)(1 - \beta)(2 - \beta)\dots(n - 1 - \beta)}{n!} z^n.\\
    &= \frac{A}{\Gamma(-\beta)} \sum_{n = 0}^{\infty}z_*^{-n} \frac{\Gamma (n -\beta)}{\Gamma(n + 1)} z^n. \label{eq:series_asymp_form}
\end{align}
Comparing \eqref{eq:series_asymp_form} to \eqref{eq:power_series}, we find the asymptotic value of $c_n$:
\begin{equation}
    c_n \sim \frac{A z_*^{-n}}{\Gamma(-\beta)} \frac{\Gamma (n -\beta)}{\Gamma(n + 1)}.
\end{equation}
In the large-$n$ limit, one can use Stirling's approximation
\begin{equation}
    \Gamma(n+1) \approx \sqrt{2\pi n}\left(\frac{n}{e}\right)^n
\end{equation}
to obtain
\begin{equation}
    c_n \sim \frac{A}{\Gamma (-\beta)} n^{-\beta - 1} z_*^{-n}.
\end{equation}
Finally, determining the inverse of the ratio of coefficients, we get
\begin{align}
    \left|\frac{c_{n+1}}{c_{n}} \right| &= \left(1 + \frac{1}{n}\right)^{-\beta - 1} \frac{1}{|z_*|}\\
    &= \frac{1}{R} - \frac{1 + \beta}{R} \frac{1}{n} + \mathcal{O}\left(\frac{1}{n^2}\right).
\end{align}
Thus, plotting $|c_{n+1}/c_n|$ against $1/n$ yields a straight line for large $n$. The $y$-intercept of this line is the inverse of the radius of convergence and the gradient of this line depends on the type of singularity $\beta$ closest to the origin.

\section{\texorpdfstring{$U(1)^2$}{U(1)2} solid model\label{app:solid}}
Beyond the lattice scale, the mechanical deformation of solids is well described by an effective field theory of phonons, the Goldstone modes of spontaneously broken translations. This framework has proven versatile: it provides the theoretical foundation for standard elasticity theory (see, for example, Refs.~\cite{Alberte:2018doe,Esposito:2018sdc}) and has more recently been extended to cosmological settings, where it provides the basis for solid-inflation scenarios (see, for example, Ref.~\cite{Endlich:2012pz}). 
%Given the wide applicability of this setting, it is tempting to use the EFT of solids as a further testing ground for the microcausality bounds derived in Sections \ref{sec:bounds_omega} and \ref{sec:omega2_bounds}. \WPMcB{Need to check}

As a simplified version of the model analysed in \cite{Esposito:2020wsn}, we consider a solid in $2+1$ spacetime dimensions described by two complex scalars $\Phi_i$ with $i =1,2$, charged under an internal $U(1)^2$ symmetry and related by the permutation symmetry $\Phi_1\leftrightarrow\Phi_2$. The most general Poincaré-invariant action consistent with these symmetries is
\begin{equation}
S = \int \mathrm{d}^3 x \left\{\sum_{i = 1}^{2} \left[-\left|\partial \Phi_i\right|^2 + m^2 \left|\Phi_i\right|^2 - \lambda_1 \left|\Phi_{i}\right|^4\right] - \lambda_2 \left|\Phi_1\right|^2\left|\Phi_2\right|^2\right\},
\label{eq:action_u1sq}
\end{equation}
where $\lambda_1 > 0$ and $2\lambda_1 + \lambda_2 > 0$ in order to have a Hamiltonian bounded from below. We can then decompose the scalar fields $\Phi_{i}$ in polar form as
\begin{equation}
\Phi_i = \frac{\rho_i}{\sqrt{2}} e^{i\theta_{i}/v},
\end{equation}
where $v$ denotes the common vacuum expectation value of the radial modes $\rho_{i}$, equal for both fields as a consequence of the permutation symmetry $\Phi_1\leftrightarrow\Phi_2$. Compression and dilation of the solid can then be introduced by expanding around the space-dependent background $\theta_i = \alpha\, x_i + \pi_{i}$, where $\alpha$ is a parameter measuring the level of compression/dilation, satisfying $\alpha^2>0$ by definition. Notice that the mass term in Eq.~\eqref{eq:action_u1sq} has been chosen with $m^2>0$ so as to trigger spontaneous symmetry breaking already in the absence of dilatation/compression, i.e.~at $\alpha=0$. 

After expanding the radial modes around the vev, namely $\rho_i = v + h_i$, we can rewrite the quadratic part of the action in Eq.~\eqref{eq:action_u1sq} as 
\begin{equation}
    S^{(2)} = \int \mathrm{d}^3x \left[-\frac{1}{2} (\partial \vec{h})^2 - \frac{1}{2} \vec{h}^T M^2 \vec{h} - \frac{1}{2}(\partial \vec{\pi})^2 - \frac{\alpha}{v} (\vec{h}^T \vec{\nabla}\cdot \vec{\pi} - \vec{\pi}^T\vec{\nabla} \cdot \vec{h})\right].
\end{equation}
The mass matrix of the radial modes reads
\begin{equation}
 M^2 = \begin{bmatrix}
        2\lambda_1 & \lambda_2\\
        \lambda_2 & 2\lambda_1
    \end{bmatrix}v^2.
\end{equation}
Diagonalizing it as $M^2 = O\,\Delta^2\,O^T$, with
\begin{equation}
\Delta^2 = \text{diag} (\Delta_1^2, \Delta_2^2), \quad O = \frac{1}{\sqrt{2}}\begin{bmatrix}
        1 & 1\\
        1 & -1
    \end{bmatrix},
    \label{eq:rotation_matrix}
\end{equation}
yields the radial mode masses:
\begin{equation}
\begin{split}
\Delta^2_1 &= (2\lambda_1 + \lambda_2)v^2,\\[2mm]
\Delta^2_2 &= (2\lambda_1 - \lambda_2)v^2.
\end{split}
\end{equation}
For the configuration $\rho_1=\rho_2=v$ to be a minimum, both masses must be positive, which requires $2\lambda_1-\lambda_2>0$.
We can write the Fourier transform of the quadratic action, in terms of the mass eigenbasis $\vec{H} = O^T\vec{h}$ and the phonon fields $\vec\pi$, as 
\begin{align}
    S^{(2)} &= \frac{1}{2} \int \frac{\mathrm{d}\omega\mathrm{d}^2k}{(2\pi)^3}  \begin{bmatrix}
        \Tilde{\vec{H}}^T(-\omega, -k)&\Tilde{\vec{\pi}}^T(-\omega, -k)
    \end{bmatrix} K \begin{bmatrix}
        \Tilde{\vec{H}}(\omega,k)\\
        \Tilde{\vec{\pi}}(\omega, k)
    \end{bmatrix},
\end{align}
where the kinetic matrix is given by
\begin{equation}
     K = \begin{bmatrix}
        (\omega^2 - k^2) - \Delta_1^2 & 0 & -i \frac{\sqrt{2}\alpha}{v}k & 0\\
        0 & (\omega^2 - k^2) - \Delta_2^2 & -i \frac{\sqrt{2}\alpha}{v}k & 0\\
        i \frac{\sqrt{2}\alpha}{v}k  & i \frac{\sqrt{2}\alpha}{v}k & (\omega^2 - k^2) & 0 \\
        0 & 0& 0 & (\omega^2 - k^2)
    \end{bmatrix}.
\end{equation}
The first block captures the mixing between the longitudinal phonon $\pi_1$ and the two radial mass eigenstates $H_1$ and $H_2$, which we will focus on, as it is the interesting sector for our analysis. The second block refers to the transverse phonon $\pi_2$ which decouples from the rest of the fields. Since $K$ is a block diagonal matrix, the physical spectrum of the longitudinal sector is obtained by setting the determinant of the first block to zero, which yields a cubic equation in $\omega^2-k^2$:
\begin{equation}
    (\omega^2 - k^2)^3 - (\Delta_1^2 + \Delta_2^2)(\omega^2 - k^2)^2 + \left[\Delta_1^2\Delta_2^2 - \frac{4\alpha^2k^2}{v^2}\right](\omega^2 - k^2) + \frac{2\alpha^2 k^2}{v^2}(\Delta_1^2 + \Delta_2^2)  = 0. \label{eq:cubic_dispersion}
\end{equation}
The sound speed of the Goldstone can be extracted directly from this equation, without first solving explicitly for the gapless branch of $\omega^2(k)$. We differentiate Eq.~\eqref{eq:cubic_dispersion} with respect to $k^2$ and, using \eqref{eq:omega2_series}, evaluate the result in the limit $\omega^2-k^2\to0$ as $k^2\to0$. Doing so, we find
\begin{equation}
c_s^2 = 1 - \frac{2\alpha^2}{v^2} \left(\frac{1}{\Delta_1^2} + \frac{1}{\Delta_2^2}\right).
\end{equation}

We now proceed to solve \eqref{eq:cubic_dispersion} explicitly. The discriminant of this cubic equation is positive for any $\gamma \in (0,1/4]$ and real $k$. This places us in the \emph{casus irreducibilis} of Cardano's formula, according to which we have three real distinct roots, which take a trigonometric form:
\begin{equation}
\omega_n^2(k) = k^2 + \frac{2}{3}\Delta^2 + 2\sqrt{-\frac{p}{3}} \cos \left[\frac{1}{3}\arccos \left(\frac{3q}{2p}\sqrt{-\frac{3}{p}}\right)-\frac{2n\pi}{3}\right], \quad n =0,1,2,
\label{eq:omega_u1^2}
\end{equation}
where we have introduced
\begin{align}
p &= -\frac{4\Delta^4}{3}(1-3\gamma) - 4(1-c_s^2)k^2\gamma\Delta^2, \label{eq:p_u1^2_redef}\\
q &= \frac{4(1-c_s^2)k^2\gamma\Delta^4}{3} - \frac{8\Delta^6}{27}(2-9\gamma),\label{eq:q_u1^2_redef}
\end{align}
together with the compact notation
\begin{equation}
\begin{split}
\Delta^2 &\equiv \frac{\Delta_1^2 + \Delta_2^2}{2}, \\
\gamma &\equiv \frac{\Delta_1^2 \Delta_2^2}{(\Delta_1^2 + \Delta_2^2)^2} = \frac{1}{4} \frac{\Delta_1^2 \Delta_2^2}{\Delta^4}.
\end{split}
\end{equation}
Notice that for physical values of the masses, $\gamma$ ranges from $0$ (one mass vanishes or the other diverges) to $1/4$ (degenerate masses). By sending $k \to 0$, we find that $n = 2$ corresponds to the gapless mode, while $n = 1$ corresponds to $\min (\Delta_1^2, \Delta_2^2)$ and $n = 0$ corresponds to  $\max (\Delta_1^2, \Delta_2^2)$.
%Written like this, it is not obvious a priori which branch is gapless. However, at $k=0$ Eq.~\eqref{eq:cubic_dispersion} factorizes as
%\begin{equation}
%\omega^2(\omega^2-\Delta_1^2)(\omega^2-\Delta_2^2)= 0,
%\end{equation}
%so the three roots are exactly $\omega^2=0,\Delta_1^2,\Delta_2^2$. By definition, the gapless mode is the one that vanishes at $k=0$; since $\Delta_1^2,\Delta_2^2>0$, this branch is unambiguously identified, and corresponds to $n=2$ in Eq.~\eqref{eq:omega_u1^2}. The remaining two roots, $n=0,1$, correspond to the massive radial modes.

Taking the square root of the gapless mode's dispersion relation, $\omega_2^2(k)$ in \eqref{eq:omega_u1^2}, we find
\begin{equation}
      \omega_2(k) = \pm \sqrt{k^2 + \frac{2}{3}\Delta^2 +  2\sqrt{-\frac{p}{3}} \cos \left[\frac{1}{3}\arccos \left(\frac{3q}{2p}\sqrt{-\frac{3}{p}}\right)-\frac{4\pi}{3}\right]}.
      \label{eq:omega_u1sq}
\end{equation}
In what follows we investigate the analytic structure of $\om_2(k)$ and $\om_2^2(k)$.
\paragraph{\texorpdfstring{$\om^2(k)$}{omega2(k)}-singularities}
\mbox{}\\
The branch points of $\omega_2^2(k)$ arises from the arccosine which occurs when
\begin{equation}
\frac{3q}{2p}\sqrt{-\frac{3}{p}} = \pm 1. \label{eq:argument_arccos}
\end{equation}
Squaring and substituting \eqref{eq:p_u1^2_redef}, \eqref{eq:q_u1^2_redef}, we find
\begin{equation}
     16 Z^3 + 9 Z^2 - 6(27\gamma - 7)Z + (27\gamma - 7)^2 = 0, \label{eq:cubic_Z}
\end{equation}
where we have defined 
\begin{equation}
    Z \equiv (3 \gamma - 1) - 3\gamma (1 - c_s^2) \frac{k^2}{\Delta^2}.
\end{equation}
Therefore, every solution of Eq.~\eqref{eq:cubic_Z} corresponds to a branch point at
\begin{equation}
k^2 = \frac{(3\gamma - 1)-Z}{3\gamma} \frac{\Delta^2}{ (1 - c_s^2)}. \label{eq:ki^2_u1^2}
\end{equation}
The discriminant of Eq.~\eqref{eq:cubic_Z} is given by
    \begin{equation}
        D_Z = -46656 (-1 + 4\gamma) (-7 + 27\gamma)^3
        \label{eq:discriminant_cubic_Z}
    \end{equation}
   which means that \eqref{eq:cubic_Z} has three real roots for $\gamma=1/4$ ($D_Z=0$) and one real root, together with a complex-conjugate pair, for $0<\gamma<1/4$ ($D_Z<0$). Each root is labeled by the pair of colliding branches: $Z_{21}$ denotes the real root associated with the collision of $\omega_2^2$ and $\omega_1^2$, while $Z_{10}$ and its complex conjugate $\bar Z_{10}$ denote the roots associated with the collision of $\omega_1^2$ and $\omega_0^2$.
   For later convenience, we just report the value of the real root:
   \begin{equation}
         Z_{21} = -\frac{3}{16} + \sqrt[3]{-\frac{x}{2} + \frac{1}{512}\sqrt{-\frac{D_Z}{27}}} + \sqrt[3]{-\frac{x}{2} - \frac{1}{512}\sqrt{-\frac{D_Z}{27}}},
   \end{equation}
   with $D_{Z}$ defined as in \eqref{eq:discriminant_cubic_Z} and 
   \begin{equation}
        x \equiv \frac{729}{16}\gamma^2 - \frac{2781}{128}\gamma + \frac{5291}{2048}.
   \end{equation}
Two limiting cases of $\gamma$ are worth highlighting:
\begin{enumerate}
\item $\gamma=1/4$, i.e.~the degenerate mass limit: Eq.~\eqref{eq:cubic_Z} factorizes as
\begin{equation}
\frac{1}{16}(1+4Z_{10})^2(1+16Z_{21}) = 0,
\end{equation}
giving $Z_{10}=\bar Z_{10}=-1/4$ and $Z_{21}=-1/16$. The first solution corresponds to $k_{10}^2=0$, which is not a branch point. The second corresponds to $k_{21}^2 = -\Delta^2/[4(1-c_s^2)]$, which is the branch point of the $U(1)$ theory lying on the imaginary axis (See Section \ref{subsec:u1_superfluid} for more details).

\item \textbf{$\gamma\to0$:} Eq.~\eqref{eq:cubic_Z} factorizes as
\begin{equation}
    (1+Z_{21})(16Z_{10}^2-7Z_{10}+49)=0,
\end{equation}
with $Z_{21}=-1$ giving a branch point on the imaginary axis. Expanding $Z_{21}\approx-1+3\gamma+\mathcal{O}(\gamma^2)$ shows $k_{21}^2\propto\mathcal{O}(\gamma)$, so this branch point smoothly approaches $0$ as $\gamma\to0$; the remaining branch points, for which $k_{10}^2\propto\mathcal{O}(1/\gamma)$, instead, recede to infinity.
\end{enumerate}

To conclude, notice that the square roots $\sqrt{-p/3}$ and $\sqrt{-3/p}$ may appear to develop branch points at $p=0$, but this is not the case. Indeed, at $p=0$ Eq.~\eqref{eq:cubic_dispersion} still has three distinct roots in $\omega^2-k^2$. The apparent singularity at $p=0$ is therefore an artefact of the explicit trigonometric solution, not a genuine feature of $\omega_2(k)$: taking the limit $p\to0$ directly in Eq.~\eqref{eq:omega_u1^2} shows that $\omega_2^2$ remains analytic there.
\paragraph{\texorpdfstring{$\om(k)$}{omega(k)}-singularities}
\mbox{}\\
These occur when $\omega_2^2(k) = 0$. Looking at the gapless branch in Eq.~\eqref{eq:omega_u1^2}, this condition can be rewritten as
    \begin{equation}
         \cos \left\{\frac{1}{3}\left[\arccos \left(\frac{3q}{2p}\sqrt{-\frac{3}{p}}\right) - 4\pi\right]\right\} = \frac{1}{2} \sqrt{-\frac{3}{p}} \left(-k^2 - \frac{2}{3}\Delta^2\right)
    \end{equation}
Using the triple-angle formula 
\begin{equation}
    \cos(3\theta) = 4\cos^3\theta -3\cos\theta, \quad \theta \equiv \frac{1}{3}\left[\arccos \left(\frac{3q}{2p}\sqrt{-\frac{3}{p}}\right) - 4\pi\right]
\end{equation}
and Eqs.~\eqref{eq:p_u1^2_redef},~\eqref{eq:q_u1^2_redef}, we find 
\begin{equation}
     k^2 \left\{k^4 + 2\Delta^2 k^2 + 4\gamma \Delta^2 \left[\Delta^2 - (1 - c_s^2)k^2 \right] - 4 \gamma \Delta^4  (1 - c_s^2) \right\} = 0.
     \label{eq:branch_point_omega_eq}
\end{equation}
Apart from the trivial $k^2 = 0$ solution, which is not a genuine branch point, but simply reflects the point around which $\omega_2(k)$ is expanded, the solutions of \eqref{eq:branch_point_omega_eq} are 
\begin{equation}
      k_{\pm}^2 = -\Delta^2 \left[1- 2 \gamma (1 - c_s^2)\right]\left[1 \pm \sqrt{1 - \frac{4c_s^2 \gamma}{[1 - 2\gamma(1 - c_s^2)]^2}}~\right]. \label{eq:k_pm_u1^2}
\end{equation}
Notice that at $\gamma=1/4$ the two radial masses become degenerate and, as a consequence, the field $h_2$ decouples completely\footnote{Although the rotational matrix $O$ of Eq.~\eqref{eq:rotation_matrix} defines $H_1,H_2$ as the mass eigenstates for generic $\gamma$, at $\gamma=1/4$ the mass matrix becomes proportional to the identity and is left invariant by any further rotation of $(H_1,H_2)$. This residual freedom is what allows the coupling to the phonon to be diagonalized as well and the rotation that achieves this is exactly the inverse of $O$, which restore the original fields $h_1,h_2$. In this basis, $h_1$ is the only field coupled to the phonon, while $h_2$ decouples.}: $k^2_{-}=-c_s^2\Delta^2$ is then the branch point of the Goldstone mode itself, matching the single-field $U(1)$ case, discussed in Section \ref{subsec:u1_superfluid}, while $k^2_{+}=-\Delta^2$ is simply where the decoupled $h_2$'s own free dispersion vanishes.
\newcommand{\RtwoPanel}[7]{%
\begin{tikzpicture}[x=0.6\linewidth, y=0.42\linewidth, >=latex]
  \def\g{#1}\def\F{#2}\def\cs{#3}\def\rs{#4}\def\xcut{#5}
  \def\xmax{1.15}\def\ymax{1.65}
 
  % axes
  \draw[->, thick] (0,0) -- (\xmax,0) node[right] {$c_s$};
  \draw[->, thick] (0,0) -- (0,\ymax) node[above] {$R_2/\Delta,\ \tilde R_2/\Delta$};
 
  % dashed guide lines to the tangency point (c_s^*, R_2^*)
  \draw[gray, dashed] (0,\rs) -- (\cs,\rs) -- (\cs,0);
 
  % bottom branch of R_2: 0 <= c_s <= c_s^*
  \draw[very thick] plot[domain=0:\cs, samples=80]
    (\x,{sqrt(max(0,(1-2*\g*(1-\x*\x)) - sqrt(max(0,(1-2*\g*(1-\x*\x))^2 - 4*\x*\x*\g))))});
 
  % top branch of R_2: c_s^* <= c_s <= cut-off (two segments for the steep rise)
  \draw[very thick] plot[domain=\cs:0.9, samples=60] (\x,{\F/sqrt(1-\x*\x)});
  \draw[very thick] plot[domain=0.9:\xcut, samples=60] (\x,{\F/sqrt(1-\x*\x)});
 
  % tilde R_2: top-branch function over the full range 0 <= c_s <= cut-off
  \draw[very thick, red, dash pattern=on 9pt off 6pt]
    plot[domain=0:0.9, samples=80] (\x,{\F/sqrt(1-\x*\x)})
    -- plot[domain=0.9:\xcut, samples=60] (\x,{\F/sqrt(1-\x*\x)});
 
  % tangency point
  \fill (\cs,\rs) circle[radius=1.6pt];
 
  % axis labels
  \node[below left] at (0,0) {$0$};
  \node[below] at (1,0) {$1$};
  \node[below] at (\cs,0) {$#6$};
  \node[left] at (0,\rs) {$#7$};
\end{tikzpicture}}
 
\begin{figure}[H]
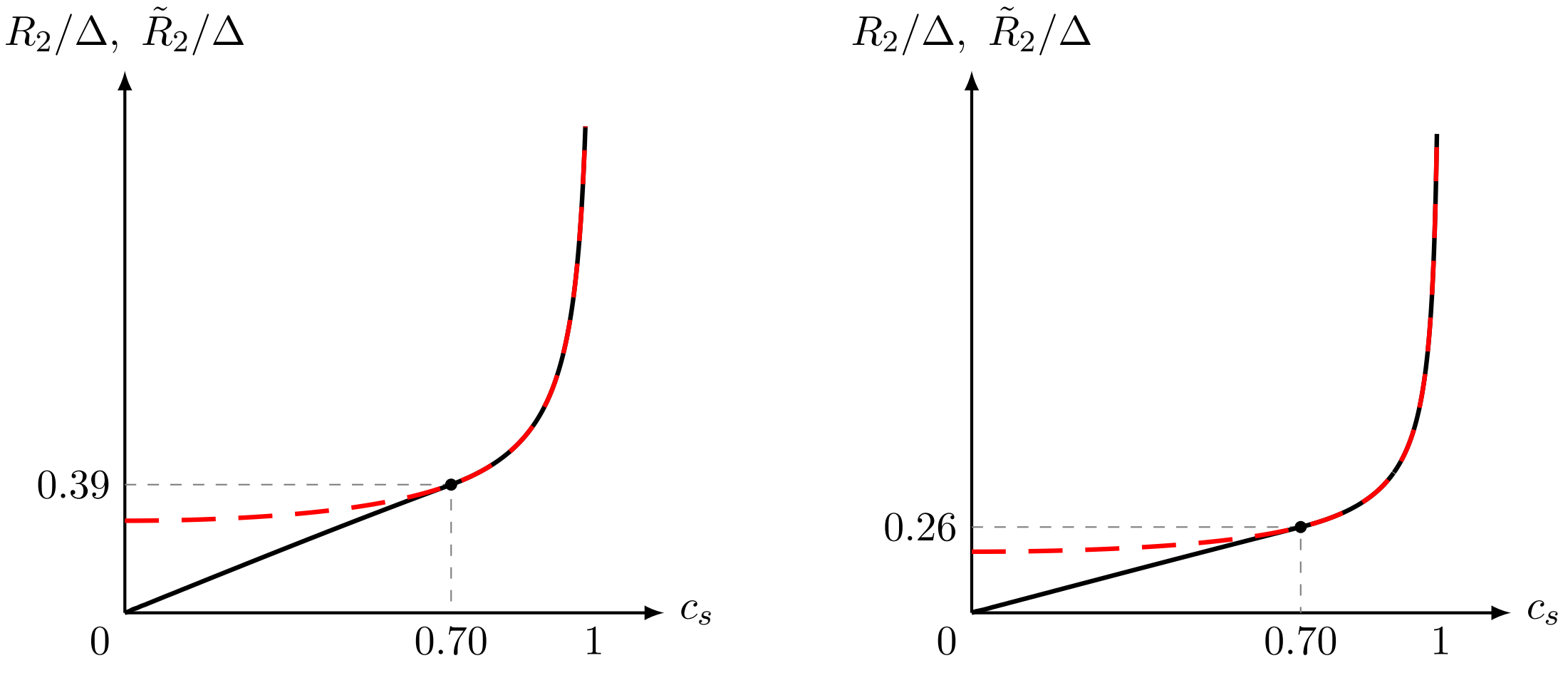

  \centering
    \begin{subfigure}[b]{0.48\textwidth}
    \centering
    \RtwoPanel{0.125}{0.280262}{0.695621}{0.390115}{0.9824}{0.70}{0.39}
    \caption{$\gamma=1/8$}
  \end{subfigure} \hfill
    \begin{subfigure}[b]{0.48\textwidth}
    \centering
    \RtwoPanel{0.0625}{0.186065}{0.701411}{0.261049}{0.9923}{0.70}{0.26}
    \caption{$\gamma=1/16$}
  \end{subfigure}
  \caption{Radius of convergence of the gapless dispersion relation $\omega_2(k)$ ($R_2$, solid black) and of its square $\omega_2^2(k)$ ($\tilde R_2$, dashed red), as a function of the sound speed $c_s$ for selected values of $\gamma$. Note that when $\gamma = 1/4$, the radius of convergence reduces back to the $U(1)$ case as shown in Figure \ref{fig:radius_u1}.}
  \label{fig:R2_vs_tilde_R2}
\end{figure}
\begin{figure}[p]   % 'p' = float page: only floats, no text
  \centering
  \begin{subfigure}[t]{\textwidth}
    \centering
    \caption{Bounds on $\omega(k)$}
    \label{fig:om_u1^2checks}
    \includegraphics[width=0.9\textwidth]{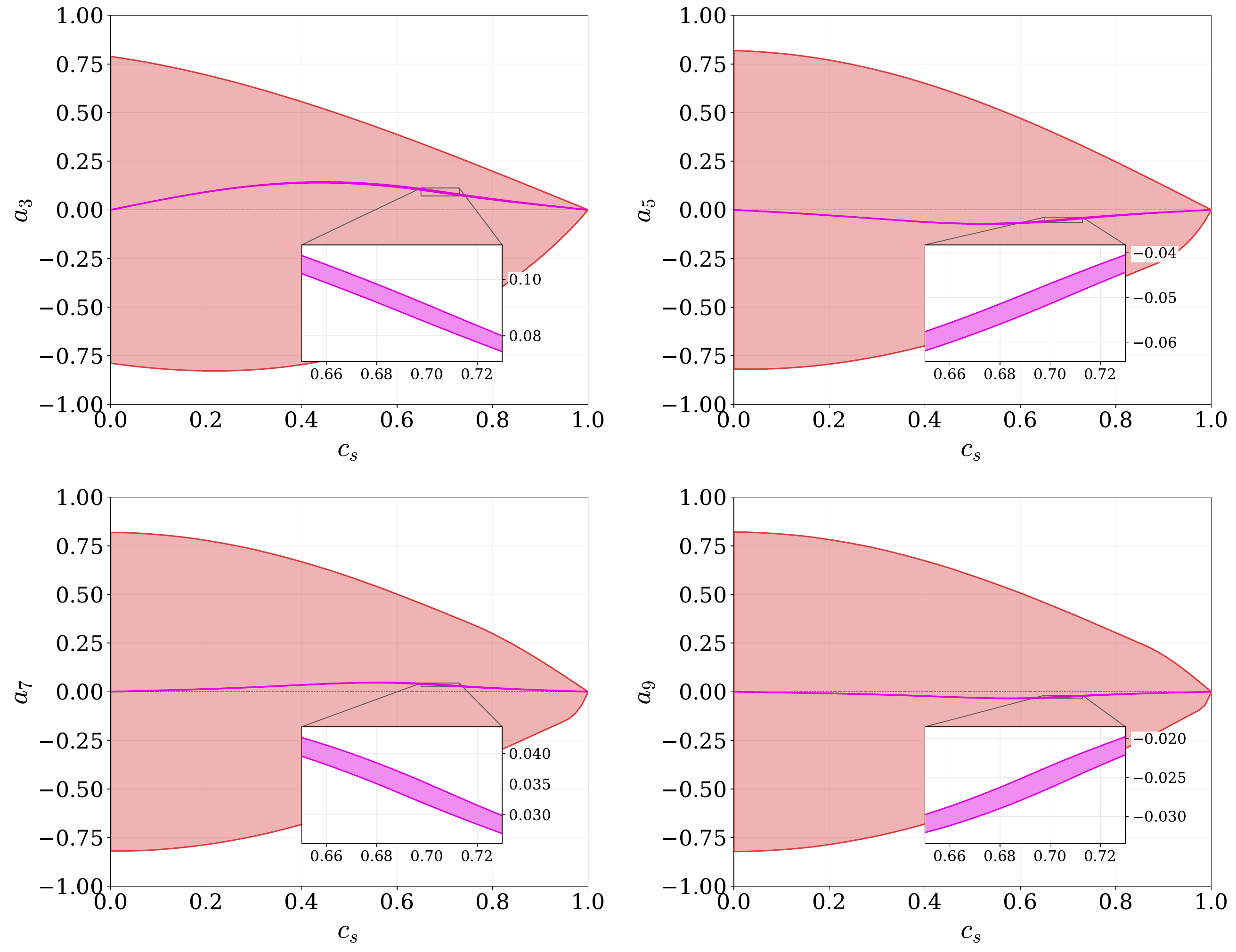}
  \end{subfigure}
  \begin{subfigure}[t]{\textwidth}
    \centering
       \caption{Bounds on $\omega^2(k)$}
    \label{fig:om2_u1^2_checks}
    \includegraphics[width=0.9\textwidth]{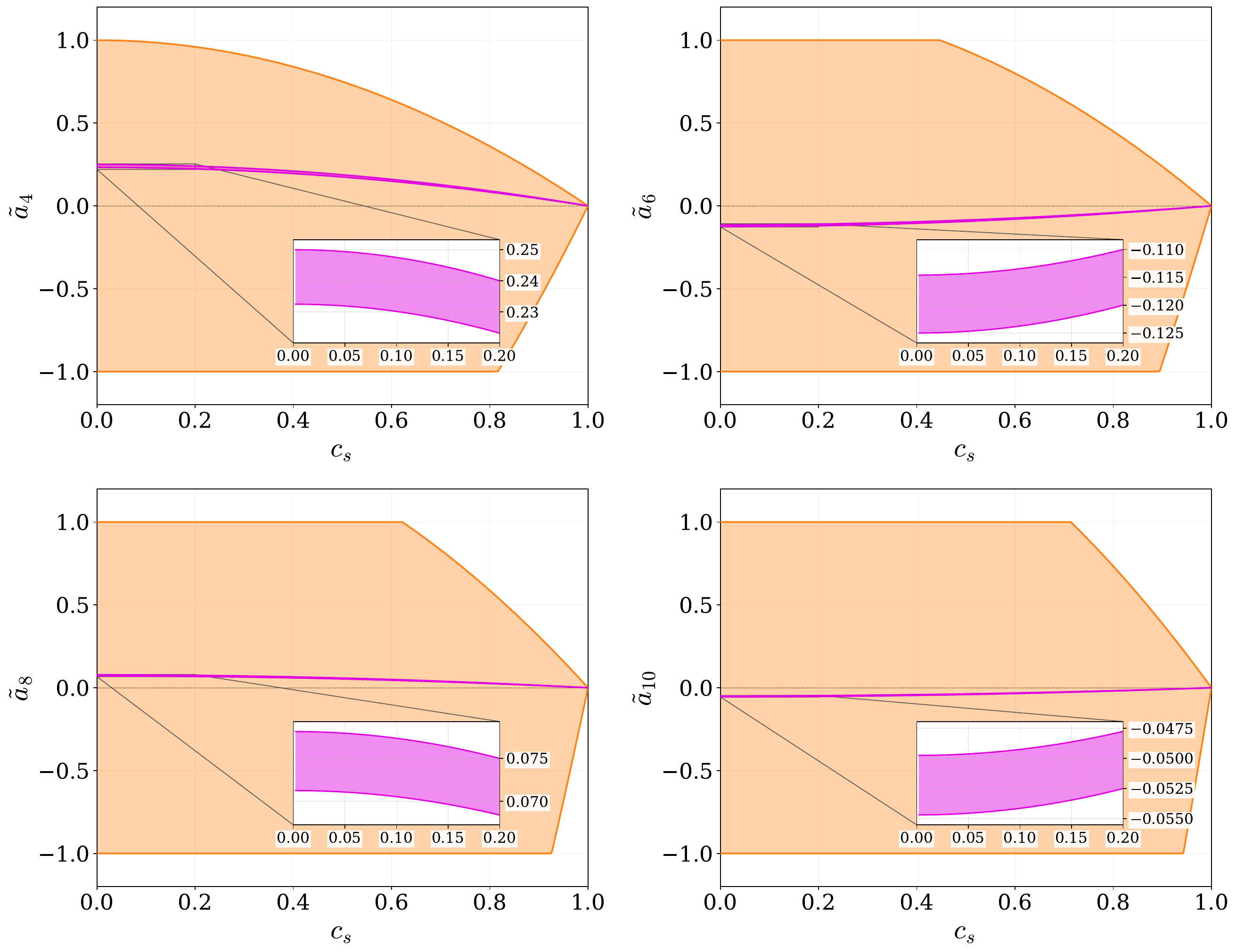}
  \end{subfigure}
  \caption{Bounds on the $U(1)^2$ solid.}
  \label{fig:tested_bounds_u1^2_all}
\end{figure}

%The radius of convergence of $\omega_2(k)$ around $k=0$ is set by the distance, in the complex $k$-plane, from the origin to the nearest singularity. Two branch points compete for this role: $k_{21}$, the value of $k^2$ in Eq.~\eqref{eq:ki^2_u1^2} evaluated at $Z=Z_{21}$, and $k_-$, defined in Eq.~\eqref{eq:k_pm_u1^2}. The other root $k_+$ in Eq.~\eqref{eq:k_pm_u1^2} is out of the game: it always lies farther from the origin than $k_{-}$ and $k_{21}$. Equating $|k_-|$ and $|k_{21}|$ we find that the two coincide at $c_s=c_s^*$, defined by
The radius of convergence of the dispersion relation is given by
\begin{equation}
R = \begin{cases}
\Delta \sqrt{1 - 2\gamma (1 - c_s^2) - \sqrt{[1 - 2\gamma (1 - c_s^2)]^2 - 4c_s^2 \gamma }}, & 0 \leq c_s \leq c_s^{*}\\[2mm]
\sqrt{\frac{Z_{21} - (3\gamma -1)}{3\gamma}} \frac{\Delta}{\sqrt{1 - c_s^2}}, & c_s^{*} < c_s \leq 1.
\end{cases} \label{eq:radius_u1^2}
\end{equation}
in which
\begin{equation}
c_s^{*} \equiv \sqrt{1 - \frac{2(Z_{21} + 1) + \sqrt{2(27\gamma - 7)+2Z_{21}(2Z_{21} - 5)}}{18\gamma}},
\end{equation}
and 
\begin{equation}
\tR = \sqrt{\frac{Z_{21} - (3\gamma -1)}{3\gamma}} \frac{\Delta}{\sqrt{1 - c_s^2}}, \qquad 0 \leq c_s \leq 1.
\end{equation}
We illustrate this behaviour in Figure \ref{fig:R2_vs_tilde_R2}. The bounds are checked in Figure \ref{fig:tested_bounds_u1^2_all}. The uppermost part of the band corresponds to $\gamma = 1/4$, but also for $\gamma \to 0$ as well. In this limit, one of the masses becomes very large, and we get (without loss of generality we take $\Delta_2 \ll \Delta_1$)
\begin{align}
    \Delta &= \sqrt{\frac{\Delta_1^2 + \Delta_2^2}{2}} \to \frac{\Delta_1}{\sqrt{2}} \label{eq:Delta_gamma0}\\
    \gamma &= \frac{1}{4} \frac{\Delta_1^2 \Delta_2^2}{\Delta^4} \to \frac{\Delta_2^2}{\Delta_1^2} \label{eq:gamma_gamma0}
\end{align}
and in this limit, one sees that $\alpha_{2n+1}$ and the radius of convergences $R$ and $\Tilde{R}$ reduces back to $U(1)$ case with the mass gap determined by the lighter mass. In other words, the heavier mass $\Delta_1$ decouples from the system in this limit. This also explains why the coefficients form a very narrow band.

\section{Check of the gapped bounds\label{app:checkmassive}}
We here test the bounds in \eqref{eq:massive_bounds_inter} for the radial mode of the $U(1)$ superfluid model. 

Consider the square of the gapped dispersion relation in Eq.~\eqref{eq:dispersion_u1}. Expanding it for $k/\Delta \ll 1$, we obtain
\begin{equation}
    \omega^2(k) = m^2 + c_g^2k^2 + \sum_{n = 2}^{\infty}\tilde{\alpha}_{2n}k^{2n}, \quad \tilde{\alpha}_{2n} \equiv (-1)^{n+1}C_{n-1}\frac{(c_g^2-1)^n}{m^{2n-2}}, \quad c_g^2 \equiv 2-\cs^2,
\end{equation}
where we have relabeled the gap $\Delta$ with $m$ for the sake of clarity and $C_{n}$ are the Catalan numbers\footnote{The $n$-th Catalan number is defined in terms of the central binomial coefficients as $C_n = \frac{1}{n+1}\begin{pmatrix}
    2n\\n
\end{pmatrix} = \frac{(2n)!}{(n+1)!n!}, \, n \geq 0$.}.
In this model the gapped speed is a function of the sound speed $\cs \in [0,1]$ and, as a consequence, $c_{g}^2 \in [1,2]$. This is consistent with the bound in Eq.~\eqref{eq:cg_bound}. Indeed, using the $\om^2$-radius of convergence $\tilde{R}$ in Eq.~\eqref{eq:radius_u1} and the definition of $c_g^2$ in terms of $c_s^2$, \eqref{eq:cg_bound} reduces to 
\begin{equation}
    c_g^2 \geq 1.
\end{equation}
Given that $\tilde{a}_{2n} \equiv \tilde{\alpha}_{2n}\tR^{2n-2}$ and, for this specific model, $c_*^2 = 4c_g^2-3$, we can rewrite the two-sided bounds in Eq.~\eqref{eq:massive_bounds_inter} for the U(1) superfluid gapped mode as 
\begin{equation}
    \max[-3(2n-1)(c_g^2-1),7-8c_g^2] \leq \frac{(-1)^{n+1}}{4^{n-1}}C_{n-1}(c_g^2-1) \leq \min[3k_{2n-1}(c_g^2-1),8c_g^2-7],
\end{equation}
where we recall the definition of $k_{2n-1}$ in \eqref{eq:kappa_2n+1}. Before proceeding, it is easy to check that the Catalan numbers satisfy 
\begin{equation}
    \frac{C_{n-1}}{4^{n-1}} \leq \frac{C_1}{4} = \frac{1}{4}.
\end{equation}
Given this, we can write 
\begin{equation}
    |\tilde{a}_{2n}| = \frac{C_{n-1}}{4^{n-1}}(c_g^2-1) \leq \frac{c_g^2-1}{4}.
\end{equation}
It is thus sufficient to prove that for any $n \geq 2$:
\begin{align}
    \frac{c_g^2-1}{4} &\leq \min[3k_{2n-1}(c_g^2-1),8c_g^2-7], \label{eq:upper_bound_gapped}\\
    \frac{1-c_g^2}{4} &\geq \max[-3(2n-1)(c_g^2-1),7-8c_g^2] \label{eq:lower_bound_gapped}.
\end{align}
\paragraph{Upper bound}
Let us first start proving the bound in \eqref{eq:upper_bound_gapped}.
Take for instance $\theta_0 = 3\pi/[2(2n-1)]$, which satisfies $0 < \theta_0 \leq \pi/2$ for any $n \geq 2$. Given this, we can write 
\begin{equation}
    k_{2n-1} \geq -\frac{\sin[(2n-1)\theta_0]}{\sin(\theta_0)} \geq 1. 
\end{equation}
With this result in hand, it is easy to verify that \eqref{eq:upper_bound_gapped} is satisfied for any $n \geq 2$ and $c_g^2 \in [1,2]$.

\paragraph{Lower bound}
For what concerns the bound in Eq.~\eqref{eq:lower_bound_gapped}, we can first of all notice that $1-2n \leq -3$ for any $n \geq 2$ and, as a direct consequence, \eqref{eq:lower_bound_gapped} is verified for any $c_g^2 \in [1,2]$ and $n \geq 2$.
Hence, we have proven analytically that the U(1) superfluid gapped mode satisfies the bounds in \eqref{eq:massive_bounds_inter}.

%\paragraph{Note added.} This is also a good position for notes added
%after the paper has been written.

% Bibliography
%% [A] Recommended: using JHEP.bst file
\bibliographystyle{JHEP}
\bibliography{biblio.bib}

%% or
%% [B] Manual formatting (see below)
%% (i) We suggest to always provide author, title and journal data or doi:
%% in short all the informations that clearly identify a document.
%% (ii) please avoid comments such as "For a review'', "For some examples",
%% "and references therein" or move them in the text. In general, please leave only references in the bibliography and move all
%% accessory text in footnotes.
%% (iii) Also, please have only one work for each \bibitem.
\end{document}